\documentclass{aa}  

\usepackage{pgffor}
\usepackage{amssymb,times,graphicx, subfigure}
\usepackage[english]{babel}
\usepackage[varg]{txfonts}
\usepackage{xspace}
\usepackage{longtable,lscape,supertabular}
\usepackage{rotating}

\usepackage{natbib,afterpage,twoopt}
\bibpunct{(}{)}{;}{a}{}{,} %% natbib format for A&A and ApJ
\def\irc{IRC\,+10\,216\xspace}
\def\iktau{IK~Tau\xspace}
\def\vycma{VY~CMa\xspace}
\def\cit6{CIT~6\xspace}

\def\rdor{R~Dor\xspace}
\def\um{$\mu$m\xspace}
\def\soo{SO$_2$\xspace}
\def\water{H$_2$O\xspace}

\def\nutwo{$\nu_2$\xspace}

\def\rstar{$R_{\star}$\xspace}
\def\msun{$M_{\sun}$\xspace}
\def\lsun{$L_{\sun}$\xspace}
\def\msunyr{$M_{\sun}$\,yr$^{-1}$\xspace}
\def\kms{km\,s$^{-1}$\xspace}

\def\vlsr{$\varv_{\mathrm{LSR}}$\xspace}

\def\irames{IRAM~30\,m\xspace}
\def\fwhm{$FWHM$\xspace}

\begin{document}

   \title{Circumstellar environment of the M-type AGB star \rdor}
   \subtitle{APEX spectral scan at 159.0\,--\,368.5\,GHz \thanks{This publication is based on data acquired with the Atacama Pathfinder Experiment (APEX). APEX is a collaboration between the Max-Planck-Institut f\"{u}r Radioastronomie, the European Southern Observatory, and the Onsala Space Observatory.}\fnmsep\thanks{The FITS file containing the fully reduced spectrum presented in this paper is available at the CDS via anonymous ftp to cdsarc.u-strasbg.fr (130.79.128.5) or via http://cdsweb.u-strasbg.fr/cgi-bin/qcat?J/A+A/.}}

   \author{E. De Beck
          \and H. Olofsson
          }

   \institute{Department of Space, Earth and Environment, Chalmers University of Technology, Onsala Space Observatory, 43992, Onsala, Sweden
                  \\ \email{elvire.debeck@chalmers.se}
              }

   \date{Received 15 December 2017 / Accepted 17 January 2018}

  \abstract
  % context heading (optional)
   {
Our current insights into the circumstellar chemistry of asymptotic giant branch (AGB) stars are largely based on studies of carbon-rich stars and stars with high mass-loss rates. 
   }
  % aims heading (mandatory)
   {
In order to expand the current molecular inventory of evolved stars we present a spectral scan of the nearby, oxygen-rich star R Dor, a star with a low mass-loss rate ($\sim2\times10^{-7}$\,\msunyr).
   }
  % methods heading (mandatory)
   {
   We carried out a spectral scan in the frequency ranges $159.0-321.5$\,GHz and $338.5-368.5$\,GHz (wavelength range $0.8-1.9$\,mm) using the SEPIA/Band-5 and SHeFI instruments on the APEX telescope and we compare it to previous surveys, including one of the oxygen-rich
AGB star \iktau, which has a high mass-loss rate ($\sim5\times10^{-6}$\,\msunyr).
   }
  % results heading (mandatory)
   {The spectrum of \rdor is dominated by emission lines of \soo and the different isotopologues of SiO. We also detect CO, \water,  HCN, CN, PO, PN, SO, and tentatively TiO$_2$, AlO, and NaCl. Sixteen out of approximately 320 spectral features remain unidentified. Among these is a strong but previously unknown maser at 354.2\,GHz, which we suggest could pertain to H$_2$SiO, silanone. With the exception of one, none of these unidentified lines are found in a similarly sensitive survey of \iktau performed with the \irames telescope. We present radiative transfer models for five isotopologues of SiO ($^{28}$SiO, $^{29}$SiO, $^{30}$SiO, Si$^{17}$O, Si$^{18}$O), providing constraints on their fractional abundance and radial extent. We derive isotopic ratios for C, O, Si, and S and estimate that, based on our results for $^{17}$O/$^{18}$O, \rdor likely had an initial mass in the range $1.3-1.6$\,\msun, in agreement with earlier findings based on models of H$_2$O line emission. From the presence of spectral features recurring in many of the measured thermal and maser emission lines we tentatively identify up to five kinematical components in the outflow of \rdor, indicating deviations from a smooth, spherical wind.
   }
  % conclusions heading (optional)
  {}
  

   \maketitle

% % % % % % % % % % % % % % % % % % % % % % % % % % % % % % % % % % % % % % % % % % % % % % % % % % % %

\begin{figure*}[!h]
\includegraphics[width=\linewidth,trim={0.8cm 0 0 0},clip]{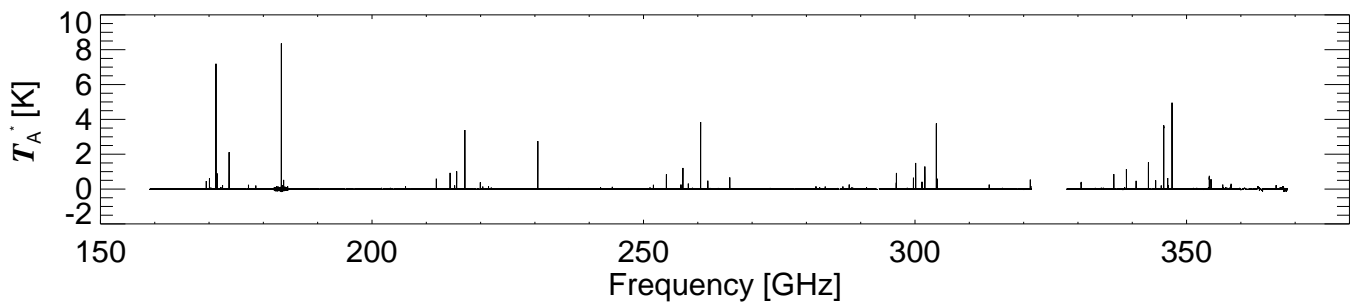}%{RDOR_COMBO}
\caption{APEX survey of \rdor in the range $159.0-368.5$\,GHz. \label{fig:fullscan_summary}}
\end{figure*}

\section{Introduction}\label{sect:introduction}
In order to get a comprehensive view of the physical and chemical properties of the circumstellar envelopes (CSEs) of asymptotic giant branch (AGB) stars and to quantify their return to the interstellar medium, it is necessary to characterise the gas and dust contents of these outflows. The set-up of an inventory of chemical species present in the CSEs of evolved stars has predominantly used, as reference, the nearby, high-mass-loss-rate, carbon-rich AGB star \irc (also commonly referred to as \object{CW Leo}), skewing the community's knowledge towards the carbon-rich chemistry, both observationally and theoretically. A similar wealth of observational constraints on the circumstellar chemistry of M-type AGB stars, often called oxygen-rich given their atmospheric C/O$<1$, is currently still lacking.

Unbiased spectral surveys, scanning broad frequency ranges without pre-selecting particular molecules of interest, are excellent tools to set up such inventories. Several such surveys using single-dish telescopes have been presented for high-mass-loss-rate, carbon-rich AGB stars:  \irc \citep{cernicharo2000_irc10216_survey_2mm, cernicharo2010_irc10216_survey_hifi}, CIT~6 \citep[\object{RW LMi};][]{zhang2009_cit6}, and CRL~3068  \citep[\object{LL Peg};][]{zhang2009_crl3068}. \citet{patel2011} presented an interferometric survey of \irc obtained with the Submillimeter Array (SMA). \citet{kaminski2013_vycma_sma} reported on a similar SMA line-imaging survey of the red supergiant \object{VY CMa}, a star with an oxygen-rich composition, an extremely high mass-loss rate of several $10^{-4}$\,\msunyr, and a geometrically and chemically complex CSE \citep[e.g.][]{humphreys2007_vycma_3dmorphology_kinematics, richards2014_alma_vycma, ziurys2007_vycma_complexity}.  \citet{debeck2015_AGBconfproceedings} presented an overview of an interferometric spectral survey of the Mira-type, high-mass-loss-rate, M-type star \object{IK Tau} in the range $279-355$\,GHz carried out with the SMA. A forthcoming publication (De Beck et al., \emph{in prep.}) will report on this survey in its entirety, including a discussion on the extent and geometry of the imaged line emission. Based partially on this survey, \citet{debeck2013_popn} presented first results on phosphorus-bearing molecules in the CSE of an M-type AGB star and highlighted the need for updated chemical models for these CSEs.  \citet{velillaprieto2017_iktau_iram} recently presented a spectral scan of \iktau obtained with the IRAM 30\,m telescope covering the ranges $79-116$, $128-175$, and $202-356$\,GHz. They reported a very rich chemical content, including several carbon-bearing molecules, such as H$_2$CO and HCO$^+$, and several nitrogen-bearing molecules, such as NS and NO. The abundances they derive for several of the detected species further underscore the need for improved models of the chemistry around oxygen-rich stars. 

We present here for the first time a spectral scan of a low-mass-loss-rate M-type AGB star, \rdor, enabling a direct comparison of the chemical contents of CSEs  representative of two significantly different density regimes. \rdor is a nearby \citep[59\,pc;][]{knapp2003} AGB star with a luminosity of about 6500\,\lsun \citep{maercker2016_water}. It shows a semi-regular pulsation pattern with two periods of 175 and 332\,days, respectively \citep{bedding1998}. Studies of its circumstellar environment have mainly focussed on CO and \water\/ line emission \citep[e.g.][]{ramstedt2014_12co13co,maercker2016_water}, constraining its mass-loss rate to $1-2\times10^{-7}$\,\msunyr. Recently, \citet{vandesande2017_rdor} reported on a detailed abundance analysis of SiO and HCN.

We list the details of the observations in Sect.~\ref{sect:observations}, present and discuss the results in Sect.~\ref{sect:results}, and provide a list of unidentified spectral features in Sect.~\ref{sect:unidentified}. We compare our results to those obtained from other line surveys, in particular of \iktau, in Sect.~\ref{sect:comparison}, discuss possible deviations in the outflow from spherical symmetry in Sect.~\ref{sect:kinematics}, and present our conclusions in Sect.~\ref{sect:conclusion}. Appendix~\ref{sect:COvariability} addresses possible time variability of the CO line emission. Appendix~\ref{sect:maserobs} addresses maser line variability and includes a list of detected masers. Appendix~\ref{sect:overview} provides an overview of the full survey.

\section{Observations}\label{sect:observations}
We used the Swedish Heterodyne Facility Instrument \citep[SHeFI;][]{vassilev2008_shefi} and band 5 of the Swedish-ESO PI Instrument \citep[SEPIA;][]{billade2012_band5} on the Atacama Pathfinder Experiment (APEX) telescope to carry out a spectral survey covering the frequency range $159.0-368.5$\,GHz. Due to excessive interference of water vapour in the atmosphere the survey does not cover the range $321.5-328.0$\,GHz. The SHeFI observations were carried out over a long period, spanning from May 2011 up to November 2015. All SEPIA observations were carried out on 22 and 23 November, 2015.

SHeFI is a single-sideband (sideband-separating), single-polarisation heterodyne receiver, whereas SEPIA is a sideband-separating, dual-polarisation heterodyne receiver. The backends for SHeFI changed during the observing campaign and were extended from an initial  $2\times1.0$\,GHz bandwidth (FFTS) to $2\times2.5$\,GHz (XFFTS), corresponding to instantaneous bandwidths of 1.9\,GHz and 4.0\,GHz, respectively, significantly increasing the observing efficiency. The SEPIA backends consist of $2\times2\times2\times2.5$\,GHz, covering both lower and upper sidebands simultaneously, in two polarisations. The SHeFI spectra were recorded at a nominal resolution of 122\,kHz when the FFTS was used, and at 76\,kHz when the XFFTS was used. The SEPIA spectra were recorded at 38\,kHz nominal resolution. Consecutive tunings were separated by 1.5\,GHz or 3.5\,GHz, depending on the backend in use, to ensure sufficient overlap.

The observations were carried out using wobbler switching with a standard beam throw of 50\arcsec. Due to some technical issues with the wobbler, some settings were observed in position switching mode, without compromising the final result. The main beam half power beamwidth of APEX varies in the range $17-39$\arcsec\/ over the observed frequency range. 

Data reduction is done using the \textsc{gildas/class}\footnote{\texttt{http://www.iram.fr/IRAMFR/GILDAS/}} package. We inspect scans individually and ignore those with very unstable baselines. Bad channels, which commonly appear in the outer $30-50$\,MHz of the band\footnote{\texttt{http://www.apex-telescope.org/backends/ffts/}} are blanked.  We subtract first-degree polynomial baselines from the individual scans, after masking relevant spectral features, and combine all data into one final spectrum.

In a few cases, we find aliases of strong lines of CO and SiO, in the outer $\sim60$\,MHz of the receiver backend. These can occur when the strong signal is very close to the edge of the XXFTS unit and gets folded into the band. Given the proximity to the band edge, and the overlapping observing settings, these aliases are removed from the spectra without losing spectral coverage.

The SHeFI receivers on APEX work in single-sideband mode with reported image-sideband suppression of typically $>15$\,dB.  The sideband-separating SEPIA instrument has an average image-sideband suppression of $18.5$\,dB. Some strong lines can, therefore, leak into the one sideband from the other for some frequency settings.  In case there is no overlapping spectral coverage from another frequency tuning available, we do not blank these lines in our spectrum, but clearly mark these leakages in the overview tables and plots. The leakage levels in our survey range from 0.3\% (25\,dB suppression of an 11\,K strong signal) up to 54\% (3\,dB) in the most extreme case, but are overall well in line with the reported suppression values. 

We assume overall calibration uncertainties of 20\% on the survey data, but argue in Sect.~\ref{sect:sio} that the internal uncertainties, that is, line-to-line uncertainties, are most likely lower than that. 

Unless stated otherwise, we show all spectra in $T_{\mathrm{A}}^*$, that is, the antenna temperature scale corrected for atmospheric losses and, for example, antenna spillover. Typical rms noise values on this scale vary from 3\,mK to 10\,mK at 2\,\kms resolution, with higher rms locally induced by the interference of atmospheric H$_2$O at 183\,GHz and 321\,GHz.  For conversion into flux units one can employ point-source sensitivities  $S_{\nu}/T_{\rm A}^*$ of 38\,Jy/K, 39\,Jy/K, and 41\,Jy/K for the SEPIA/band-5 ($159-211$\,GHz), SHeFI-1 ($213-275$\,GHz), and SHeFI-2 ($267-378$\,GHz) observations, respectively. Conversion to main-beam brightness temperature, $T_{\rm mb} = T_{\mathrm{A}}^*/ \eta_{\rm mb}$, used in this work for comparison to modelling results,  uses main-beam efficiencies $\eta_{\rm mb}$ of 0.68, 0.75, and 0.73 for the different bands, respectively. The values for SEPIA/band-5 are preliminary, but currently the most up-to-date  \citep{immer2016_sepia}. 

We claim detection of a spectral feature at signal-to-noise ratios (S/Ns) of at least 3 at a spectral resolution that leaves a minimum of five spectral bins covering the line. Additionally, tentative detections of lines at low S/N can be claimed in the case where a stacked spectrum reaches this S/N criterion (e.g. for PN, see below) or if one, or more, other emission lines of the same molecule are present elsewhere in the spectrum.

\section{Results}\label{sect:results}
Figure~\ref{fig:fullscan_summary} shows a low-resolution overview of the APEX line survey of \rdor. Figure~\ref{fig:fullscan} in the appendix shows the entire survey at 3\,MHz frequency resolution. Table~\ref{tbl:lineID} gives an overview of all emission features in the scan, while Table~\ref{tbl:permol} gives an overview of the identified lines per molecule. The spectrum is dominated by emission lines of SO$_2$ (134 lines), SiO (50 lines), and SO (28 lines), including isotopologues.

For line identification we use the Cologne Database for Molecular Spectroscopy\footnote{\texttt{https://www.astro.uni-koeln.de/cdms/}} \citep[CDMS;][]{mueller2001_cdms,mueller2005_cdms} and the catalogue\footnote{\texttt{http://spec.jpl.nasa.gov/}} for molecular line spectroscopy hosted by the Jet Propulsion Laboratory \citep[JPL;][]{pickett1998_jpl} as primary reference catalogues, with priority given to CDMS when entries are present in both. 

\begin{figure}
\includegraphics[width=\linewidth]{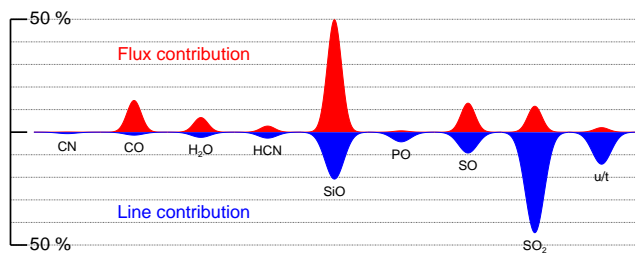}%{LineFluxStats_gauss}
\caption{Overview of the relative contribution per molecule (including isotopologues) to the survey in terms of flux (\emph{top}) and number of lines (\emph{bottom}). Unidentified and tentatively identified spectral features are grouped under the label ``u/t''. }\label{fig:contributions}
\end{figure}

\begin{table*} 
\caption{Overview of molecules in the APEX survey of \rdor. \label{tbl:permol}}
\centering
\begin{tabular}{lrp{14cm}}
\hline\hline\\[-2ex]
Molecule & Number  & Line numbers \\
 & of lines &  \\
\hline\\[-2ex]
CN	&	           2	&	         268,          270\\
CO	&	           2	&	          86,          289\\
$^{13}$CO	&	           2	&	          80,          243\\
H$_2$O	&	           2	&	          37,          239\\
H$_2$O, $v_2=1$	&	           5	&	          87,          141,          192,          202,          256\\
H$_2$O, $v_2=2$	&	           1	&	         149\\
HCN	&	           3	&	          34,          144,          303\\
H$^{13}$CN	&	           3	&	          27,          134,          286\\
PO	&	          12	&	          48,           49,           50,           51,           94,           95,           96,           97,          173,          174,          175,          176\\
SiO	&	           5	&	          29,           78,          136,          220,          295\\
SiO, $v=1$	&	           5	&	          26,           75,          132,          216,          284\\
SiO, $v=2$	&	           3	&	          22,          127,          212\\
SiO, $v=3$	&	           4	&	          20,           68,          203,          269\\
SiO, $v=5$	&	           1	&	         115\\
$^{29}$SiO	&	           5	&	          23,           71,          129,          213,          280\\
$^{29}$SiO, $v=1$	&	           4	&	          21,          123,          205,          272\\
$^{29}$SiO, $v=2$	&	           2	&	         120,          262\\
$^{29}$SiO, $v=3$	&	           2	&	         193,          254\\
$^{30}$SiO	&	           5	&	          19,           67,          121,          200,          265\\
$^{30}$SiO, $v=1$	&	           3	&	          18,          118,          257\\
Si$^{17}$O	&	           5	&	          16,           65,          112,          190,          251\\
Si$^{18}$O	&	           5	&	           5,           56,          100,          169,          331\\
$^{29}$Si$^{18}$O	&	           1	&	         161\\
SO	&	          16	&	          25,           36,           63,           74,           79,          117,          130,          138,          199,          215,          221,          226,          266,          273,          281,          293\\
$^{34}$SO	&	          12	&	          32,           57,           76,          108,          119,          126,          195,          206,          250,          260,          267,          306\\
SO$_2$	&	         116	&	           1,            4,            9,           11,           13,           14,           31,           43,           45,           46,           47,           52,           54,           55,           58,           60,           61,           64,           69,           72,           77,           81,           82,           83,           85,           88,           90,           91,           92,           98,           99,          102,          105,          106,          109,          111,          113,          114,          122,          124,          125,          128,          131,          133,          135,          140,          142,          145,          146,          152,          157,          158,          163,          164,          165,          167,          168,          171,          179,          186,          197,          198,          208,          211,          214,          217,          223,          227,          228,          229,          230,          231,          235,          236,          240,          244,          248,          252,          255,          258,          263,          264,          271,          274,          275,          276,          277,          279,          285,          292,          294,          296,          298,          299,          304,          305,          309,          311,          312,          313,          315,          316,          317,          318,          319,          320,          321,          323,          325,          327,          328,          329,          332,          333,          334,          336\\
SO$_2$, $v_2=1$	&	           1	&	         297\\
$^{34}$SO$_2$	&	           5	&	           7,           30,          104,          310,          330\\
SO$^{17}$O	&	           2	&	         233,          290\\
SO$^{18}$O	&	           9	&	         177,          178,          183,          184,          185,          218,          219,          283,          291\\
u	&	          16	&	          28,           44,           59,           84,          148,          156,          162,          172,          196,          201,          207,          225,          234,          237,          245,          302\\
\hline\\[-2ex]
\end{tabular}
\tablefoot{The columns list the molecule (vibrational states and isotopologues are listed separately), the number of lines per species, and the line numbers from Table~\ref{tbl:lineID} corresponding to the identified lines. Tentative identifications (e.g. of PN and TiO$_2$) and features resulting from data issues or image contamination are not listed in this table.}
\end{table*}
%{PerMolecule}

\begin{figure}[ht]
\includegraphics[width=\linewidth]{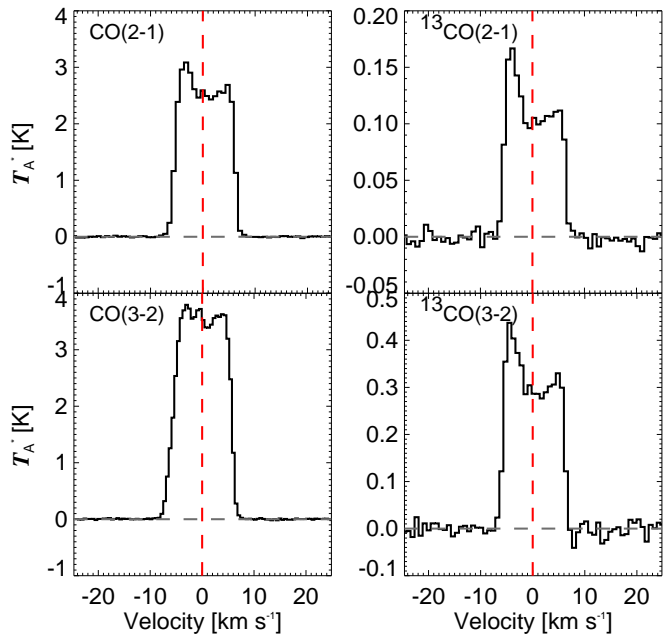}%{CO_zoom}
\caption{CO line emission.  \emph{Left:} $^{12}$CO, \emph{right:} $^{13}$CO. \label{fig:CO_zoom}}
\end{figure}
\subsection{Molecular line luminosity}\label{sect:luminosity}
The molecular line emission observed in the survey is dominated by only a few molecules. Approximately 50\% of the total line flux comes from SiO and its isotopologues, which accounts for 21\% of the emission features. Another 14\% of the flux comes from the four (1.3\%) CO lines. SO accounts for 13\% of the flux and 9\% of the lines. SO$_2$ accounts for another 12\% of the flux and 45\% of the lines. H$_2$O accounts for 6\% of the flux and 2\% of the lines. Other species each contribute less than 3\% of the total flux. Figure~\ref{fig:contributions} presents a visual summary of this.

\subsection{Molecular species}
Throughout the paper, we assume that the source's systemic velocity, \vlsr, is 6.5\,\kms, and plot profiles with respect to the stellar \vlsr, that is, centered around 0.0\,\kms. This value for the systemic velocity corresponds to individual peaks in, for example, the CO lines, the 321\,GHz H$_2$O maser, and many of the SO$_2$ lines already presented by \citet{danilovich2016_sulphur}. It also matches the central depression in the 183\,GHz H$_2$O maser and appears appropriate based on, for example, the selection of emission features pertaining to different molecules and spanning a large range of excitation conditions as discussed in Sect.~\ref{sect:kinematics}. Finally, it is also in agreement with the radiative-transfer modelling efforts from earlier publications \citep[e.g.][]{maercker2016_water}.

\subsubsection{Carbon monoxide}
The estimated radial extent of the CO envelope of \rdor, that is, the $e$-folding radius of the CO abundance profile, is $1.6\times10^{16}$\,cm, corresponding to an angular (diametric) size of 36\arcsec\/ at a distance of 59\,pc  \citep{maercker2016_water}. The $^{12}$CO and $^{13}$CO emission lines measured towards \rdor in this survey are hence slightly spatially resolved (see e.g. Fig.~\ref{fig:bumps}) and the spectra do not recover all of the emission as these are single-pointing observations. On the contrary, for all the other molecular line emissions the loss of flux is estimated to be small.

Our observations of  the $^{12}$CO($J=2-1,3-2$) and $^{13}$CO($J=2-1,3-2$) line emission (Fig. \ref{fig:CO_zoom}) agree within 15\% (in intensity) with the independently performed APEX observations of \citet{ramstedt2014_12co13co}. We discuss possible variability over time of the CO line emission in App.~\ref{sect:COvariability}. We refer to \citet{ramstedt2014_12co13co} and \citet{maercker2016_water} for detailed radiative transfer models of CO.

\begin{figure*}\centering
\subfigure[$^{28}$Si$^{16}$O]{\includegraphics[width=5cm]{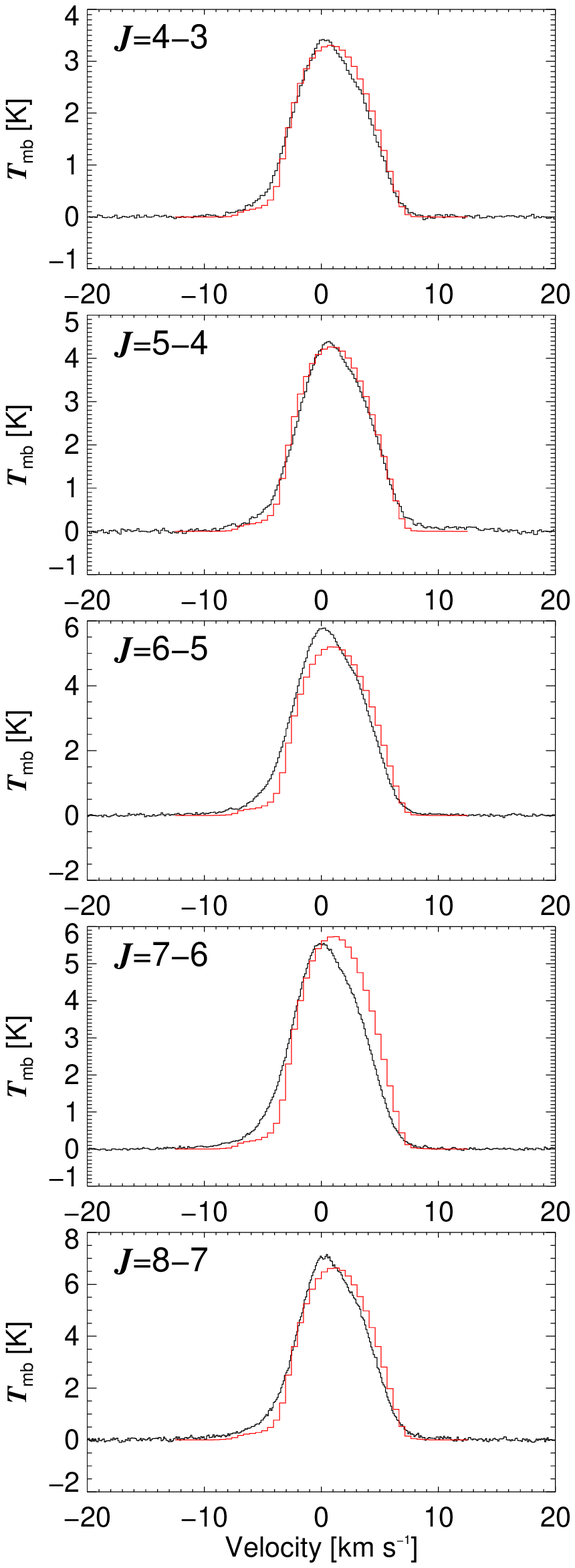}}%{SiOgrid5_73}}
\subfigure[$^{29}$Si$^{16}$O]{\includegraphics[width=5cm]{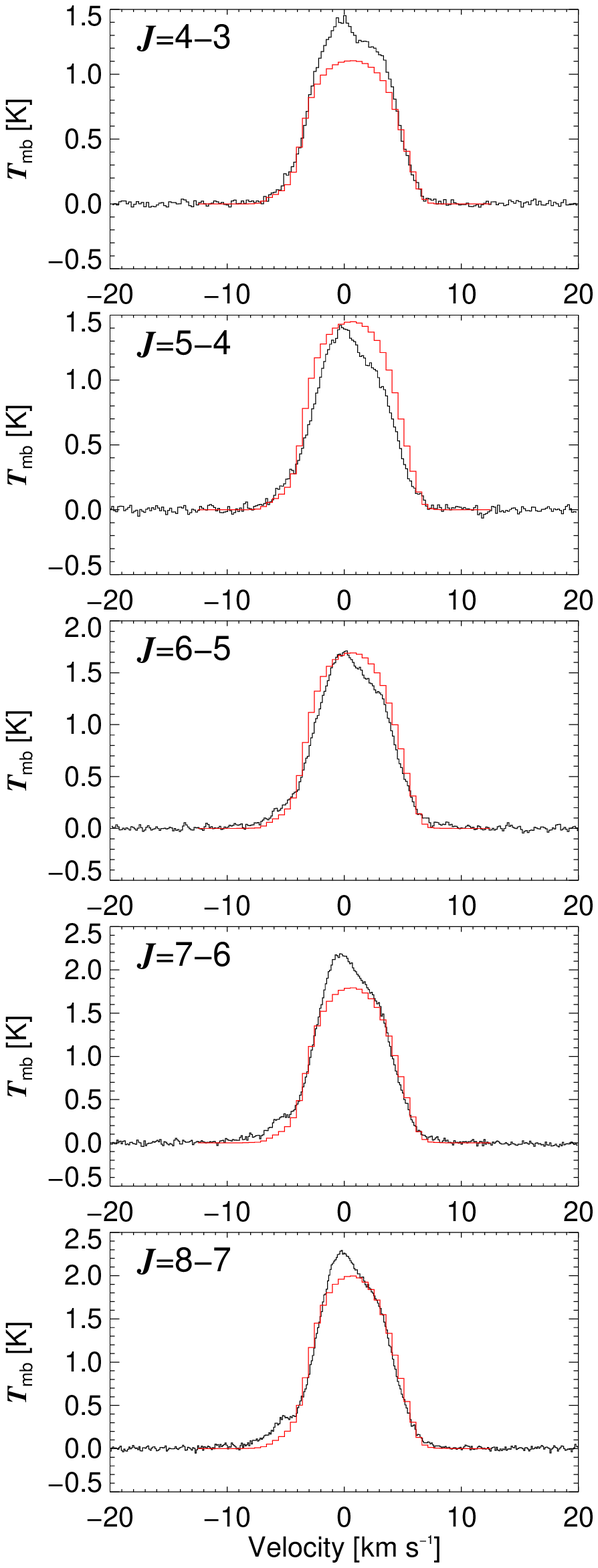}}%{29SiOgrid4_107}} 
\subfigure[$^{30}$Si$^{16}$O]{\includegraphics[width=5cm]{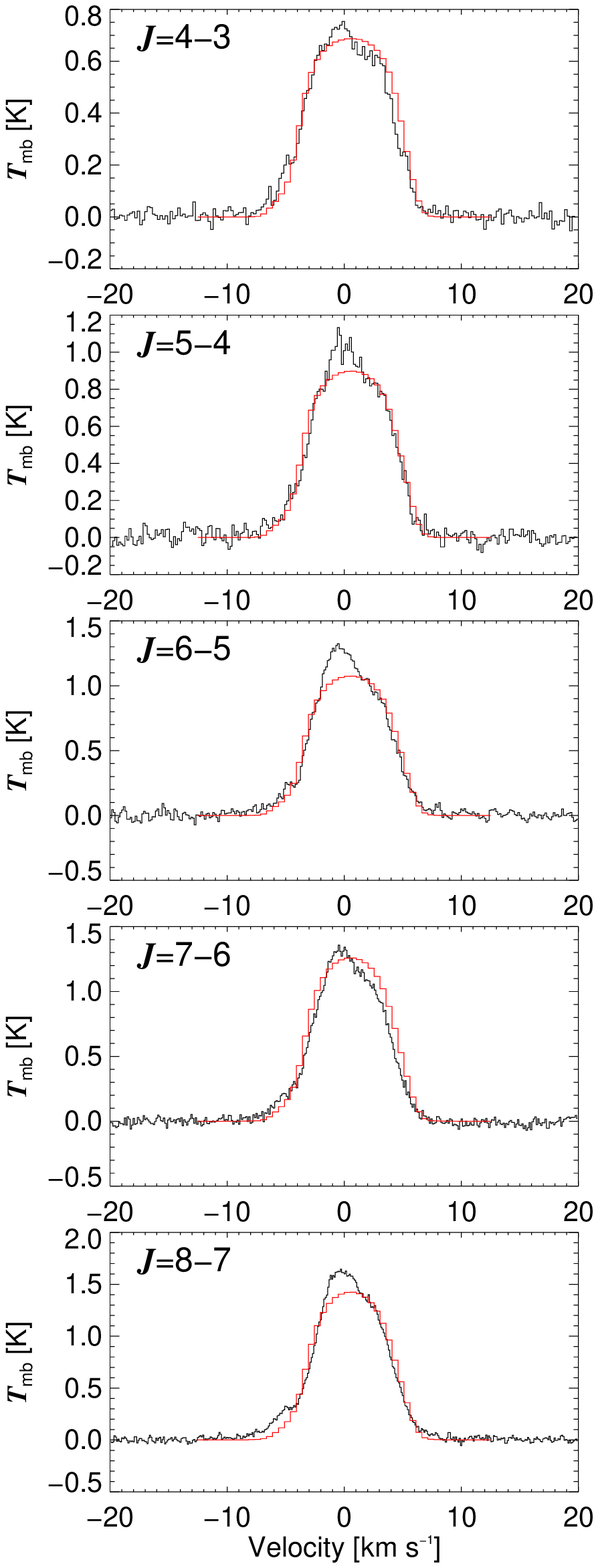}}%{30Si16Ogrid1_89}}
\caption{SiO isotopologue ($^{28,29,30}$Si$^{16}$O) $v=0$ line emission \emph{(black)} and the predictions from our best-fit radiative transfer models \emph{(red)}. We note that the line intensity is given as main-beam brightness temperature $T_{\rm mb}$.  
  \label{fig:SiO_v0}}
\end{figure*}

\begin{figure*}
\sidecaption
\subfigure[$^{28}$Si$^{17}$O]{\includegraphics[width=5cm]{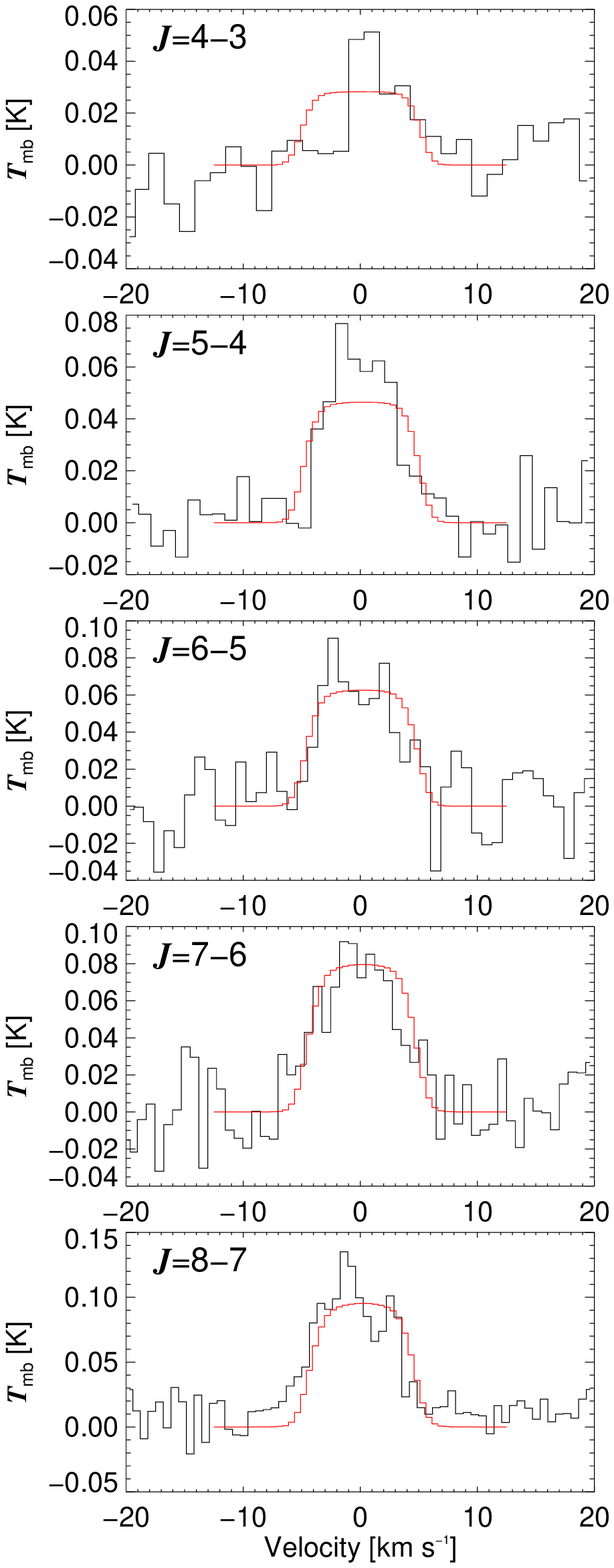}}%{28Si17Ogrid1_61}}
\hspace{1cm}
\subfigure[$^{28}$Si$^{18}$O]{\includegraphics[width=5cm]{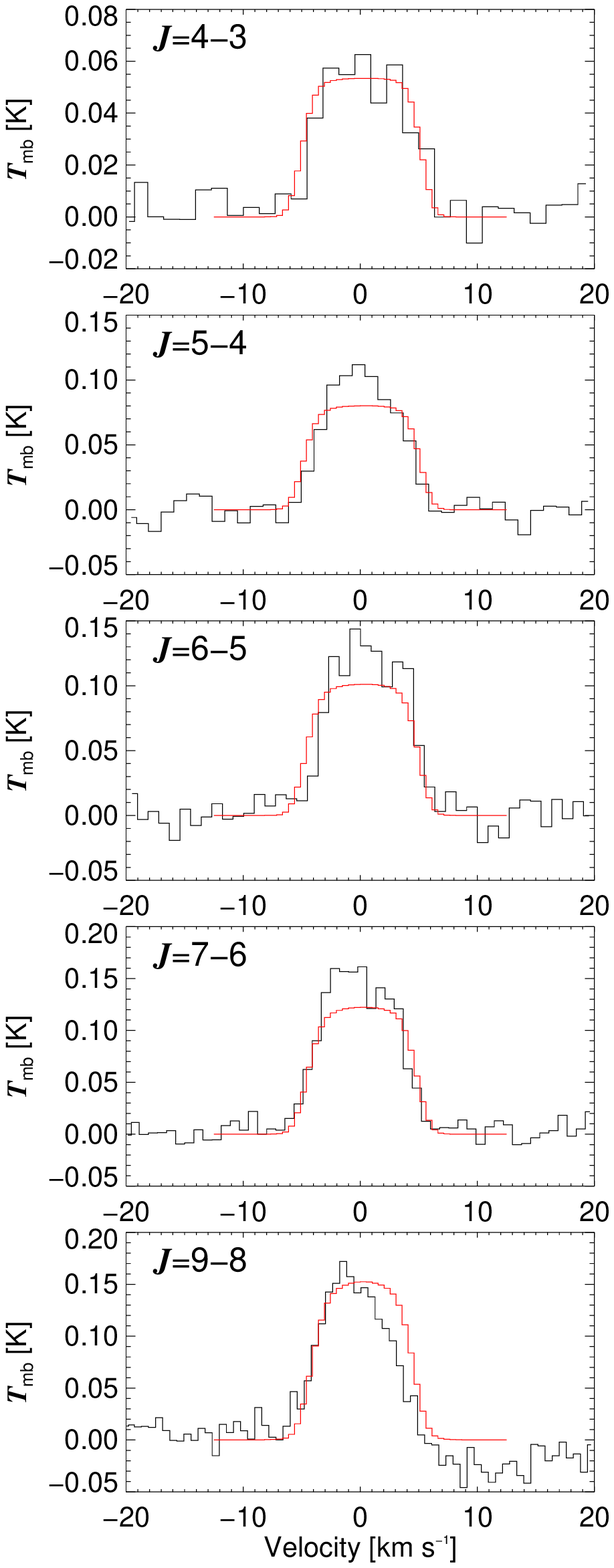}}%{28Si18Ogrid1_74}}
\caption{SiO isotopologue  ($^{28}$Si$^{17,18}$O) $v=0$ line emission  \emph{(black)} and the predictions from our best-fit radiative transfer models \emph{(red)}. We note that the line intensity is given as main-beam brightness temperature $T_{\rm MB}$. \label{fig:SiO_v0_17o_18o}}
\end{figure*}

\begin{figure*}\centering
\subfigure[$^{28}$Si$^{16}$O]{\includegraphics[width=.24\linewidth]{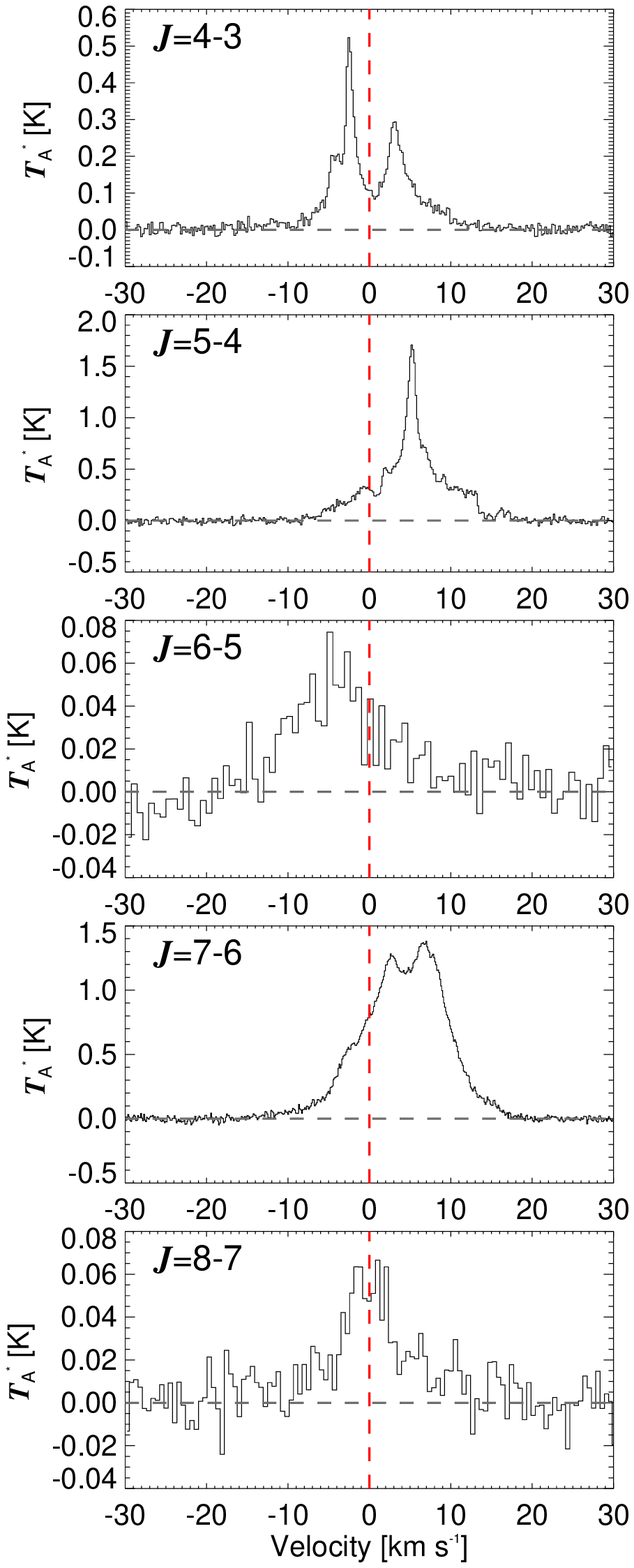}}%{SiOv1_state_zoom}}
\subfigure[$^{29}$Si$^{16}$O \label{fig:29sio_v0}]{\includegraphics[width=.24\linewidth]{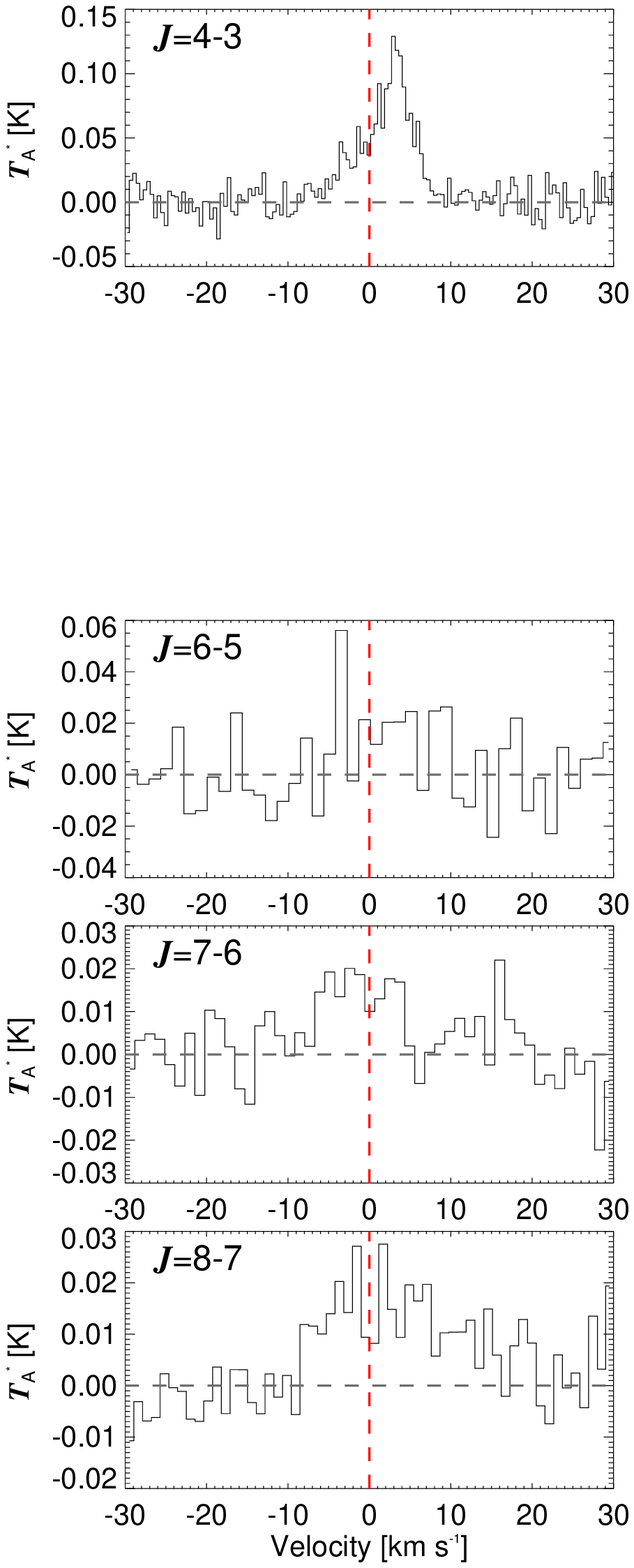}}%{29SiOv1_state_zoom}}
\subfigure[$^{30}$Si$^{16}$O \label{fig:30sio_v0}]{\includegraphics[width=.24\linewidth]{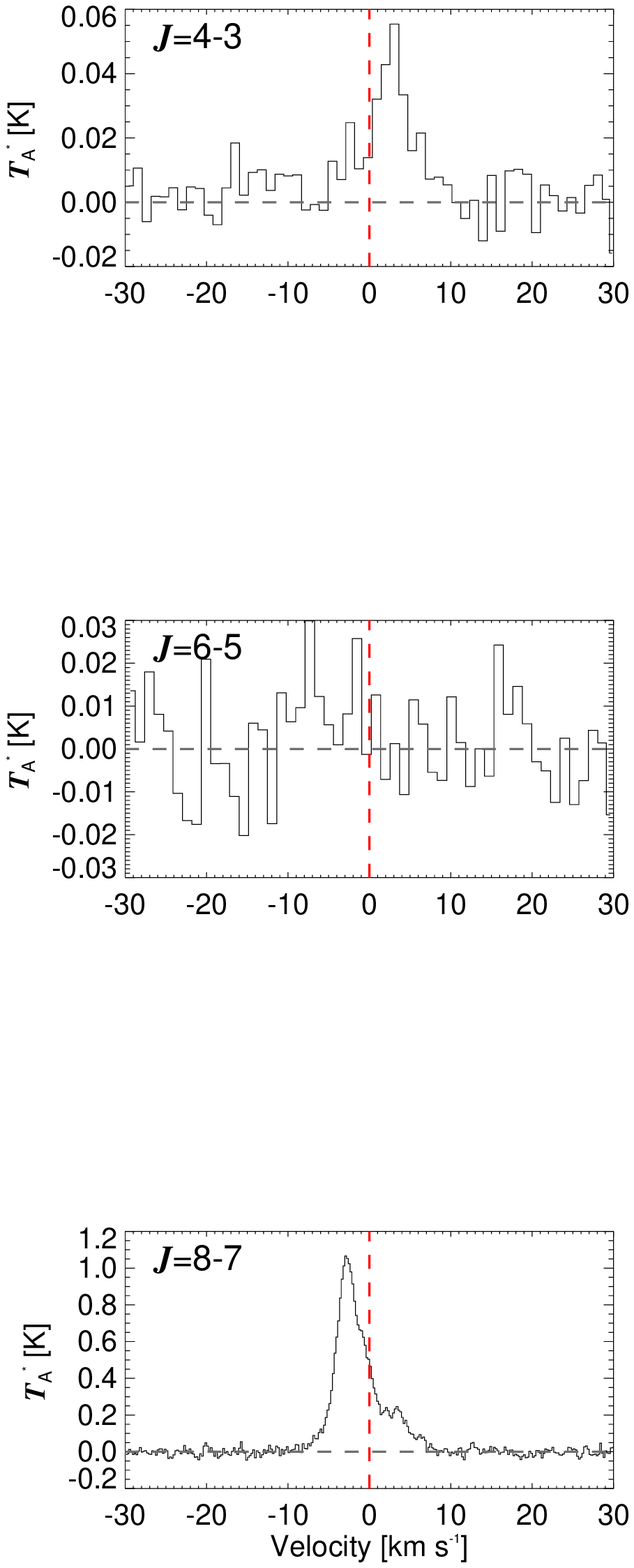}}%{30SiOv1_state_zoom}}
\caption{SiO isotopologue $v=1$ line emission. We point out that the $^{30}$SiO ($v=1,J=8-7$) emission is blended with emission from SO$_2$ ($v_2=1,J_{K_{\rm a},K_{\rm c}}=6_{5,1}-7_{4,4}$) centred at $\sim$3\,\kms in the rest frame of the SiO line. \label{fig:SiO_v1}}
\end{figure*}

\begin{figure}\centering
\subfigure[$^{28}$Si$^{16}$O]{\includegraphics[width=.48\linewidth,trim={0 10cm 0 0},clip]{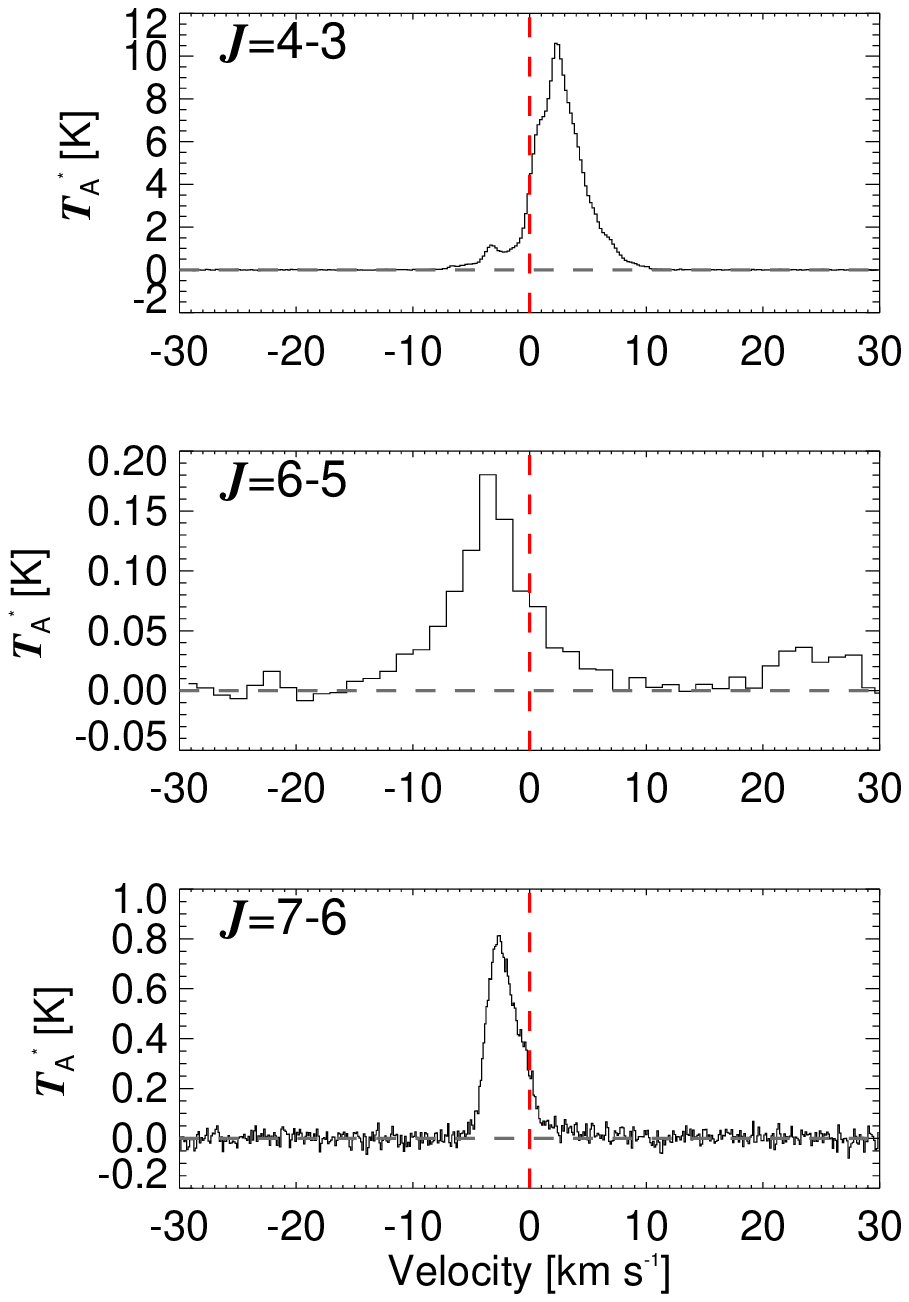}}%{SiOv2_state_zoom}}
\subfigure[$^{29}$Si$^{16}$O]{\includegraphics[width=.48\linewidth,trim={0 14.4cm 0 0},clip]{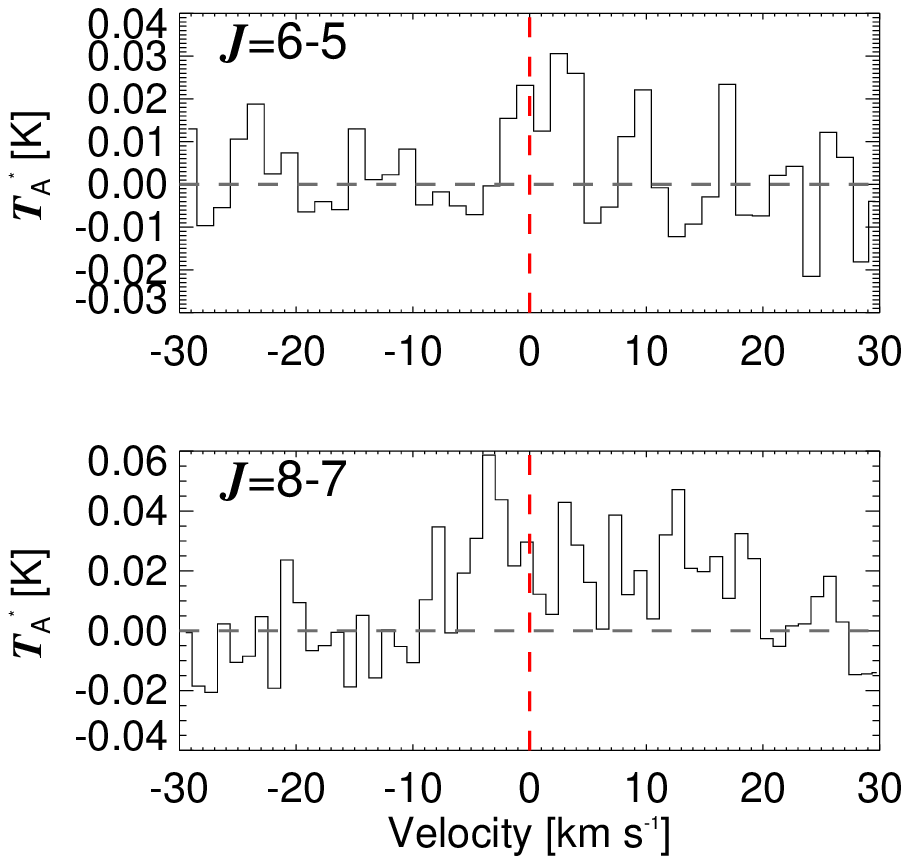}}%{29SiOv2_state_zoom}}
\caption{SiO isotopologue $v=2$ line emission. 
\label{fig:SiO_v2}}
\end{figure}
\begin{figure}\centering
\subfigure[$v=3$]{\includegraphics[width=.48\linewidth,trim={0 5.5cm 0 0},clip]{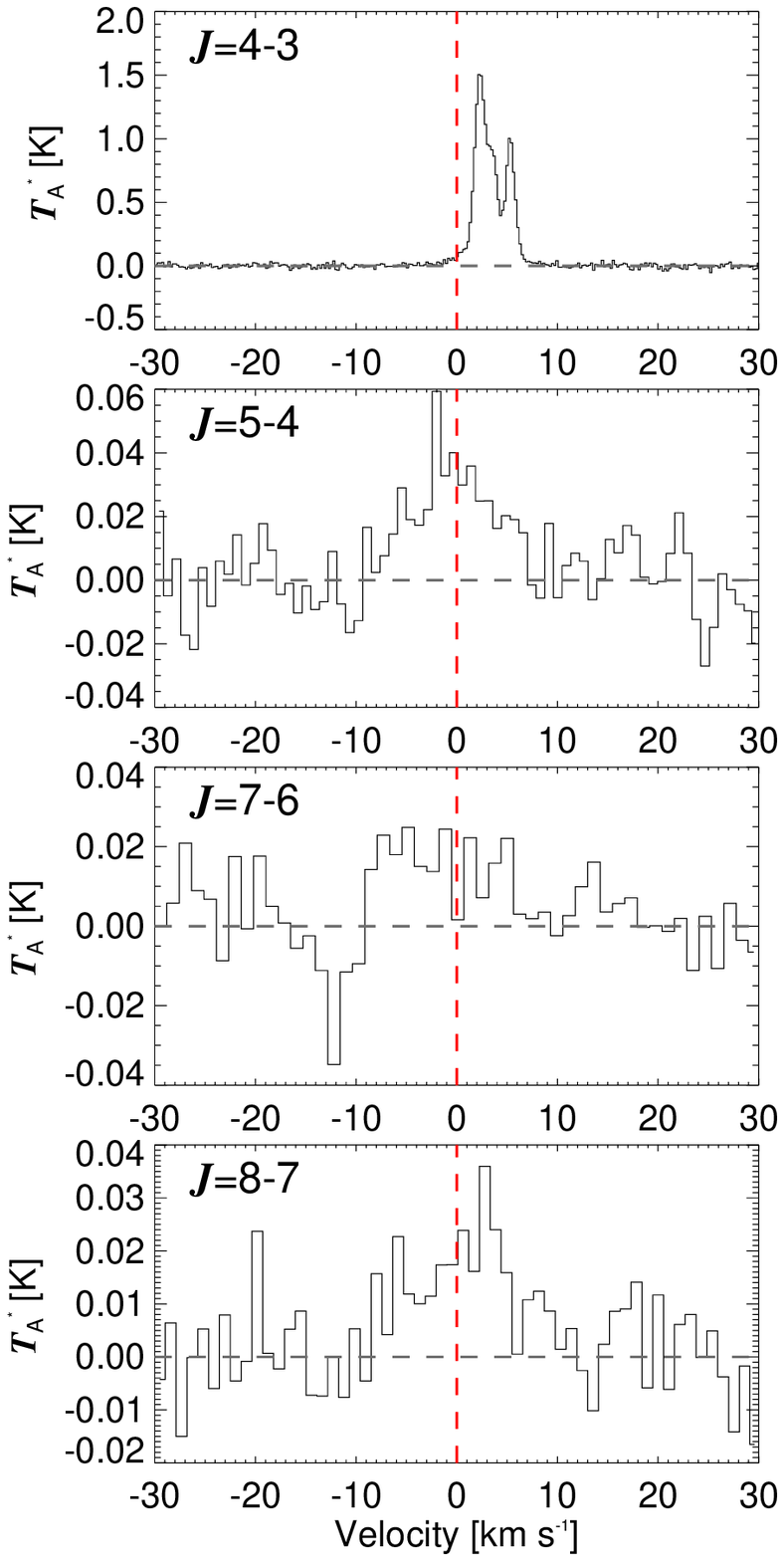}}%{SiOv3_state_zoom}}
\subfigure[$v=5$]{\includegraphics[width=.48\linewidth,trim={0 12cm 0 0},clip]{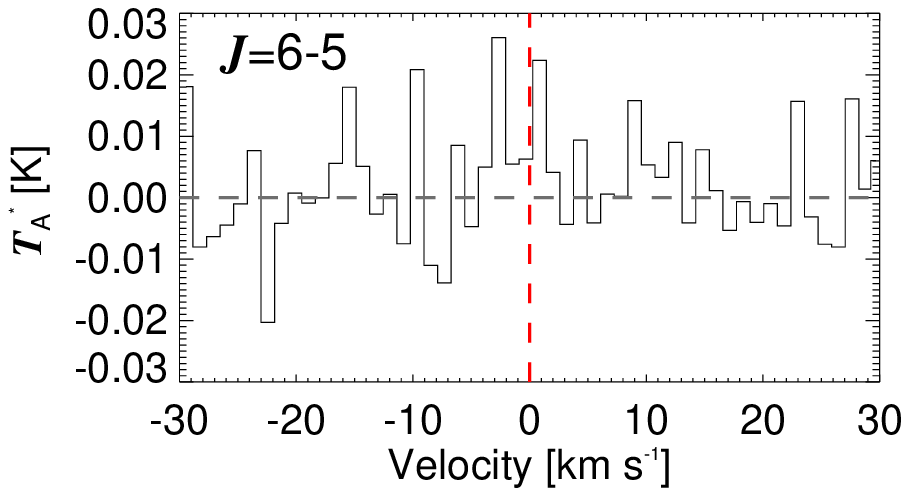}}%{SiOv5_state_zoom}}
\caption{$^{28}$Si$^{16}$O line emission in higher excited states. 
\label{fig:SiO_v3,4,5}}
\end{figure}

\begin{figure*}
\centering
\subfigure[$^{28}$SiO  \label{fig:sio_redchisquare}]{\includegraphics[width=0.33\linewidth]{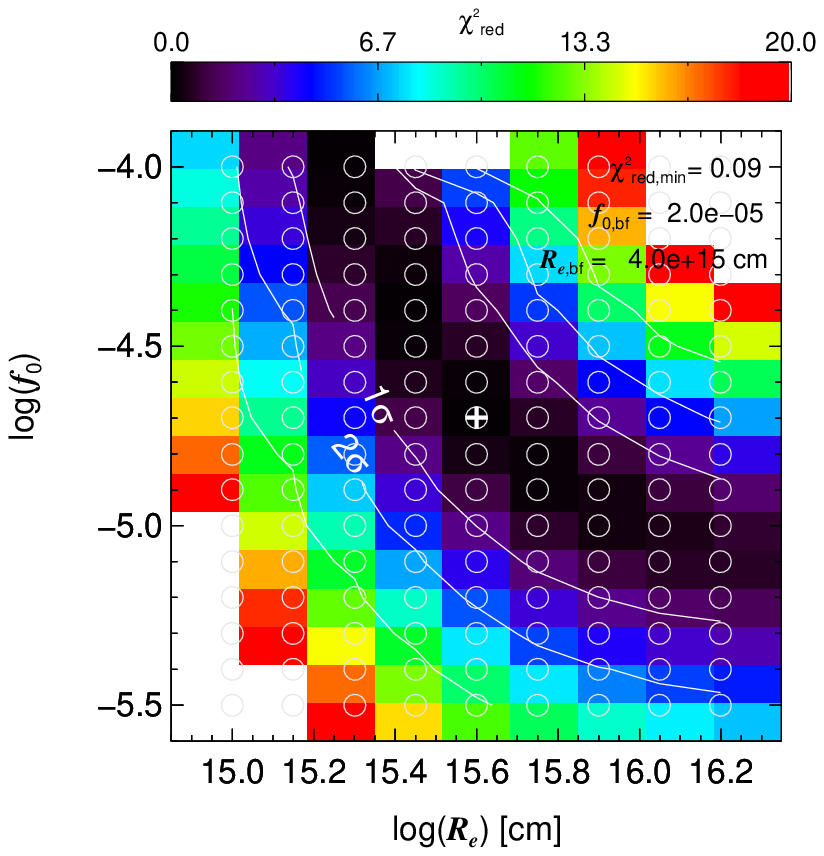}}%{SiOgrid5_redchisquare}}
\subfigure[$^{29}$SiO  \label{fig:29sio_redchisquare}]{\includegraphics[width=0.33\linewidth]{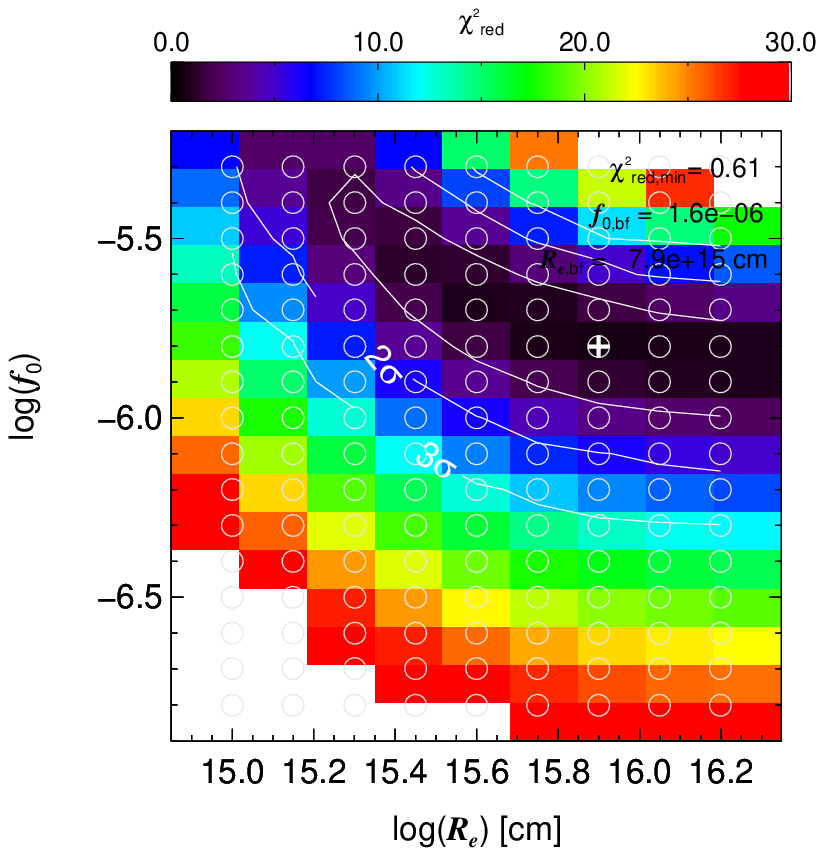}}%{29SiOgrid4_redchisquare}}
\subfigure[$^{30}$SiO  \label{fig:30sio_redchisquare}]{\includegraphics[width=0.33\linewidth]{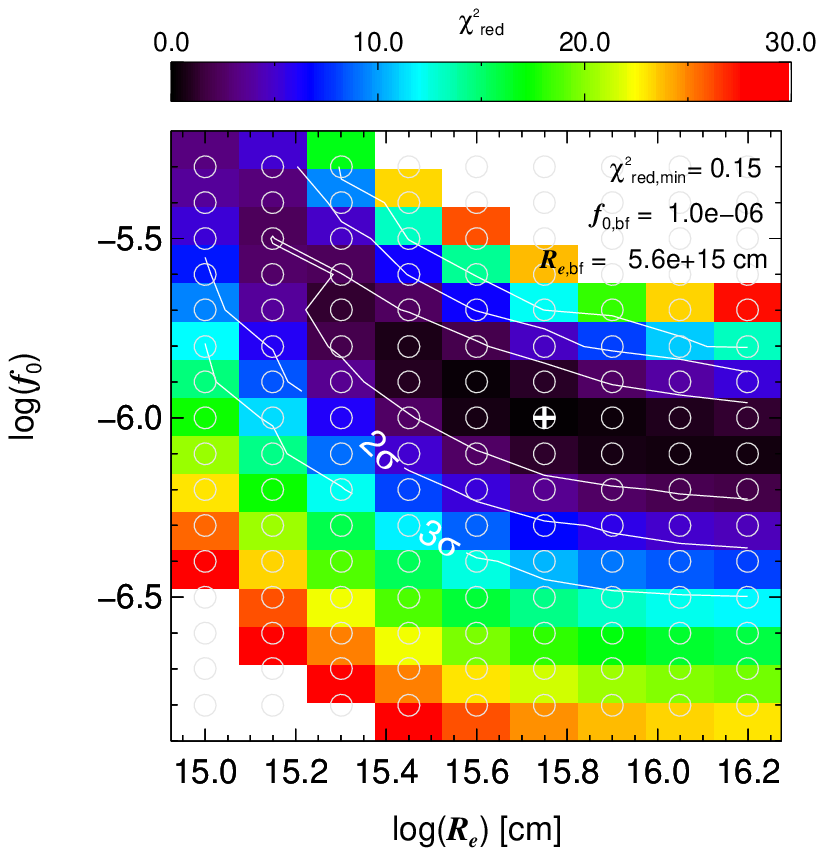}}%{30Si16Ogrid1_redchisquare}}
\\
\subfigure[$^{28}$Si$^{17}$O  \label{fig:si17o_redchisquare}]{\includegraphics[width=0.33\linewidth]{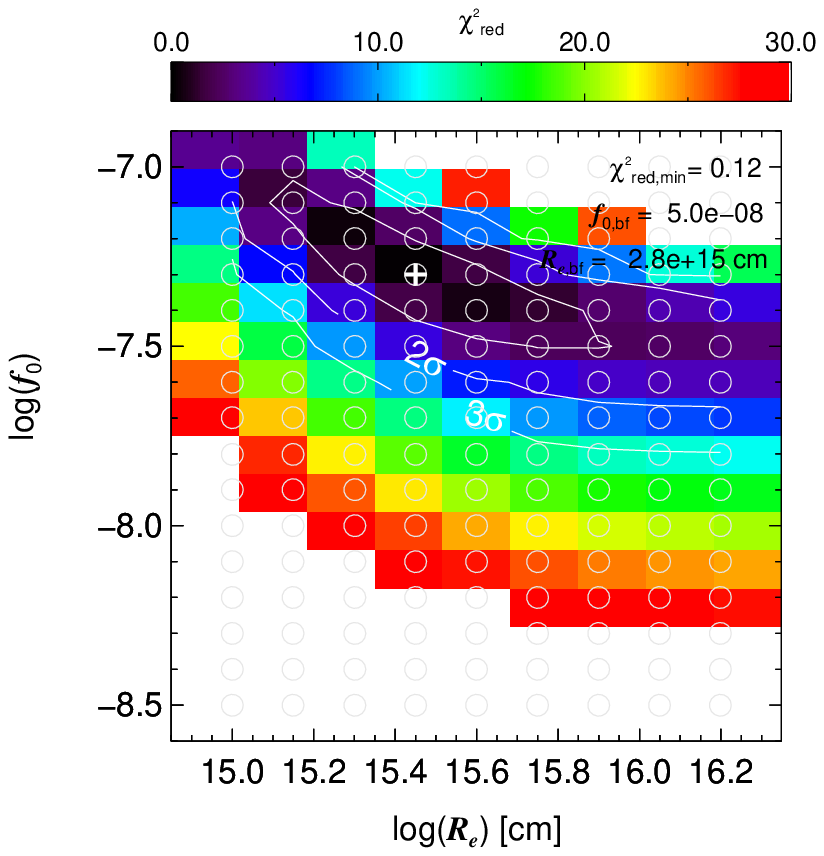}}%{28Si17Ogrid1_redchisquare}}
\subfigure[$^{28}$Si$^{18}$O  \label{fig:si18o_redchisquare}]{\includegraphics[width=0.33\linewidth]{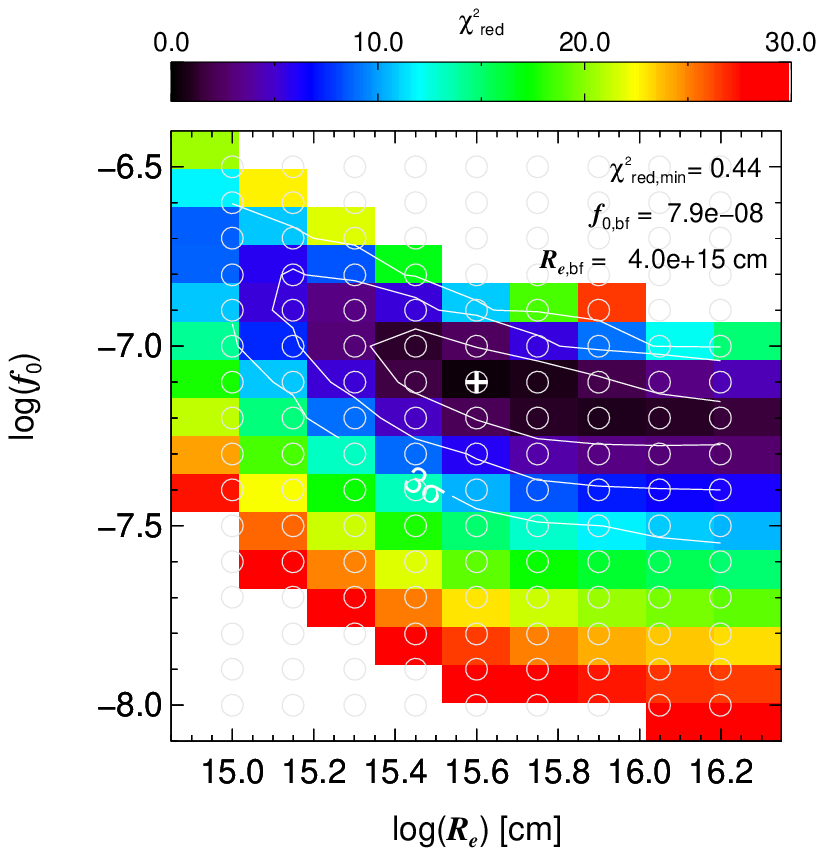}}%{28Si18Ogrid1_redchisquare}}
\caption{Reduced-$\chi^2$ maps of the model grids for the $v=0$ state of five SiO isotopologues. Circles indicate the grid points and white crosses indicate the best-fit models. Contours are given at the $\sim$68\%, $\sim$95\%, and 99.7\% confidence intervals ($1\sigma,2\sigma,3\sigma$, resp.).\label{fig:sio_redchisquare_all}}
\end{figure*}

\subsubsection{Silicon monoxide}\label{sect:sio}
We detect 50 transitions of SiO and its isotopologues. Figures~\ref{fig:SiO_v0} to \ref{fig:SiO_v3,4,5} show the spectra for the transitions $J=4-3,...,9-8$ in the vibrational states $v=0,...,5$ for the  isotopologues $^{28}$SiO, $^{29}$SiO, $^{30}$SiO, Si$^{17}$O, and Si$^{18}$O. The intensity-weighted mean velocity of the lines (in $v=0$) of the different isotopologues shifts slightly to the red with increasing $J$. This could be explained by an increasing optical thickness that causes stronger self-absorption in the blue wing with increasing $J$, shifting the peak of the emission redwards. This effect is reproduced by our radiative transfer models discussed below.

Emission in transitions in the vibrational ground state ($v=0$) is most often assumed to be of non-maser nature. However, recently, \citet{devicente2016_SiO1-0} showed that the SiO($v=0,J=1-0$) transition exhibits clear (and variable) signatures of maser nature in many of the oxygen-rich and S-type AGB stars in their sample. These maser components are thought to be excited in the very inner layers of the CSE, where no dust is present yet.  \citet{devicente2016_SiO1-0} further state that higher-$J$ transitions are typically of thermal nature. We note, however, that time-variable features are occasionally seen in the $J=2-1$ emission and that these are likely of maser nature, see \citet{nyman1985_SiO21variability}. Judging from the appearance of the emission features in this survey and in the HIFI data (Fig.~\ref{fig:bumps}; $J=4-3$ and higher) thermal excitation indeed seems to be likely for the observed $v=0$ lines. 

\begin{table*}\caption{Radiative transfer modelling results for SiO isotopologues.\label{tbl:sio_RT}}
\centering
\begin{tabular}{cccc|ccc}
\hline\hline\\[-2ex]
\textbf{Isotopologue} &\multicolumn{3}{c}{$\log(f_0)$} &\multicolumn{3}{c}{$\log(R_{e} [\mathrm{cm}])$}\\
\cline{2-4} \cline{5-7}\\ [-2ex]
                                        &       Grid range (step)       & Best-fit value & $1\sigma$-range&       Grid range (step)& Best-fit value & $1\sigma$-range\\
\hline\\[-2ex]
$^{28}$Si$^{16}$O &$[-5.5,-4.0]$ (0.1)&-4.7&$[-5.2,-4.0]$&$[15.00,16.20]$ (0.15)&15.60&$[15.15,16.20]$\\
$^{29}$Si$^{16}$O &$[-6.8,-5.3]$ (0.1)&-5.8&$[-5.9,-5.4]$&$[15.00,16.20]$ (0.15)&15.90&$[15.30,16.20]$\\
$^{30}$Si$^{16}$O &$[-6.8,-5.3]$ (0.1)&-6.0&$[-6.2,-5.5]$&$[15.00,16.20]$ (0.15)&15.75&$[15.15,16.20]$\\
$^{28}$Si$^{17}$O &$[-8.5,-7.0]$ (0.1)&-7.3&$[-7.5,-7.1]$&$[15.00,16.20]$ (0.15)&15.45&$[15.15,15.90]$\\
$^{28}$Si$^{18}$O &$[-8.0,-6.5]$ (0.1)&-7.1&$[-7.2,-7.0]$&$[15.00,16.20]$ (0.15)&15.60&$[15.45,16.20]$\\
\hline
\end{tabular}
\end{table*}

For vibrationally excited states ($v\neq0$) the emission is clearly of maser nature, often with multiple peaks in the line profiles. These complex shapes are furthermore not constant across the excitation ladder within one vibrational state. We plot a selection of the highest-S/N maser lines of SiO (any isotopologue) in Fig.~\ref{fig:SiOmasers}. As mentioned before, several of the individual maser peaks correspond to the velocities where peaks or bumps also appear in thermal emission lines.  Overall, it is clear that the SiO maser emission can differ substantially between different transitions, even within the same vibrationally excited state of a given isotopologue. Some maser lines are composed of several  peaks over the entire velocity range  (e.g. $^{28}$SiO($v=1,J=4-3$)), some are limited to only blue or red-shifted velocities  (e.g. $^{28}$SiO($v=3,J=4-3$) and  $^{28}$SiO($v=2,J=7-6$)), and some appear as very broad emission lines with barely discernible maser components (e.g. $^{28}$SiO($v=1,J=7-6$)). This is in line with both observations and simulations of SiO masers in the CSEs of red giant and supergiant stars \citep[e.g.][]{humphreys1997,humphreys2002,desmurs2014}, from which it is apparent that the excitation of different rotational transitions within the same vibrational state is not necessarily co-spatial. This, in combination with non-simultaneity of the observations, explains the variety in the profiles we observe towards \rdor. We therefore add Table~\ref{tbl:maserdates} listing the observing dates of these lines in App.~\ref{sect:maserobs}.

The wealth of emission lines from different SiO isotopologues, spanning the entire frequency range of the survey, motivates a deeper investigation through modelling. We base the radiative transfer modelling of the  SiO $v=0$ line emission on the CSE model reported by \citet{maercker2016_water}. Our molecular input covers the rotational levels  for $J=0,\dots,40$ of the $v=0,1$ states of $^{28}$SiO, $^{29}$SiO, $^{30}$SiO, Si$^{17}$O, and Si$^{18}$O. Collisional rates are included for SiO--H$_2$, adapted from the collisional rates for SiO--He of \citet{dayou2006}. The modelling is performed using the accelerated lambda iteration (ALI) radiative transfer code described and implemented by \citet{maercker2008,maercker2016_water,danilovich2014_waql}.

We assume a centrally peaked Gaussian fractional abundance distribution 
\begin{equation}
f(r) = f_0  \exp\left( -\left( \frac{r}{R_{e}}\right)^2\right),
\end{equation}
with $f_0$ the molecular abundance (w.r.t. H$_2$) at the inner radius of the CSE \citep[at 5\,\rstar, following][]{danilovich2016_sulphur}, also referred to as peak abundance in the rest of the paper, and $R_{e}$ the $e$-folding radius of the Gaussian profile. We set up a model grid for these two free parameters, $f_0$ and $R_{e}$, as summarised in Table~\ref{tbl:sio_RT}. We minimise the  reduced-$\chi^2$  
\begin{equation}
\chi^2_{\mathrm{red}}=\frac{1}{N-p}\sum_{i=1}^{N}\left( \frac{I_{\mathrm{mod}}-I_{\mathrm{obs}}}{\sigma}\right)^2
,\end{equation}
with $N$ being the number of modelled transitions (five for each isotopologue), $p$ the number of free parameters (two),  $I_{\mathrm{mod}}$ and $I_{\mathrm{obs}}$ the modelled and observed integrated line intensities, respectively, and $\sigma$ the uncertainty on  $I_{\mathrm{obs}}$. We assume uncertainties on the data at a 20\% level, accounting for predominantly the calibration uncertainties, given the high S/Ns in the modelled lines. Table~\ref{tbl:sio_RT} lists the best-fit values of $f_0$ and $R_e$ with their $1\sigma$-uncertainties for each of the isotopologues. We show the reduced-$\chi^2$ maps in Fig.~\ref{fig:sio_redchisquare_all} and the comparison of the best-fit models to the survey data in Figs.~\ref{fig:SiO_v0} and \ref{fig:SiO_v0_17o_18o}. In general, we find that there is a rather large degeneracy in the radiative transfer modelling between the two free parameters, $f_0$ and $R_e$, visible as the elongated $1\sigma$-confidence intervals. This degeneracy is most significant for the main isotopologue, $^{28}$Si$^{16}$O, where abundances in the range $0.6-10\times10^{-5}$ can reproduce the emission to within the uncertainties, according to the very clear trend that a lower $f_0$ requires a larger $R_e$ (Fig.~\ref{fig:sio_redchisquare}). This is a consequence of the high optical depth involved in the radiative transfer of this molecule. Natural upper limits to $R_e$  and $f_0$ can in this case be set by the size of the CO envelope and the Si abundance \citep[${\rm Si/H}\approx3.2\times10^{-5}$;][]{asplund2009_solarabundances} which implies ${\rm SiO/H}_2\lesssim6.4\times10^{-5}$ or $\log\left({\rm SiO/H}_2\right)\lesssim-4.2$ if all silicon is comprised in SiO and all H is in molecular form.

\citet{gonzalezdelgado2003} and \citet{schoeier2004_sio} modelled thermal SiO emission ($J=2-1,3-2,5-4,6-5$) from the CSE of \rdor. More recently, \citet[][]{vandesande2017_rdor} additionally modelled several high-$J$ transitions (up to $J=38-37$). Based on interferometric observations of SiO\,($J=2-1$), \citet{schoeier2004_sio} found that the SiO abundance profile of \rdor is possibly better represented by a compact component with a high, constant, abundance of $4\times10^{-5}$ out to $r=1.2\times10^{15}$\,cm and a component with a low abundance ($3\times10^{-6}$ and declining according to a Gaussian profile) at larger radii. The discontinuity in the abundance is thought to reflect the signature of depletion of SiO onto dust. Since we restrict our modelling to higher-excitation transitions, ignoring the possible depletion signature as derived by \citet{schoeier2004_sio} is not expected to pose a problem. Our best-fit value of $f_0$ falls in between those of the previous models and our uncertainties cover both the low-lying values of \citet{gonzalezdelgado2003} and \citet{schoeier2004_sio} and the high $f_0$ found by \citet{vandesande2017_rdor}. The range of acceptable values of $R_e$ we find is also in agreement with the results of both \citet{gonzalezdelgado2003,schoeier2004_sio} and \citet[][as presented in their Fig.~6]{vandesande2017_rdor}.

We are not aware of any earlier efforts to model the thermal emission of the less abundant isotopologues $^{29}$SiO, $^{30}$SiO, Si$^{17}$O, and Si$^{18}$O, except for the modelling by \citet{decin2010_iktau_nlte} of two $^{29}$SiO transitions measured towards \iktau.  We set up our model grids based on the one for $^{28}$SiO, covering the same values for $R_e$ and scaling those for $f_0$ with a reasonable value for the appropriate isotopic ratio. See Table~\ref{tbl:sio_RT} for the grid specifications and model results. Figures~\ref{fig:SiO_v0} and \ref{fig:SiO_v0_17o_18o} show that our model predictions reproduce the observed data very well. We note that the remarkable quality of the fit to all emission lines simultaneously, for each of the modelled isotopologues, demonstrates that the survey's internal data calibration uncertainty is actually (significantly) lower than the 20\% we quote as absolute uncertainty in Sect.~\ref{sect:observations}. To assess the variability in SiO line emission as a consequence of stellar variability, we calculated a radiative transfer model (for $^{28}$SiO) based on our best-fit model, but decreasing the luminosity to 4500\,\lsun. This significant change in luminosity leads to changes in intensity within 20\% and in integrated line intensity within 15\% for the observed lines. Such line variability would hence not be significant with respect to the observational uncertainties in the selection of our models.

To obtain isotopologue abundance ratios from these results, we consider all models, that is, combinations of $f_0$ and $R_e$, that fall within $1\sigma$ of the best-fit model for a given isotopologue. Additionally, assuming that all isotopologues are photodissociated by the interstellar radiation field at the same radius and, hence, have the same $e$-folding radius $R_e$, we can reduce the uncertainties on these ratios by only considering the abundance ratios at a given $R_e$. This approach leads to the values and uncertainties listed in Table~\ref{tbl:isotopes} and discussed in Sect.~\ref{sect:isotopes}.

\begin{figure}
\includegraphics[width=\linewidth]{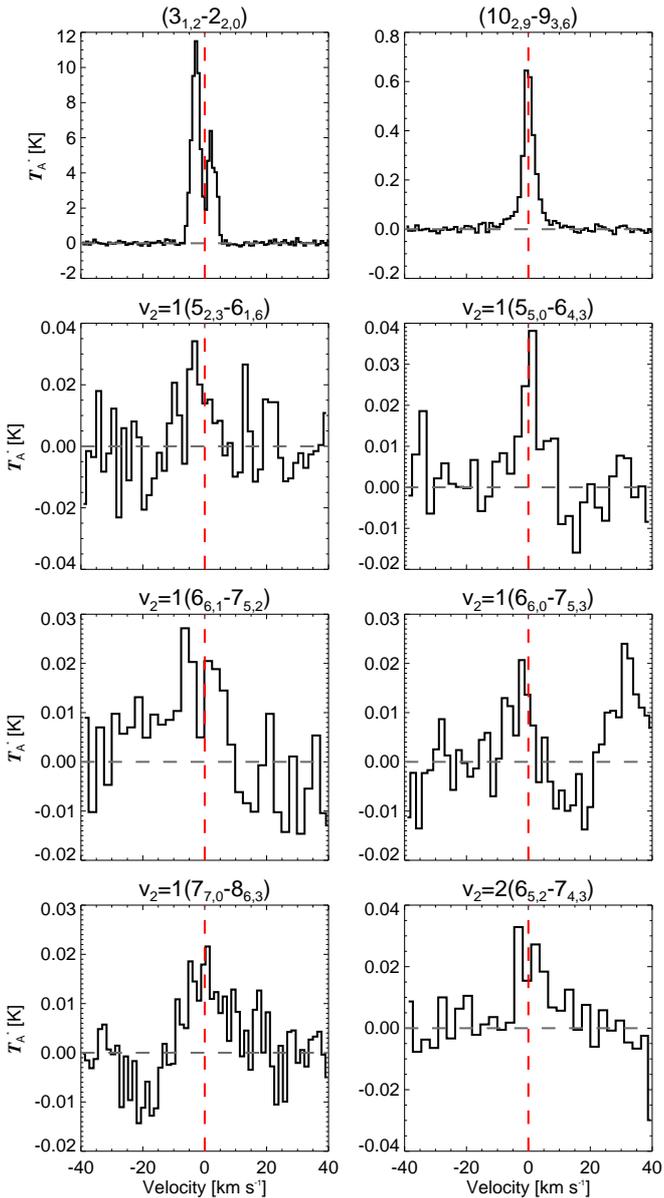}%{H2O_zoom}
\caption{\water emission in the vibrational ground state and the vibrationally excited states $v_2=1,2$. }\label{fig:H2O_zoom}
\end{figure}

\subsubsection{Water}\label{sect:water}
We detect eight emission features of \water: the well-known masers at 183\,GHz and 321\,GHz in the vibrational ground state, and five and one lines in the vibrationally excited states \nutwo and 2\nutwo, respectively, see Fig.~\ref{fig:H2O_zoom}. The maser at 325\,GHz was not observed. \citet{maercker2016_water} carried out a detailed radiative transfer study of the \water line emission observed with HIFI towards a sample of oxygen-rich stars, including \rdor. A similar study of the vibrationally excited \water lines in this survey and of several more oxygen-rich AGB stars is forthcoming and will be based on these results.

\begin{figure}
\includegraphics[width=\linewidth]{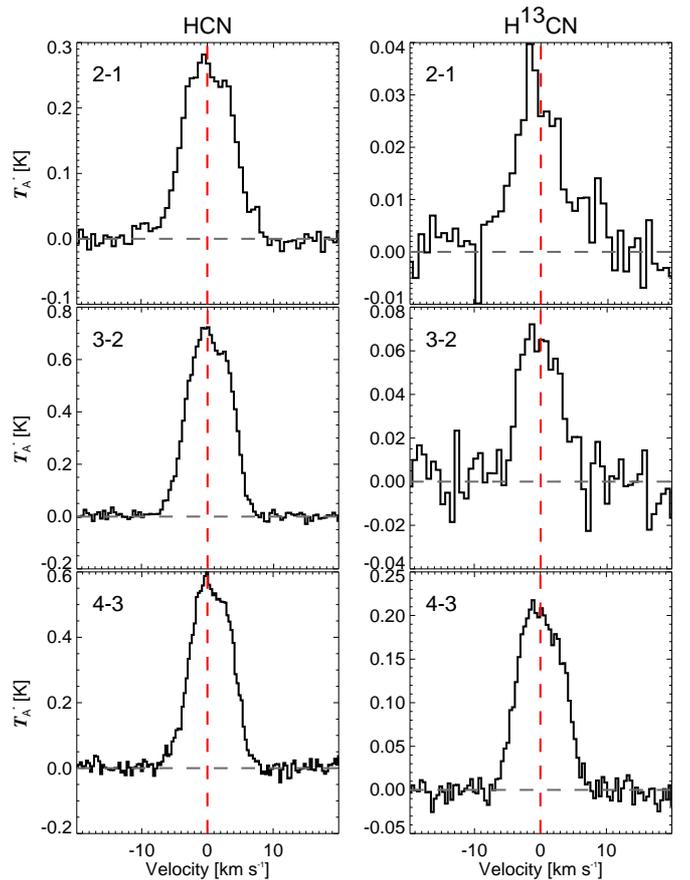}%{HCN_zoom}
\caption{HCN emission in the survey: H$^{12}$CN (\emph{left}) and H$^{13}$CN (\emph{right}). }\label{fig:HCN_zoom}
\end{figure}

\begin{figure}
\includegraphics[width=\linewidth]{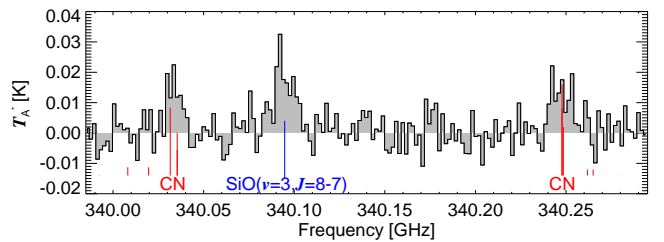}%{CN_zoom}
\caption{CN($3-2$) emission. The positions of the red vertical lines indicate the rest frequencies of the hfs components, and their lengths are proportional to the relative component strengths in LTE.}\label{fig:CN_zoom}
\end{figure}

\begin{figure}
\centering
\includegraphics[width=.75\linewidth]{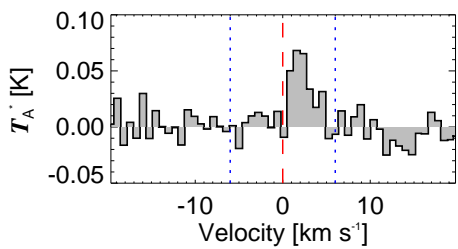}%{SO2_v2_zoom}
\caption{SO$_2$, $v_2=1$ ($J_{K_a,K_c}=69_{9,61}-70_{6,64})$ emission.   \label{fig:SO2_v2}}
\end{figure}

\begin{figure}
\includegraphics[width=\linewidth]{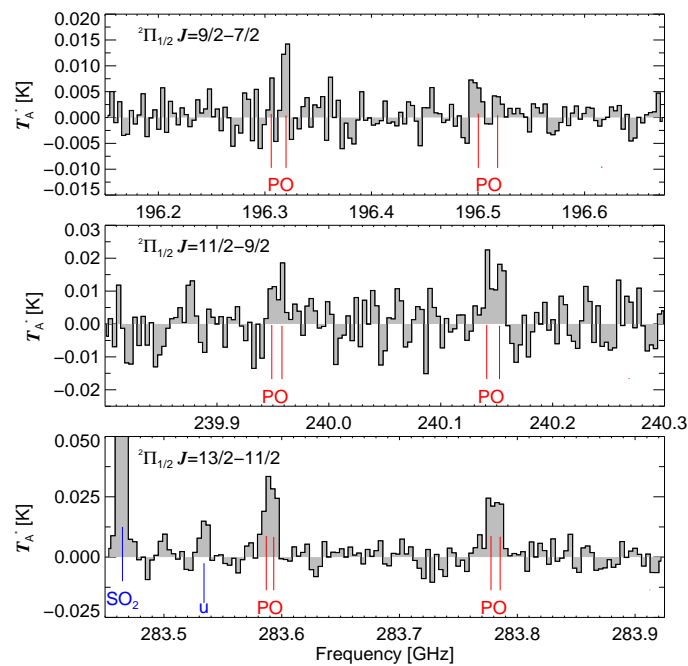}%{PO_zoom}
\caption{PO line emission. Vertical red lines mark the rest frequencies of the different hfs components of each rotational transition.} \label{fig:PO_zoom}
\end{figure}

\begin{figure}\centering
\includegraphics[width=.9\linewidth]{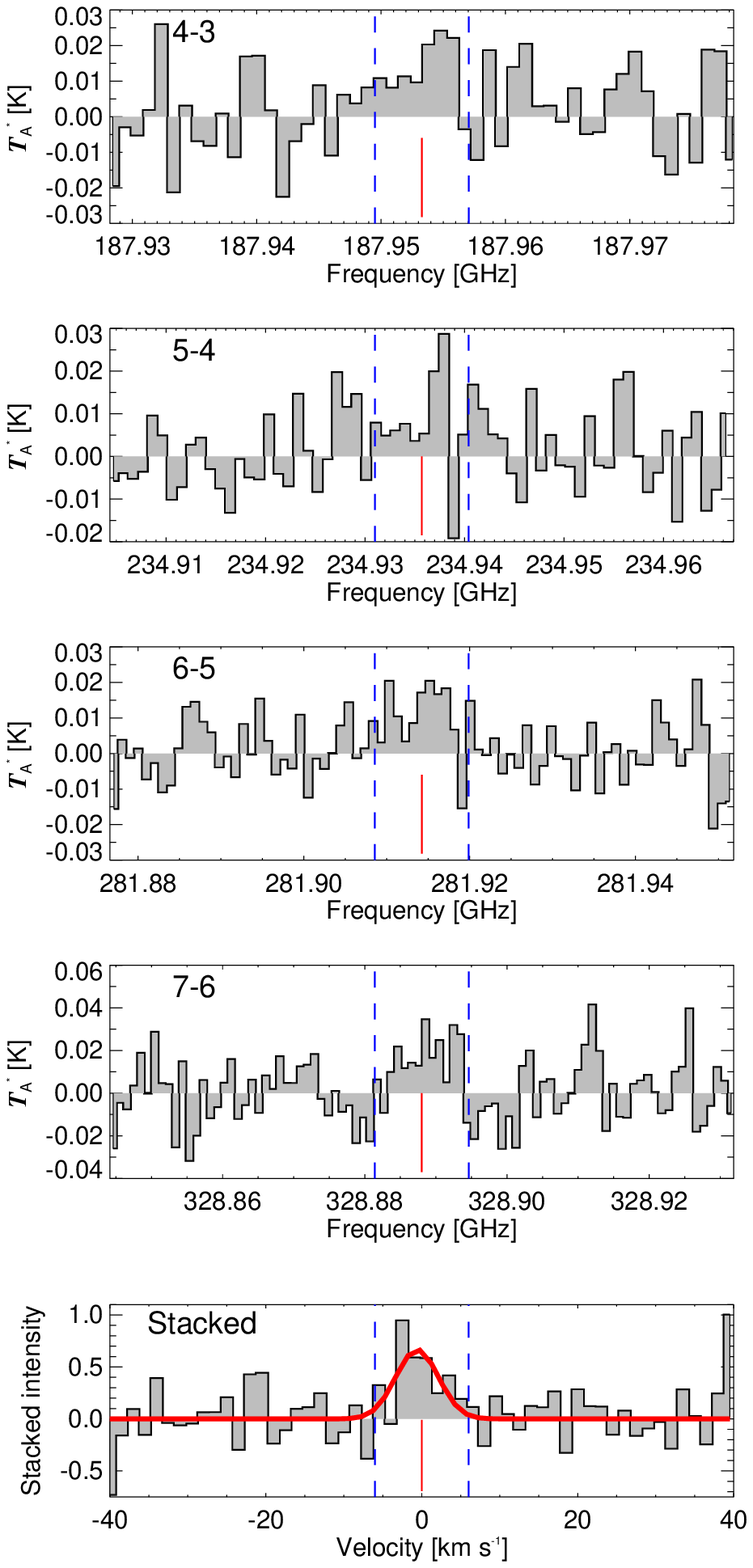}%{PN_zoom}
\caption{PN line emission. The first four panels show the spectra of the $J=4-3,\dots,7-6$ transitions; the bottom panel shows the peak-normalised result of stacking these at a velocity resolution of 1.5\,\kms, weighted with the upper level statistical weights, $2J+1$ of the respective transitions. Red lines indicate the rest frequency (or $\varv=0$\,\kms in the bottom panel) of each transition, blue lines indicate velocities $\pm6$\,\kms with respect to the stellar \vlsr. The red curve in the bottom panel represents a Gaussian fit to the stacked data. }\label{fig:PN_zoom}
\end{figure}

\begin{figure}
\centering
\includegraphics[width=.7\linewidth]{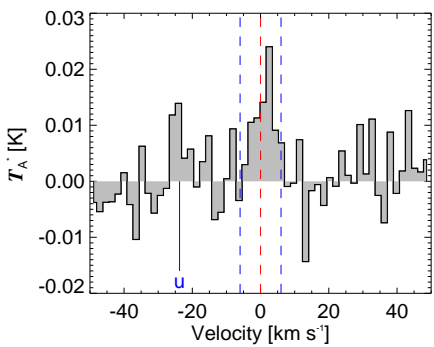}%{PNO_zoom}
\caption{Tentative identification of PNO line emission. The red dashed line corresponds to the rest frequency of the PNO($J=25-24$) transition and the blue dashed lines indicate the stellar \vlsr$\pm6$\,\kms. In blue we indicate an unidentified emission feature.}\label{fig:PNO_zoom}
\end{figure}

\begin{figure*}
\centering
\includegraphics[width=\linewidth]{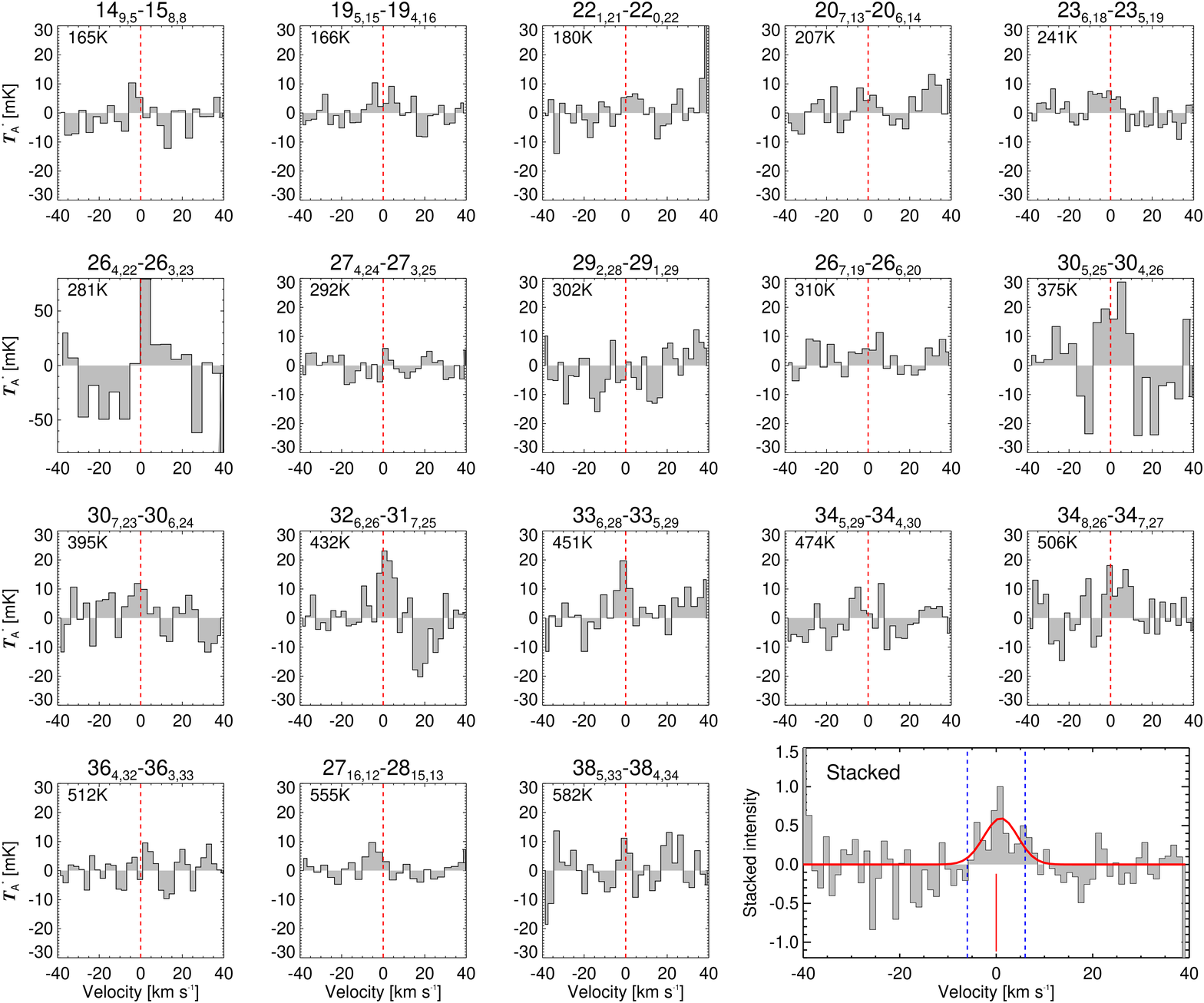}%{TiO2_zoom}
\caption{Tentative identification of TiO$_2$ line emission. Quantum numbers for each transition and upper-level energies $E_{\mathrm{up}}/k$ are given for each panel. The last panel shows the spectrum that results from stacking (in velocity space, at 1.2\,\kms resolution) the spectra in the other 18 panels, assuming equal weights. The red line represents a best-fit Gaussian curve to the stacked spectrum.  }\label{fig:TiO2_zoom}
\end{figure*}

\begin{figure*}
\centering
\includegraphics[width=\linewidth]{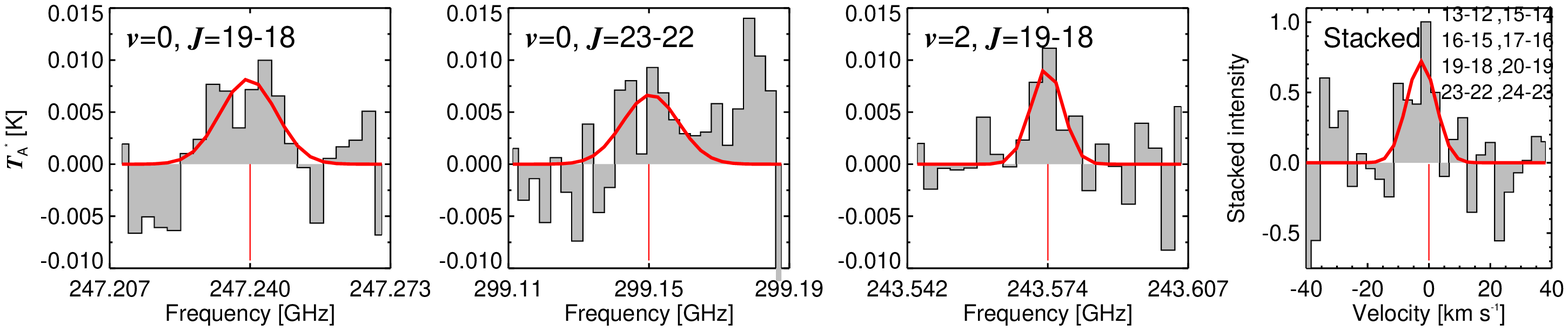}%{NaCl_zoom}
\caption{Tentative detection of NaCl line emission. The first three panels show tentative detections; for the $v=2, J=19-18$ spectrum, this is the combination of observations obtained at two different epochs (see text). The last panel shows a stacked spectrum using all rotational transitions (in $v=0$) marked in the top right-hand corner.\label{fig:NaCl_zoom}}
\end{figure*}

\subsubsection{Hydrogen cyanide, cyanide}
We detect emission from the $J=2-1,3-2$, and $4-3$ transitions of both H$^{12}$CN and H$^{13}$CN; see Fig.~\ref{fig:HCN_zoom}. Our observation of the H$^{12}$CN($3-2$) line is consistent within 10\% with the spectrum presented by \citet{schoeier2013_hcn}. For the H$^{12}$CN($4-3$) line the latter authors only listed an integrated intensity and did not show the spectrum. The integrated intensity in our survey is 77\% of that reported by \citet{schoeier2013_hcn}. This is within reasonable calibration uncertainties. We are not aware of earlier observations of the $J=2-1$ line or of any previously reported detections of H$^{13}$CN for \rdor. We do not include radiative transfer modelling of H$^{12}$CN and H$^{13}$CN in this paper. 

We do not detect CN($2-1$) emission, but we do detect two clear components of the CN($3-2$) line; see Fig.~\ref{fig:CN_zoom}. These components coincide with the intrinsically brightest hyperfine structure components of the transition.  We do not detect  $^{13}$CN, in line with the intensity of the $^{12}$CN emission and the $^{12}$C/$^{13}$C isotopic ratio derived by \citet{ramstedt2014_12co13co}.

\subsubsection{Sulphur-bearing species}
We identify 16 lines of $^{32}$SO and 12 of $^{34}$SO.  We identify 118 lines of $^{32}$SO$_2$  in the vibrational ground state, and 1 in the $v_2=1$ state (Fig.~\ref{fig:SO2_v2}), 5 lines of $^{34}$SO$_2$, 10 lines of SO$^{18}$O, and 2 lines of SO$^{17}$O.
 
\citet{danilovich2016_sulphur} presented detailed radiative transfer models of both the SO and SO$_2$ emission we observe towards \rdor, as well as towards several other oxygen-rich AGB stars. At the time of that publication, the survey had not been completed, and we have in the mean time identified more SO and SO$_2$ lines than were available then. A careful check between the lines observed in the range $159-211$\,GHz and the model predictions shows an excellent agreement for 17 SO$_2$ and 3 SO lines, and an overprediction of the SO$_2$ lines $5_{2,4}-5_{1,5}$ at 165.1\,GHz and $3_{2,2}-2_{1,1}$ at 208.7\,GHz by about a factor of 2. We note here that the SO$_2$($12_{4,8}-12_{3,9}$) transition at 355.05\,GHz modelled by \citet{danilovich2016_sulphur}  has a peak-$T_{\mathrm{A}}^{*}$ of $\sim$0.05\,K, whereas in additional data obtained in June 2016, it is 0.07\,K. This indicates possible time variability in the excitation of SO$_2$, as is also mentioned by \citet{danilovich2016_sulphur}. 

We detect emission from SiS and CS only tentatively, consistent with the idea that SO and SO$_2$ are the main sinks of sulphur in low-mass-loss-rate M-type AGB stars \citep{danilovich2016_sulphur}. We refer to the latter study for details on the modelled abundance distributions and the implications for the chemical networks.

\subsubsection{Phosphorus-bearing molecules}\label{sect:Pmols}
We detect multiple transitions of PO and PN, making \rdor the second oxygen-rich AGB star, after \iktau \citep{debeck2013_popn,velillaprieto2017_iktau_iram}, for which these species are detected. 

Our observations cover three transitions of PO in the  $\Omega=1/2$ lower-spin component: $^{2}\Pi_{1/2}\,J=9/2-7/2$, $^{2}\Pi_{1/2}\,J=11/2-9/2$, and $^{2}\Pi_{1/2}\,J=13/2-11/2$; see Fig.~\ref{fig:PO_zoom}. The tentative detection of the first two is strengthened by the definite detection of the third transition. We detect none of the higher-energy $\Omega=3/2$ upper spin component transitions. Even though the strengths of the doublets in each transition reflect excitation in, or close to, LTE, the derivation of reliable values of excitation temperature and column density through a rotational-diagram analysis is hindered by the limited sensitivity reached in the observations and the lack of spatial information. 

The survey covers four transitions of PN: $J=4-3,\dots,7-6$. We show spectra for all four in Fig.~\ref{fig:PN_zoom} and show the result of stacking these using the statistical weights of the upper levels, $2J+1$, as the weights for the transitions $J-J^{\prime}$. The resulting spectrum, stacked in velocity space at a velocity resolution of 1.5\,\kms, shows emission at peak-$S/N=3.6$. We could not identify any other species that would contribute to the velocity range $[-10;10]$\,\kms in this stacked spectrum.  The best-fit Gaussian profile to the stacked spectrum has a full width at half maximum of 6.8\,\kms, comparable to what we see for high-$S/N$ line emission in the survey. We hence claim the presence of PN in the CSE of \rdor. 

Although the peak flux densities for PO and PN for \rdor are higher than those for \iktau for similar transitions by factors of a few,  the sensitivity of our survey is unfortunately not sufficient to compare these detections of PO and PN directly to the results obtained for \iktau by \citet{debeck2013_popn} and \citet{velillaprieto2017_iktau_iram} using the SMA and the \irames telescope. Dedicated single-tuning observations at high sensitivity are needed to further investigate the phosphorus-bearing molecules in \rdor. 

We do not conclusively detect any other P-bearing molecules in our survey, but point out a tentative, and, if true, also first, detection of PNO. At 309.4\,GHz we detect an emission feature that coincides with the rest frequency of PNO($J=25-24$); see Fig.~\ref{fig:PNO_zoom}. We can rule out image band contamination and spectrometer issues, leading to the conclusion that this is an actual emission feature. However, we cannot confirm a detection for any of the other transitions of PNO covered by the survey ($J=13-12,\dots,29-28$) and cannot find a stacked spectrum with a significant S/N. Therefore, this tentative detection should be regarded with caution.

We do not detect the ground-state PH$_3$($J_K=1_0-0_0$) line at 266.9\,GHz. Based on the NH$_3$ brightness reported by \citet{justtanont2012_orich} and a P/N abundance ratio of $\sim$0.004 \citep{asplund2009_solarabundances}, we estimate its peak intensity at $<2$\,mK, whereas the sensitivity of our survey reaches 5.6\,mK rms noise at a 2\,\kms resolution at this frequency. However, this is a zeroth-order estimate which assumes similar behaviour of NH$_3$ and PH$_3$, and an analogous estimate for \iktau leads to a peak antenna temperature of $\sim$30\,mK for the IRAM 30\,m telescope, which is invalidated by the observations presented by \citet{velillaprieto2017_iktau_iram}. 

We have successfully applied for ALMA Cycle~5 observations to observe PO, PN, and PH$_3$ in the CSEs of \rdor and \iktau, but at the date of submission of this manuscript, no data have been obtained yet.

\begin{table*}[!ht]
\caption{Isotopic ratios retrieved from line-intensity ratios, with $R_{\mathrm{c}}$ the frequency-corrected ratio: per transition this corresponds to $R_{\mathrm{c}}=R_{\mathrm{a/b},J-J^{\prime}}$ and overall to $R_{\mathrm{c}}=R_{\mathrm{a/b}}$, the error-weighted mean ratio. When available, we also list the isotopologue abundance ratios $R_{\mathrm{RT}}$ obtained from radiative transfer modelling. Solar (elemental) isotopic ratios $R_{\odot}$ are taken from \citet[][see their Table 3]{asplund2009_solarabundances}. See text for details. \label{tbl:isotopes}}
\centering
\begin{tabular}{lccrccp{6cm}}
\hline\hline\\[-2ex]                                                                                    
        &       Isotopologues   &       Transition      &       $R_{\mathrm{c}}$        & $R_{\mathrm{RT}}$&      $R_{\odot}$     &       Remarks \\
\hline\\[-2ex]                                                                                  
\textbf{C}      &       $^{12}$CO/$^{13}$CO     &       $2-1$   &       $       17.0\pm3.6$     &&              &               \\
        &               &       $3-2$   &       $       9.1\pm1.9                       $       &&              &               \\
        &               &               &       $       10.9    \pm     1.7     $       &10&            &               $R_{\mathrm{RT}}$ from \citet{ramstedt2014_12co13co}.\\
        &       H$^{12}$CN/H$^{13}$CN   &       $2-1$   &       $       8.4\pm1.8                       $       &&              &               \\
        &               &       $3-2$   &       $       9.0\pm1.9                       $       &&              &               \\
        &               &       $4-3$   &       $       2.4\pm0.5                       $       &&              &       H$^{13}$CN($J=4-3$) is blended.     \\
        &               &               &$      8.7     \pm     1.3     $       &&              &       Excluding the $4-3$ transitions.  \\
        &               &      &                                                                        &&$^{12}$C/$^{13}$C = 89.40         & \\\hline\\[-2ex]
\textbf{Si}     &       $^{28}$SiO/$^{29}$SiO   &       $4-3$   &       $       2.3     \pm 0.5             $       &&              &               \\
        &               &       $5-4$   &       $       3.2\pm0.7                       $       &&              &               \\
        &               &       $6-5$   &       $       3.2\pm0.7                       $       &       &       &               \\
        &               &       $7-6$   &       $       2.5\pm0.5                       $       &&      &               \\
        &               &       $8-7$   &       $       3.1\pm0.7                       $       &&              &               \\
        &               &               &       $       2.8     \pm     0.3     $       &$12.6_{3.98}^{39.8}$ &       $^{28}$Si/$^{29}$Si = 19.69     &               \\
        &       $^{28}$SiO/$^{30}$SiO   &       $4-3$   &       $       4.0\pm0.8                       $       &&              &               \\
        &               &       $5-4$   &       $       4.3\pm0.9                       $       &&              &               \\
        &               &       $6-5$   &       $       4.2\pm0.9                       $       &&              &               \\
        &               &       $7-6$   &       $       4.1\pm0.9                       $       &&              &               \\
        &               &       $8-7$   &       $       4.0\pm0.8                       $       &&              &               \\
        &               &               &       $       4.1     \pm     0.4     $       &$20.0_{6.31}^{63.1}$&  $^{28}$Si/$^{30}$Si = 29.87 &               \\
        &       $^{29}$SiO/$^{30}$SiO   &       $4-3$   &       $       1.7\pm0.4                       $       &&              &               \\
        &               &       $5-4$   &       $       1.4\pm0.3               $       &&              &               \\
        &               &       $6-5$   &       $       1.3\pm0.3                       $       &&      &               \\
        &               &       $7-6$   &       $       1.6\pm0.3                       $       &&              &               \\
        &               &       $8-7$   &       $       1.3\pm0.3                       $       &&              &               \\
        &               &               &       $       1.4     \pm     0.1     $       &$1.58_{1.00}^{2.51}$&  $^{29}$Si/$^{30}$Si = 1.517 &               \\\\\hline\\[-2ex]
\textbf{O}      &       $^{28}$Si$^{16}$O/$^{28}$Si$^{17}$O     &       $4-3$   &       $       76.3\pm16.2                     $       &&              &               \\
        &               &       $5-4$   &       $       65.7\pm13.9                     $       &&              &               \\
        &               &       $6-5$   &       $       54.0\pm11.5                     $       &&              &               \\
        &               &       $7-6$   &       $       57.1\pm12.1                     $       &&              &               \\
        &               &       $8-7$   &       $       61.6\pm13.1             $       &&              &               \\
        &               &               &       $       61.2    \pm     5.8     $       &$398_{200}^{2000}$&    $^{16}$O/$^{17}$O = 2632  &       \\
        &       $^{28}$Si$^{16}$O/$^{28}$Si$^{18}$O     &       $4-3$   &       $       38.1\pm8.1                      $       &&              &               \\
        &               &       $5-4$   &       $       47.9\pm10.1                     $       &&              &               \\
        &               &       $6-5$   &       $       30.6\pm6.5                      $       &&              &               \\
        &               &       $7-6$   &       $       24.9\pm5.3                      $       &&              &               \\
        &               &               &       $       31.5    \pm     3.4     $       &$251_{100}^{1000}$&    $^{16}$O/$^{18}$O = 498.8 &       \\
        &       $^{28}$Si$^{17}$O/$^{28}$Si$^{18}$O     &       $4-3$   &       $       0.5\pm0.1                       $       &&              &               \\
        &               &       $5-4$   &       $       0.7\pm 0.2                      $       &&              &               \\
        &               &       $6-5$   &       $       0.6\pm 0.1                      $       &&              &               \\
        &               &       $7-6$   &       $       0.4\pm 0.1                      $       &&              &               \\
        &               &               &       $       0.5     \pm     0.1     $       &$0.631_{0.398}^{0.794}$&       $^{17}$O/$^{18}$O = 0.190 &               \\

&C$^{16}$O/C$^{17}$O    &$2-1$ &$-$&&&\\
&                                                                       &$3-2$ &$-$&&&\\
&                                                                       &                         &$-$&$>200$&& Non-detection of C$^{17}$O emission.\\
&C$^{16}$O/C$^{18}$O & $2-1$&$-$&&&\\
&                                                                       &$3-2$ &$-$&&&\\
&                                                                       &                         &$-$&$>200$&&Non-detection of C$^{18}$O emission.\\
\\\hline\\[-2ex]
\textbf{S}      &       $^{32}$SO$_2$/$^{34}$SO$_2$     & \multicolumn{2}{c}{Overall spectrum}         &      $21.6 \pm 8.5$&$^{32}$S/$^{34}$S = 22.1&               $R_{\mathrm{RT}}$ from  \citet{danilovich2016_sulphur}\\  
\hline                                                                                  

\end{tabular}
\end{table*}

\subsubsection{Heavy metal species: titanium, aluminium, sodium}\label{sect:othermols}
Titanium- and aluminium-bearing molecules could be critical in the dust-condensation process, and much effort has recently gone into searching for such species \citep{debeck2015_tio2,debeck2017_alo,decin2017_aluminium,kaminski2013_vycma_titanium,kaminski2016_mira_aluminium,kaminski2017_mira_titanium}.

We do not detect any emission from TiO in this survey, but claim a possible detection of TiO$_2$, based on a peak S/N of $\sim$3 in the stacked spectrum of a set of 18 low-$S/N$ tentative detections; see Fig.~\ref{fig:TiO2_zoom}. The sensitivity of the observations is, however, not sufficient to set up a relevant abundance or rotational-diagram analysis. One emission feature in the spectrum agrees with a component of TiN\,($8-7$), which, if confirmed, would be a first detection of this molecule in space. However, if true, we only detect one of the two doublet components of the $8-7$ transition at 297.4\,GHz clearly, although their theoretical strengths are identical. Furthermore, we do not detect emission of any other TiN transition in this survey ($5-4,\dots,9-8$). Image sideband contamination is ruled out and we can identify no other candidate carrier for this feature. Additionally, investigation of the survey of \iktau from \citet{velillaprieto2017_iktau_iram} shows a low-level peak at the same position, 297.4\,GHz, further supporting that this is a genuine spectral feature (Sect.~\ref{sect:unidentified}). 

We detect no emission from the aluminium-bearing species AlO, AlOH and AlCl with certainty, although 4, 6, and 14 transitions, respectively, are covered by the survey (AlO: $6-5,\dots,9-8$; AlOH: $6-5,\dots,11-10$; AlCl: $11-10,\dots,25-24$). In the case of AlO, only a tentative detection could possibly be claimed for the $9-8$ transition \citep{debeck2017_alo}. Stacking of these lines is difficult owing to the hyperfine structure of the individual transitions not lining up in velocity space. Recent ALMA observations at much higher sensitivity detect these species in the CSEs of \rdor and \iktau \citep{decin2017_aluminium}, with peak flux densities consistent with the non-detections in our APEX observations.

We tentatively detect emission from NaCl in the vibrational ground state $v=0$ ($J=19-18,23-22$) and in the $v=2$ vibrationally excited state ($J=19-18$) at $S/N$ of about 3, 2.5, and 5, respectively; see Fig.~\ref{fig:NaCl_zoom}. The part of the spectrum with the $v=2$ emission was observed at two different dates, one time in September 2011 and the other in November 2015. We see a change in both the intensity and width of the measured profile between the two observations. The aforementioned $S/N$ is for the combined spectrum. Given the likely radiative excitation of such vibrationally excited transitions, this line variability is possibly linked to stellar variability. Given the sensitivity of our survey data we cannot conclusively identify other lines of NaCl in the spectrum at an acceptable S/N. Figure~\ref{fig:NaCl_zoom} additionally shows a stacked spectrum combining all parts of the spectrum where NaCl (in the vibrational ground state $v=0$) has rotational transitions and that are not obviously contaminated by line emission from other species or by high noise. The resulting Gaussian fit has a $S/N\approx3$, further supporting the tentative identification of NaCl emission in the spectrum. Multiple emission lines of NaCl, in the vibrational ground state as well as in vibrationally excited states, have been detected towards \iktau and \vycma \citep{milam2007_NaCl,decin2016_vycma_nacl,velillaprieto2017_iktau_iram}.

\begin{table}\caption{Unidentified emission features. \label{tbl:ulines}}
\centering
\begin{tabular}{crrrp{2.2cm}}
\hline\hline\\[-2ex]
Frequency &$T_{A\mathrm{,peak}}^{*}$ & \fwhm & $\delta\varv$&  Remarks \\
(MHz) & (mK) & (\kms)&(\kms)&\\
\hline\\[-2ex]
173514 & 56 & 7.4&1.1\\
189362 & 10 & 13.2&2.9& \\ 
204008 & 9 & 8.2&2.7\\ 
225998 & 10 & 11.1&2.4\\
267993 & 16 & 5.1&2.0\\
273284 & 61 & 7.6&1.1\\
280051 & 14 & 5.0&1.6\\
283535 & 13 & 9.3&1.9\\
295629 & 25 & 9.0&1.9\\
297404 & 17 & 16.0 & 1.8& Also seen towards \iktau. TiN only tentative candidate.\\ 
298464 & 16 & 7.5&1.8\\
309449 & 13  &3.9&1.4 \\ 
319253 & 10 & 8.0&1.8\\
320204 & 46 & 7.8&0.9\\
332217 & 42 & 6.3&0.9\\
354195 & 1600 & --\tablefootmark{$\dagger$} &0.5& Maser. H$_2$SiO? See Sect.~\ref{sect:h2sio}.\\
\hline
\end{tabular}
\tablefoot{The listed frequency, peak temperature, and \fwhm are obtained from fitting a Gaussian line profile at the velocity resolution $\delta\varv$ --- see Fig.~\ref{fig:ulines}.
\tablefoottext{$\dagger$}{We do not fit a Gaussian line profile to this emission feature. }
}
\end{table}

\begin{figure*}
\centering
\includegraphics[width=\linewidth]{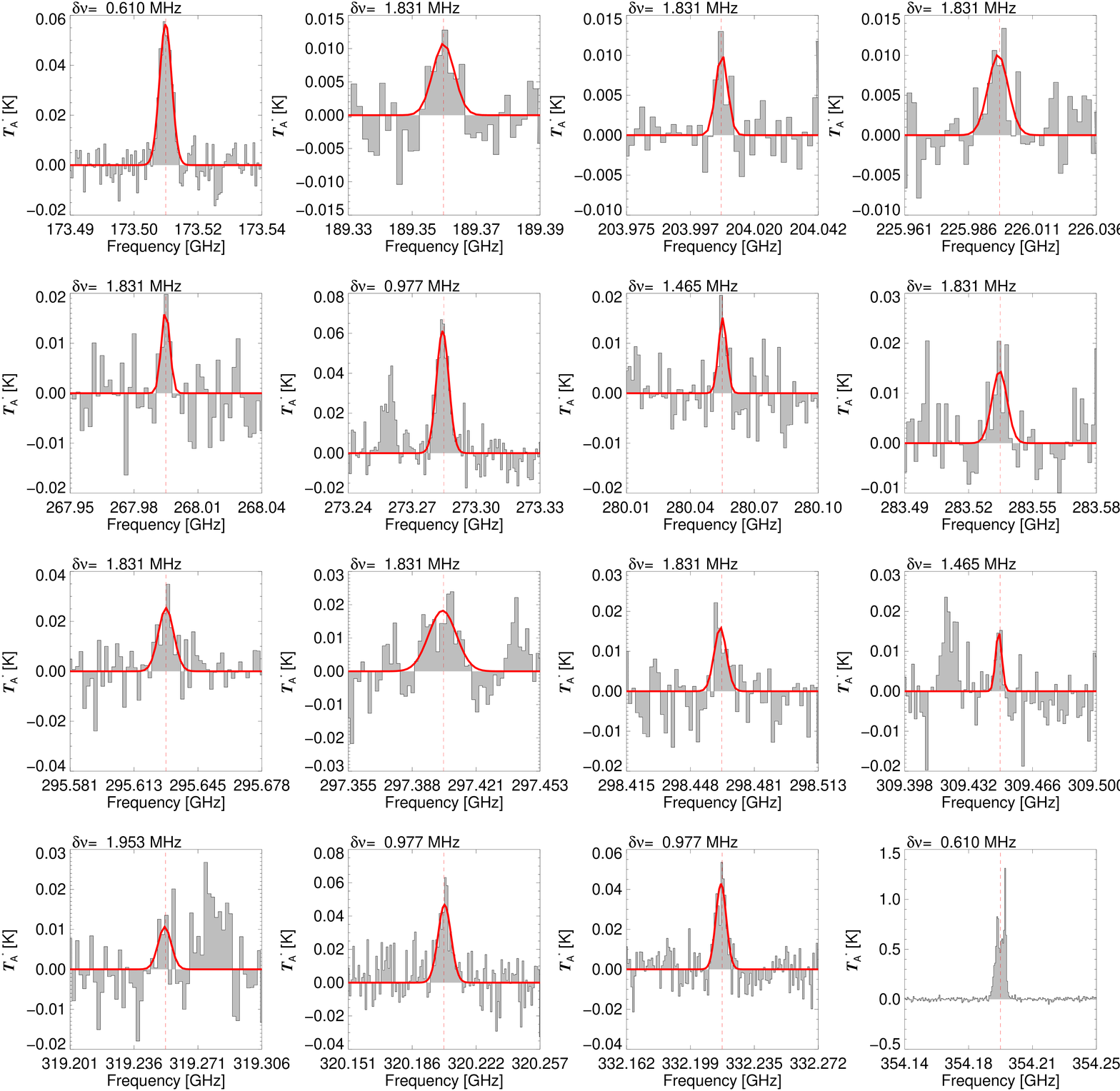}%{ulines_together}
\caption{Unidentified line emission features. We fit a single Gaussian line profile to each of these, except the suspected maser line at 354\,GHz, and list the retrieved central frequency, peak intensity, and FWHM at the given velocity resolution in Table~\ref{tbl:ulines}. The corresponding frequency resolution $\delta\nu$ of the spectra is given at the top left of each panel.}\label{fig:ulines}
\end{figure*}

\begin{figure*}
\centering
\includegraphics[width=.85\linewidth]{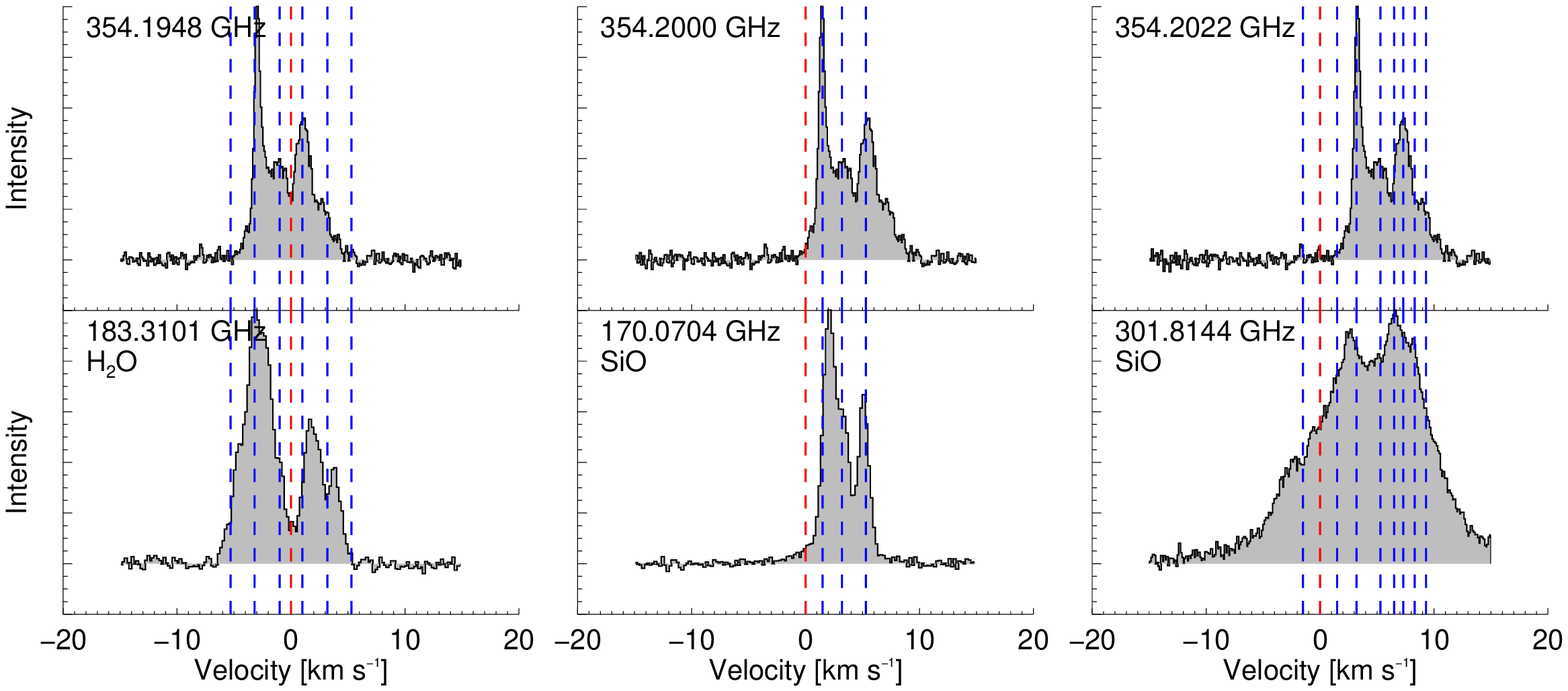}%{u354_zoom}
\caption{Alignment in velocity space of unidentified emission at $\sim$354.2\,GHz \emph{(top)} with known maser emission \emph{(bottom)}. The spectra shown in the top panels, from left to right, assume rest frequencies of 354.1948\,GHz, 354.2000\,GHz, and 354.2022\,GHz, respectively. The bottom panels show, from left to right, H$_2$O\,($J_{K_{\rm a},K_{\rm c}}=3_{1,3}-2_{2,0}$) at 183.3\,GHz, SiO\,($v=3,J=4-3$) at 170.1\,GHz, and SiO\,($v=1,J=7-6$) at 301.8\,GHz. Vertical dashed lines indicate emission components similar to those mentioned in the discussion in Sect.~\ref{sect:results}. } \label{fig:u354}
\end{figure*}

\subsection{Isotopes}\label{sect:isotopes}
Our survey covers emission from isotopologues containing different isotopes of carbon ($^{12}$C, $^{13}$C), oxygen ($^{16}$O, $^{17}$O, and $^{18}$O), silicon ($^{28}$Si, $^{29}$Si, $^{30}$Si), and sulphur ($^{32}$S, $^{34}$S). 

Accurate isotopic ratios of different elements carry information on the evolutionary stage and/or the initial mass of an AGB star, based on the assumption that the circumstellar (molecular) isotopologue ratios are representative of the atmospheric (elemental) isotopic ratios. This assumption is reasonable, since the only currently known chemical processes involved in changing the isotopologue ratio are fractionation in the very coldest parts of the CSE and isotope-selective photodissociation, for example, of CO \citep{visser2009}.  \citet{saberi2017_rscl_hcn} discuss the difference between $^{12}$CO/$^{13}$CO and H$^{12}$CN/H$^{13}$CN in the CSE of the AGB star R~Scl as a result of isotope-selective photodissociation in the case of CO, and how, consequently, H$^{12}$CN/H$^{13}$CN is likely more representative of the stellar $^{12}$C/$^{13}$C in certain cases. Since the dissociation of SiO is in the continuum, we expect our derived isotopologue ratios to be representative also for the elemental isotopic ratios of silicon and oxygen. 
  
We summarise intensity ratios for a large set of isotopologue emission lines in our survey in Table~\ref{tbl:isotopes}. These transition-specific values, $R_{\mathrm{a/b},J-J^{\prime}}$, are obtained as
\begin{equation}
R_{\mathrm{a/b},J-J^{\prime}} = \frac{I_{\mathrm{a},J-J^{\prime}}}{I_{\mathrm{b},J-J^{\prime}}} \times \left(\frac{\nu_{\mathrm{a},J-J^{\prime}}}{\nu_{\mathrm{b},J-J^{\prime}}}\right)^{-3}
,\end{equation} 
where $\nu_{J-J^{\prime}}$ is the rest frequency of each rotational transition $J-J^{\prime}$, $I_{J-J^{\prime}}$ is its integrated line intensity, and `a' and `b' denote the different isotopic variations. The cubic frequency-dependent correction factor accounts for the difference in beam-filling (when all lines are assumed spatially unresolved) and in intrinsic line strength \citep[see e.g.][]{schoeier2000_12c13c,debeck2010_comdot}. We also list approximate isotopologue ratios $R_{\mathrm{a/b}}$, calculated as the weighted average of the transition-specific values, assuming the uncertainties as the inverse weights. We note that these ratios are sometimes obtained from optically thick line emission, which is likely when a highly abundant isotopologue is considered and is very clear in the case of $^{28}$SiO, so in these cases the obtained ratio should be considered a limiting value. Only the isotopologue ratios obtained from detailed radiative transfer modelling ($R_{\mathrm{RT}}$) and those derived from exclusively optically thin line emission are representative of the actual isotopic ratio in the CSE. All calculations of line ratios in this section are based on a 15\% uncertainty on the line intensities. This is reasonable, and maybe rather conservative, given that most lines that are compared directly have been observed (quasi-)simultaneously and are almost always observed with the same instrument. 

\subsubsection{Carbon: CO, HCN}
\citet{ramstedt2014_12co13co} determine $^{12}$CO/$^{13}$CO to be $\sim$10 for \rdor, based on radiative transfer modelling of CO. Although optical depth could affect the line ratios for both CO and HCN, the values we find from line intensity ratios are consistent with the value derived by \citet{ramstedt2014_12co13co}. We do not re-model the CO line emission, nor model the HCN line emission here, but plan a detailed radiative transfer analysis of the H$^{12}$CN and H$^{13}$CN line emission in the future. 

The derived $^{12}$C/$^{13}$C is significantly lower than any values predicted by theoretical models, an issue also discussed by \citet{ramstedt2014_12co13co}. 

\subsubsection{Oxygen: SiO}
From the line intensity ratios we find an isotopologue abundance ratio $^{28}$Si$^{17}$O/$^{28}$Si$^{18}$O$\approx0.5\pm0.1$ (Table~\ref{tbl:isotopes}). From the radiative transfer modelling presented in Sect.~\ref{sect:sio}, we derive $^{28}$Si$^{17}$O/$^{28}$Si$^{18}$O=$0.6\pm0.2$.  Assuming that the isotopologue abundance ratio derived for the CSE is equal to the elemental $^{17}$O/$^{18}$O isotopic ratio at the stellar surface we find that $^{17}$O/$^{18}$O$=0.6\pm0.2$, implying an initial mass, $M_{\rm i}$ for \rdor of $1.4_{-0.1}^{+0.2}$\,\msun  \citep[][assuming solar metallicity]{karakas2016,cristallo2015}. 

Independently from the study presented here, \citet{danilovich2017_water_isotopologues} model the emission of several transitions of H$_2^{17}$O and H$_2^{18}$O towards \rdor and derive isotopologue abundance ratios o-H$_2^{17}$O/o-H$_2^{18}$O=$0.54\pm0.26$ and p-H$_2^{17}$O/p-H$_2^{18}$O=$0.30\pm0.10$ for ortho-H$_2$O and para-H$_2$O, respectively, from which they derive an initial mass in the range $1.0-1.3$\,\msun, although a slightly larger range of $1.0-1.6$\,\msun seems to better reflect their results. These results are in agreement with our findings based on the SiO emission, increasing the reliability of the derived $^{17}$O/$^{18}$O for the CSE and of the inital mass estimate, under the assumptions made. 

We do not detect C$^{17}$O or C$^{18}$O in our survey.  Assuming the abundance ratios Si$^{16}$O/Si$^{17}$O and Si$^{16}$O/Si$^{18}$O to be representative of C$^{16}$O/C$^{17}$O and C$^{16}$O/C$^{18}$O, we run radiative transfer models for C$^{17}$O and C$^{18}$O. The predictions for the $J=2-1$ and $J=3-2$ transitions are in agreement with formal non-detections in the survey, with peak intensities well below 10\,mK and 20\,mK for the respective transitions, for both C$^{17}$O and C$^{18}$O for peak abundances $f_0<10^{-6}$. Such high values of $f_0$ were considered only because of the large uncertainties on the SiO isotopologue abundance ratios.  This upper limit to the abundances means that the uncertainty on $^{18}$O/$^{16}$O is smaller than given by our radiative-transfer modelling results in Table~\ref{tbl:isotopes}, and can be constrained to a lower limit of 200.

Based on the evolutionary models of \citet{karakas2016} and \citet{cristallo2015}, our estimate of $M_{\rm i}=1.3-1.6$\,\msun implies that if \rdor becomes a carbon-rich AGB star it will only do so in its final phases with a C/O ratio only slightly above 1. During its  lifetime of $\sim1.6$\,Myr on the thermally pulsing AGB (TP-AGB), the star would dredge up $\lesssim0.01$\,\msun. According to  \citet{karakas2016}, this would happen over the course of $\sim$16 TP cycles, whereas \citet{cristallo2015} quote only 5 TP cycles for a star with $M_{\rm i}=1.5$\,\msun.

\subsubsection{Silicon: SiO}
From the radiative transfer modelling described in Sect.~\ref{sect:sio} we derive isotopologue abundance ratios $^{28}\mathrm{SiO}/^{29}\mathrm{SiO}=12.6^{79.4}_{1.6}$, $^{28}$SiO/$^{30}$SiO=$20.0^{79.4}_{2.0}$, and $^{29}\mathrm{SiO}/^{30}\mathrm{SiO}=1.6_{0.4}^{6.3}$. The latter is in good agreement with the line intensity ratio $R_{^{29}\mathrm{SiO}/^{30}\mathrm{SiO}}=1.4\pm0.1$ (Table~\ref{tbl:isotopes}). These derived ratios are in agreement with the solar isotopic ratios $^{29}$Si/$^{30}$Si=1.5 and $^{28}$Si/$^{30}$Si=29.9 \citep{asplund2009_solarabundances}. The models of \citet{karakas2016} indeed show no, or insignificant, changes in $^{28}$Si/$^{29}$Si and $^{28}$Si/$^{30}$Si as a consequence of AGB evolution for initial masses $\leq2$\,\msun.

\subsubsection{Sulphur: SO, SO$_2$} 
\citet{danilovich2016_sulphur} derived $^{32}$S/$^{34}$S$=22\pm9$ from detailed radiative transfer models of SO and SO$_2$. This is in good agreement with the solar value $^{32}$S/$^{34}$S=22.1 \citep{asplund2009_solarabundances} and the model results that sulphur does not show significant changes in its isotopic ratios throughout AGB evolution.

\subsubsection{Evolutionary stage}
In order to constrain not only the initial mass of \rdor, but also its evolutionary stage, we need observationally determined isotopic ratios for, for example, nitrogen ($^{14}$N/$^{15}$N) and aluminium ($^{26}$Al/$^{27}$Al), based on highly sensitive observations. Furthermore, there is a strong need for better constrained theoretical models for the variation of $^{12}$C/$^{13}$C throughout the AGB evolution. We refer to \citet{ramstedt2014_12co13co} for an in-depth discussion on the low $^{12}$C/$^{13}$C found for oxygen-rich AGB stars compared to stellar evolution models.

\subsection{Unidentified lines}\label{sect:unidentified}
We list the unidentified features in Table~\ref{tbl:ulines} with the rest frequencies, peak intensities, and \fwhm:s obtained from a Gaussian line profile fitting and show the spectra and their fits in Fig.~\ref{fig:ulines}. The listed velocity resolution gives a $S/N\geq3$ for the Gaussian fit in all cases. For all unidentified features we are able to rule out instrumental effects and image contamination, thanks to the coverage of our survey. None of these lines are reported for the SMA  and \irames surveys of \iktau, \vycma, or \irc  \citep{debeck2013_popn, kaminski2013_vycma_sma, patel2011, velillaprieto2017_iktau_iram}. One unidentified feature, at 297.4\,GHz, seems to be present in the \irames survey of \iktau, but was not explicitly reported by \citet{velillaprieto2017_iktau_iram}. The fully independent detections of this feature towards two significantly different CSEs and using two different telescopes suggests that it is genuine.

\subsubsection{A new maser at 354.2\,GHz}\label{sect:h2sio}
The most striking unidentified feature appears at 354.2\,GHz and is markedly maser-like in its shape. The emission reaches a peak antenna temperature of $\sim$1.6\,K ($\sim$66\,Jy; peak-$S/N\approx55$ at 0.1\,\kms resolution), and a total integrated intensity of 4.7\,K\,\kms ($\sim$193\,Jy\,\kms). A detailed check of the collected data reveals that this feature is present in all consecutively obtained individual scans and that its distinct shape is repeated in all of these. In combination with the resemblance to other spectral lines in the survey (see discussion below) this leads us to believe that this feature is a real spectral emission line.

We compare the emission feature to three maser lines in our survey with a similar shape,  H$_2$O($J=3_{1,2}-2_{2,0}$) at 183.3\,GHz, SiO($v=3,J=4-3$) at 170.1\,GHz, and SiO($v=1, J=7-6$) at 301.8\,GHz, in Fig.~\ref{fig:u354}. In order to align the peaks in the unidentified feature with those of the former two, we need to centre the line at 354.195\,GHz and 354.200\,GHz, respectively. This shift is due to the fact that the H$_2$O line emission originates in both the blue and red sides of the wind, whereas this particular SiO maser seems to originate almost exclusively from the redshifted part of the wind. From the comparison in Fig.~\ref{fig:u354}, it seems that the emission could be "H$_2$O-like", rather than "SiO-like" when one considers that the reddest emission in the profile has no counterpart in the SiO maser emission.  However, the closest known H$_2$O maser to this frequency lies at 354.8\,GHz \citep{gray2016_h2omasers} and none of the listed transitions of the isotopologues H$_2^{17}$O and H$_2^{18}$O coincide with this frequency, either. 

Searching the spectroscopic catalogues we found only  the  ortho-H$_2$SiO (silanone) doublet $J_{K_{\mathrm{a}},K_{\mathrm{c}}}=10_{4,7}-9_{4,6}$ and $J_{K_{\mathrm{a}},K_{\mathrm{a}}}=10_{4,6}-9_{4,5}$ to be a potential candidate. The doublet is located at 354202.1540\,MHz, with lower and upper level energies of 191\,K and 208\,K, respectively. The upper limit on the uncertainty on this rest frequency listed in CDMS is 0.02\,MHz. If confirmed, this is, to our knowledge, the first detection of this molecule in space. Assuming this as the rest frequency moves the emission entirely to the red velocities, with little or no resemblance with the H$_2$O maser and the first SiO maser in Fig.~\ref{fig:u354}. However, the maser emission of the SiO($v=1,J=7-6$) line at 301.8\,GHz (rightmost bottom panel in Fig.~\ref{fig:u354}) is also dominated by red-shifted emission and shows some distinct peaks that might very well correspond to those seen in the new feature. Recently, \citet{gobrecht2016_dustformation_iktau} predicted that H$_2$SiO can reach abundances of up to $10^{-6}-10^{-5}$ in the inner wind of the oxygen-rich, high-mass-loss-rate Mira-type variable \iktau. H$_2$SiO, together with HSiO, is thought crucial in the nucleation of silicate dust clusters. Whether or not these predictions can be used as representative of the low-mass-loss-rate, semi-regular variable \rdor is not clear. We do not detect any other emission from H$_2$SiO in our survey, even though transitions over a large range of excitation energies are covered. Given the maser nature of the emission line tentatively identified as H$_2$SiO, it is difficult to assess whether the non-detections are consistent with this detection.

Additional observations at this frequency, carried out on 9 June 2016, unfortunately do not reveal any trace of the emission at a $\sim$10\,mK rms noise level at 2\,\kms resolution. Archival observations of \rdor, obtained with APEX on 26 March 2008, equally show no detectable emission at 354.2\,GHz, however at a much worse rms noise of $\sim$400\,mK. We point out that the survey observations at this particular frequency were carried out on 1 September 2011, when the light curve variations of \rdor were quite regular. Currently, however, \rdor's light curve is much more chaotic and shows smaller overall amplitude changes (source: AAVSO). Considering the maser nature of the line, this change might very well significantly affect the excitation of this unidentified line. 

Observations at this frequency and other frequencies of likely observable line emission in the H$_2$SiO spectrum should also be obtained for a larger sample of AGB stars, with an initial focus on M-type stars in order to understand the physical conditions under which this maser line could be excited. 

\section{Comparison to other line surveys}\label{sect:comparison}
Spectral surveys of CSEs of M-type evolved stars have been published only in recent years. This includes the AGB stars \iktau and  \object{OH 231.8 +4.2} \citep{debeck2013_popn,debeck2015_AGBconfproceedings, sanchezcontreras2011_iktau_oh231,sanchezcontreras2015_ions,velillaprieto2015_Nbearing,velillaprieto2017_iktau_iram}, the red supergiant \vycma \cite{kaminski2013_vycma_sma}, and the yellow hypergiant \object{IRC +10420} \citep{quintanalacaci2016_irc10420}. All of these stars lose mass at high rates \citep[roughly $\gtrsim0.5\times10^{-5}$\,\msunyr; e.g.][and references therein]{debeck2010_comdot,velillaprieto2015_Nbearing}, leading to significantly higher densities in their CSEs than is the case for \rdor, with its low mass-loss rate and low expansion velocity \citep[$\dot{M}=1-2\times10^{-7}$\,\msunyr, $v_{\rm exp}=5.7$\,\kms;][]{maercker2016_water}. Apart from \iktau and \rdor, all of these sources have been shown to have outflows that very strongly deviate from smooth, spherical, constant winds \citep[e.g.][ and references therein]{bujarrabal2002_oh231,castrocarrizo2007_hypergiants,richards2014_alma_vycma}. Given all of the above, we only compare our results to those obtained for \iktau.

Figure~\ref{fig:fullscan} shows a direct comparison of our APEX data of \rdor with the \irames data of \iktau from \citet{velillaprieto2017_iktau_iram} scaled  to account for the difference in intensity that would be an effect of the different mass-loss rate and distance of the two objects and for the difference in point-source sensitivity between the telescopes (see App.~\ref{sect:overview}). The sensitivity reached in the spectral scan of \iktau is better than that of the \rdor spectral scan presented here. However, the ratio of the rms noise to the peak intensity of the closest CO emission lines is similar when assuming a spectral resolution (2\,MHz and 0.6\,MHz for \iktau and \rdor, resp.) that results in the same number of spectral elements covering the full line width of twice the expansion velocity ($v_{\rm{exp}}$ is 18.5\,\kms and 5.7\,\kms for \iktau and \rdor, resp.). This allows us, to a certain degree, to compare the two surveys. 

Both surveys show many spectral features pertaining to SO$_2$ and SiO, including isotopologues and multiple vibrational states, with strong emission lines from SiS and CS notably missing from the \rdor spectrum. The observations of \iktau reveal several other molecules that we do not  detect towards \rdor: H$_2$S, NS,  HNC, NO, H$_2$CO, and HCO$^+$. We do not detect any molecules towards \rdor that are not seen towards \iktau, apart from (tentatively) some specific, low-abundance isotopologues. At the same time, it is remarkable that none of the \rdor u-lines (except for possibly one) are seen in the \iktau spectrum. 

\cite{danilovich2016_sulphur} already reported that SO and SO$_2$ are the main reservoirs of S in the CSE of \rdor and that their abundances roughly decrease with increasing mass-loss rate when also studying other M-type CSEs (including \iktau). \citet{danilovich2017_h2s} reported that H$_2$S is unlikely to play any significant role at mass-loss rates $\lesssim 5\times10^{-6}$\,\msunyr. This is consistent with the lack of CS, SiS, and H$_2$S in \rdor.

Assuming the simple scale factor quoted above, we find that all HNC, NO, H$_2$CO, or HCO$^+$ emission lines would fall below the detection limit of our observations. Additionally, assuming that the HNC/HCN intensity ratios found for \iktau also hold for \rdor, further supports the non-detection of all observed HNC transitions ($J=2-1,3-2,4-3$). Considering this, we cannot rule out that \rdor would show emission from these molecules in more sensitive observations and cannot claim, from our survey results, a difference in chemistry to be at the base of the absence of emission from these in the case of \rdor. However, this difference is clearly at the base of the differences seen between the sulphur-bearing species in the two CSEs. Stacking did not lead to tentative detections for any of these molecules.

The H$_2$O isotopologue model results of \citet{danilovich2017_water_isotopologues} imply $M_{\rm i}=1.0-1.6$\,\msun for both \rdor and \iktau. The integrated line intensities of Si$^{17}$O and Si$^{18}$O reported by \citet{velillaprieto2017_iktau_iram}, unfortunately, do not constrain very well the ratio Si$^{17}$O/Si$^{18}$O, making an independent estimate of $^{17}$O/$^{18}$O impossible. The fact that the two stars appear so different in terms of pulsational and mass-loss properties (\iktau is a Mira, \rdor is an SRb variable; their mass-loss rates differ by $1-2$ orders of magnitude) leads us to hypothesise that \iktau could be in a later stage of its AGB evolution than \rdor, considering the trend of increasing mass-loss rate, expansion velocity, luminosity, and pulsation period with evolution along the TP-AGB \citep[e.g.][]{vassiliadis1993}.

\begin{figure*}
\centering
\subfigure[CO \label{fig:CObumps}]{\includegraphics[width=.33\linewidth]{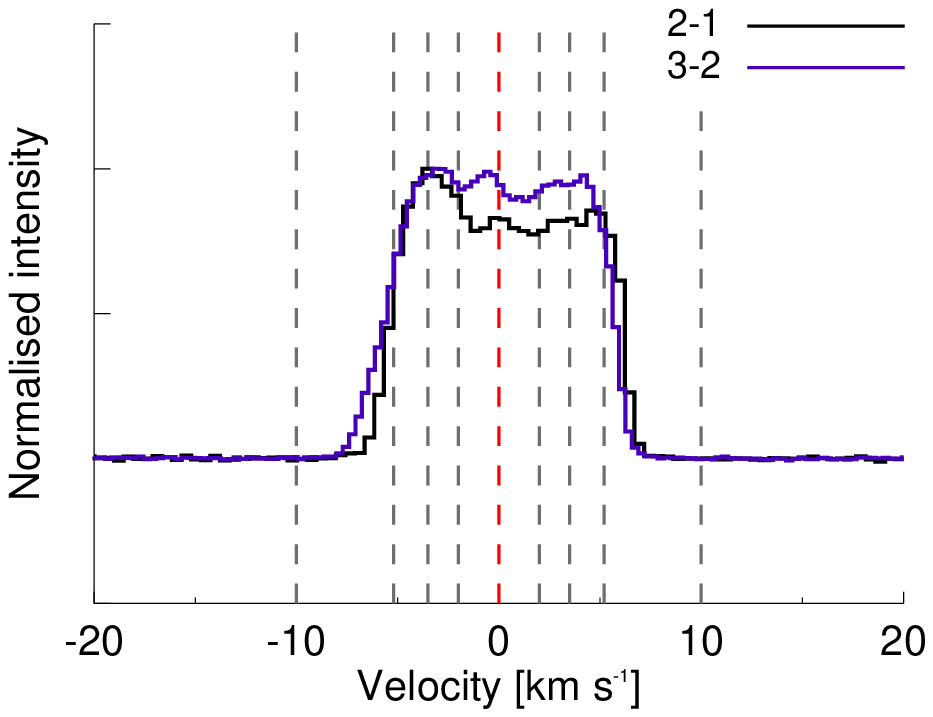}}%{CO_bump}}
\subfigure[CO, HIFI \label{fig:bumps_CO_HIFI}]{\includegraphics[width=.33\linewidth]{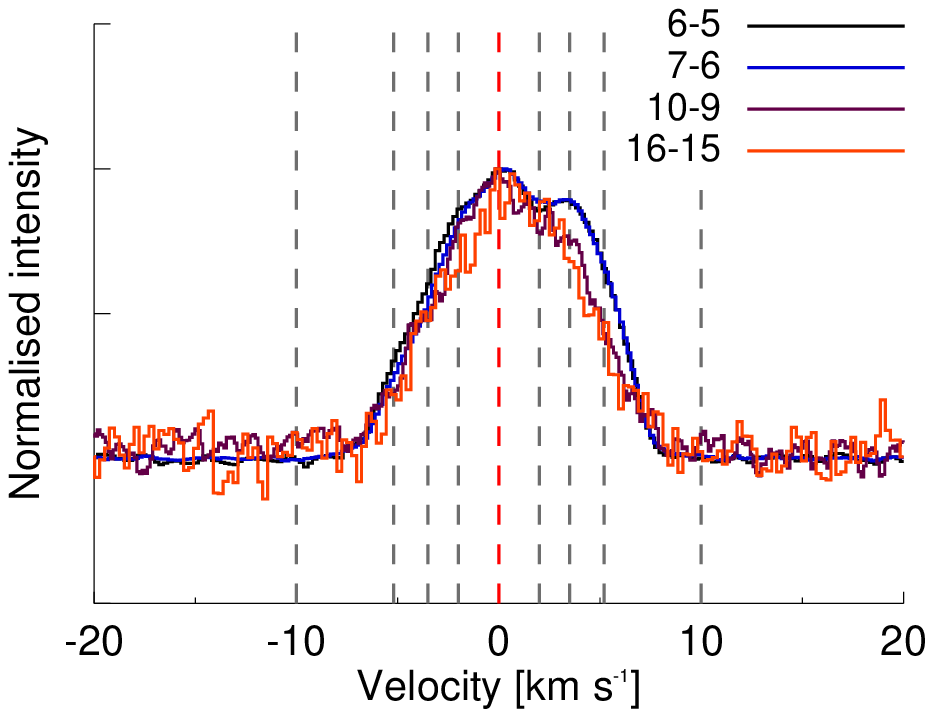}}%{CO_HIFI_bump}}
\subfigure[$^{13}$CO]{\includegraphics[width=.33\linewidth]{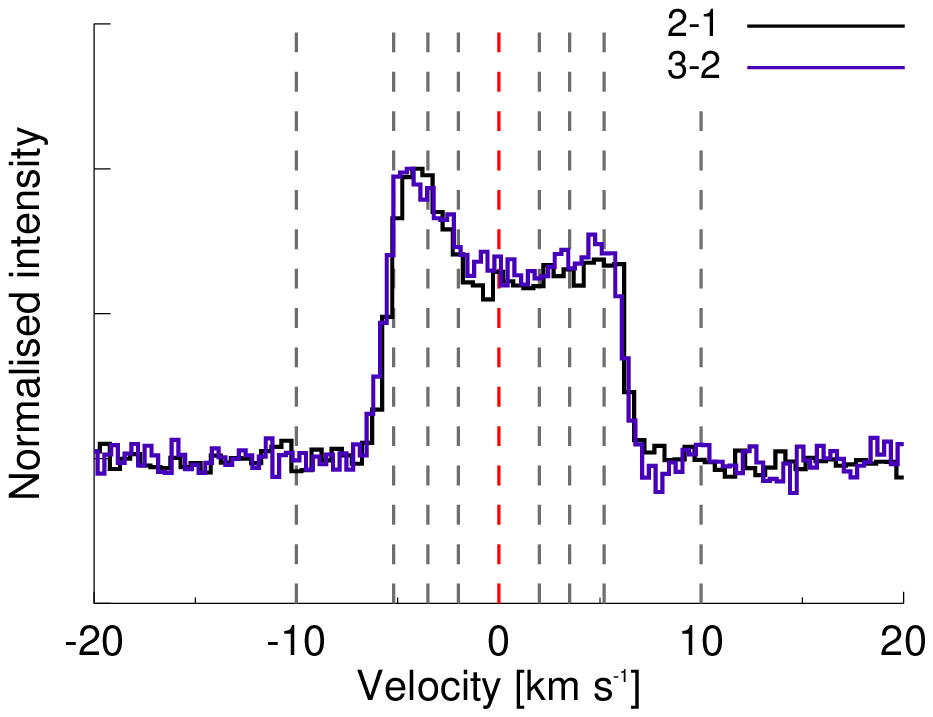}}%{13CO_bump}}
\\
\subfigure[HCN]{\includegraphics[width=.33\linewidth]{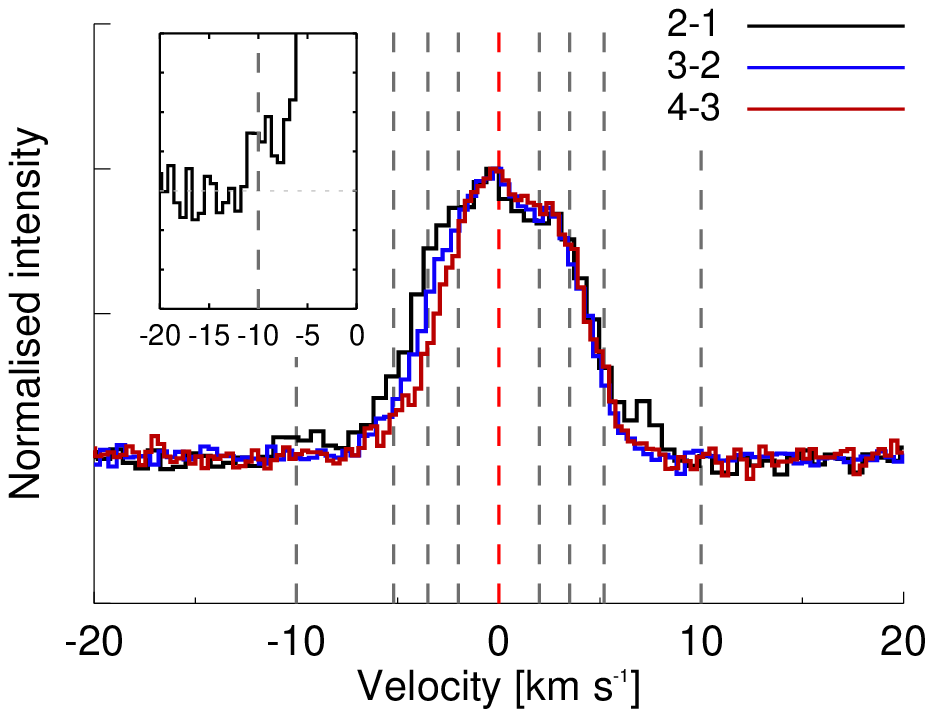}}%{HCN_bump}}
\subfigure[H$^{13}$CN]{\includegraphics[width=.33\linewidth]{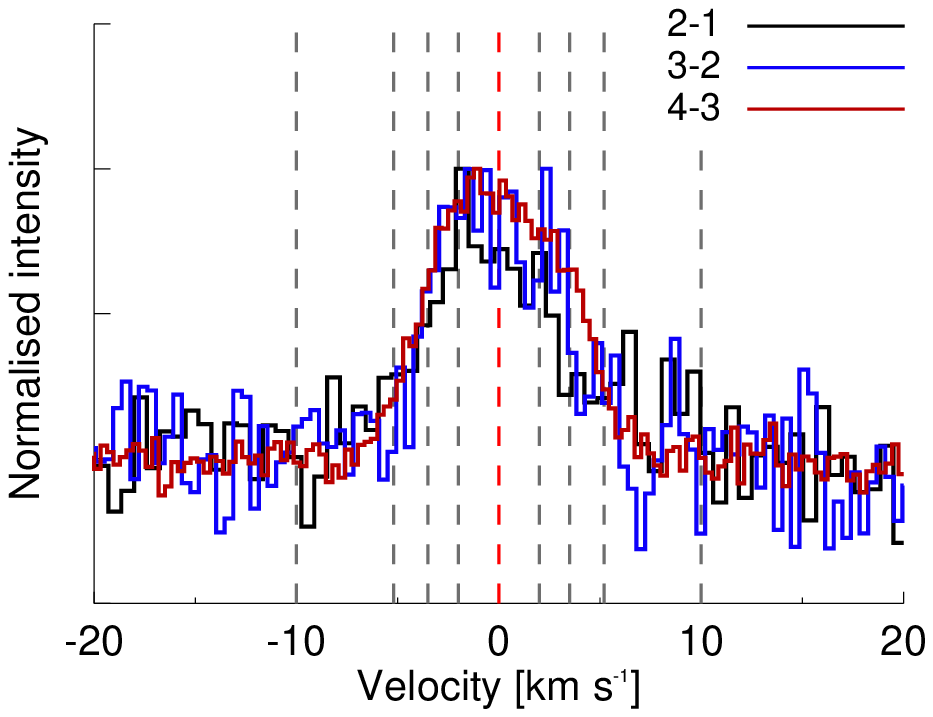}}%{H13CN_bump}}
\subfigure[H$_2$O]{\includegraphics[width=.33\linewidth]{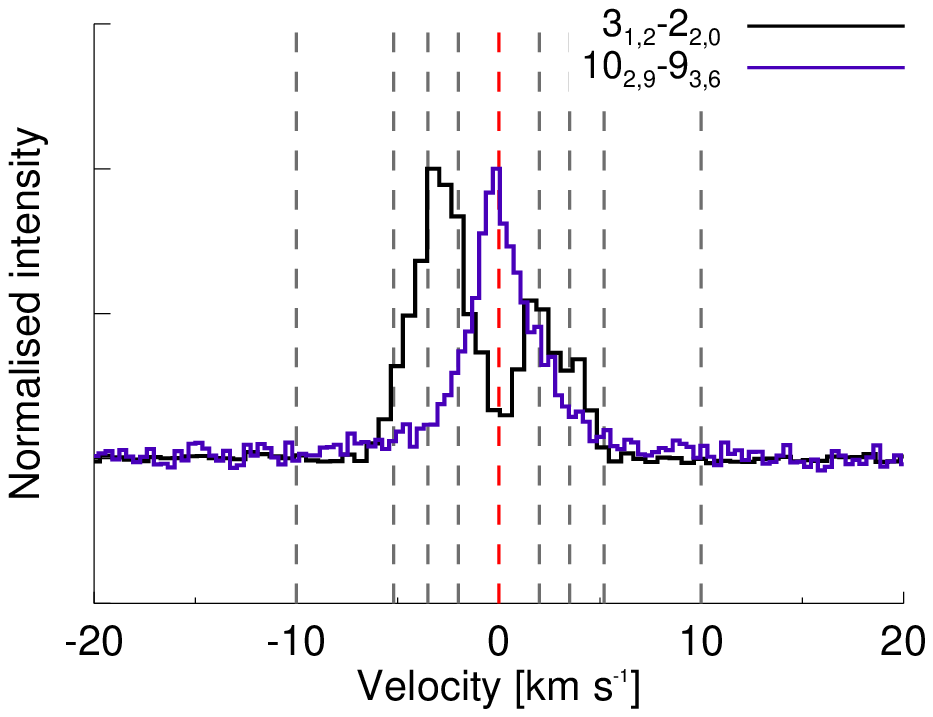}}%{H2O_bump}}
\\
\subfigure[$^{28}$SiO\label{fig:bumps_28sio}]{\includegraphics[width=.33\linewidth]{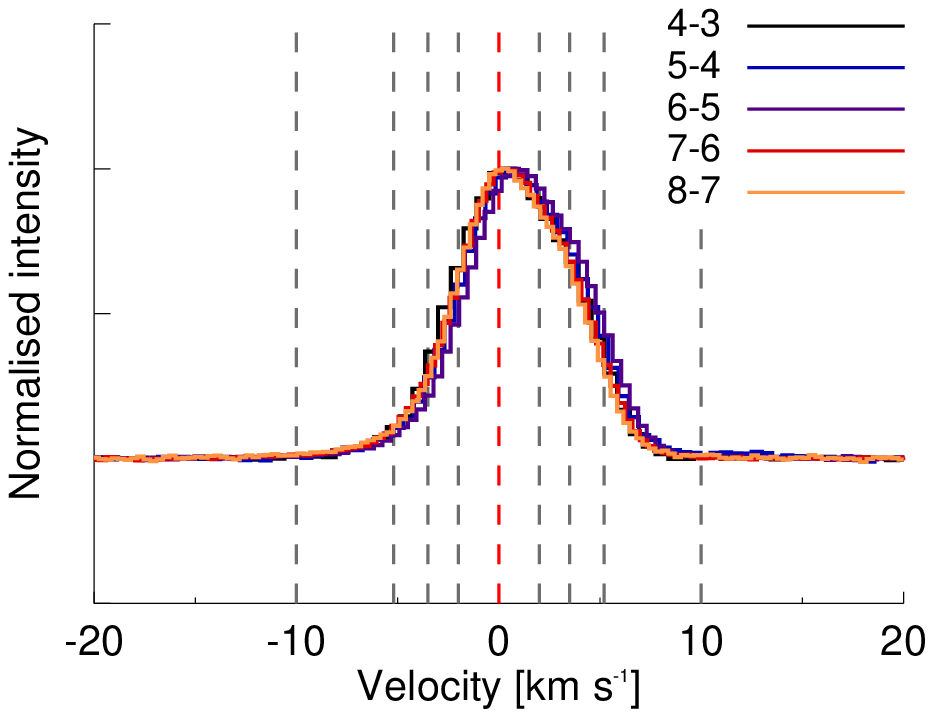}}%{SiO_bump}}
\subfigure[$^{28}$SiO, HIFI\label{fig:bumps_28sio_HIFI}]{\includegraphics[width=.33\linewidth]{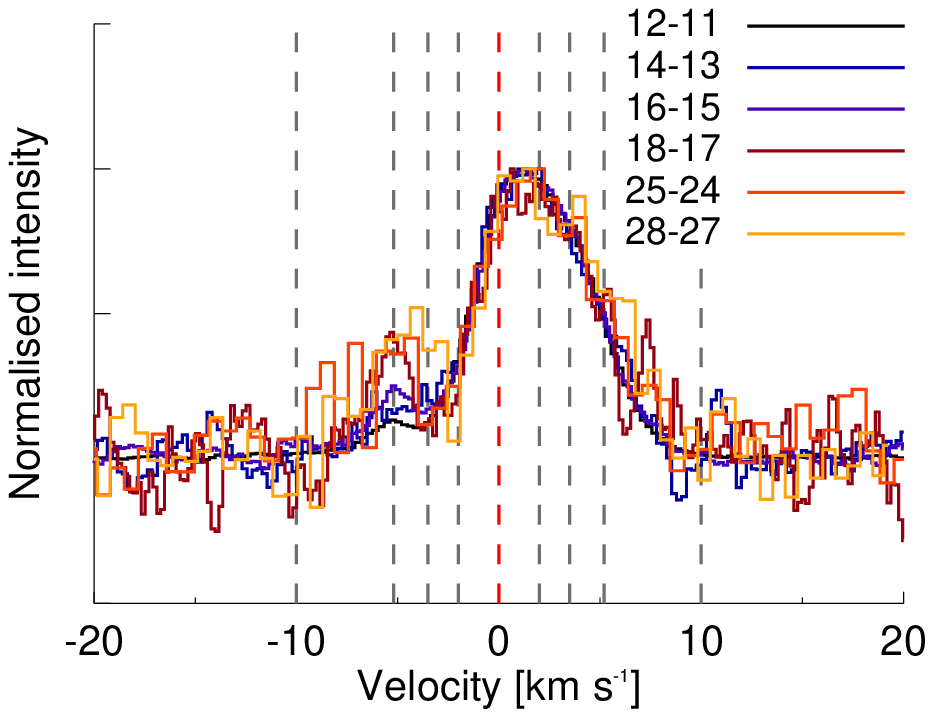}}%{SiO_HIFI_bump}}
\subfigure[$^{29}$SiO\label{fig:bumps_29sio}]{\includegraphics[width=.33\linewidth]{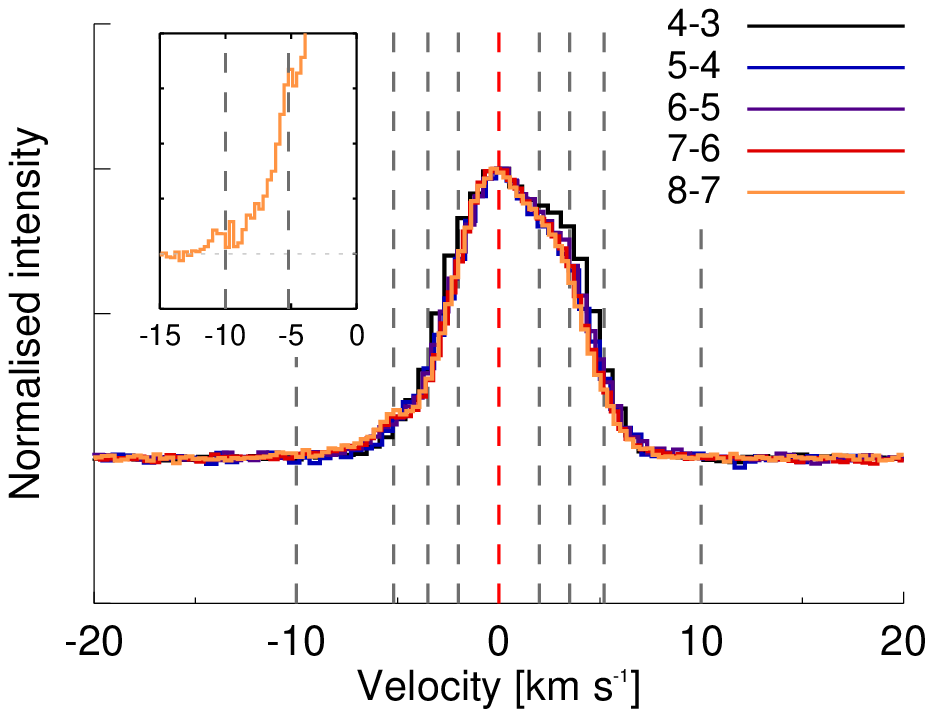}}%{29SiO_bump}}
\\
\subfigure[$^{30}$SiO\label{fig:bumps_30sio}]{\includegraphics[width=.33\linewidth]{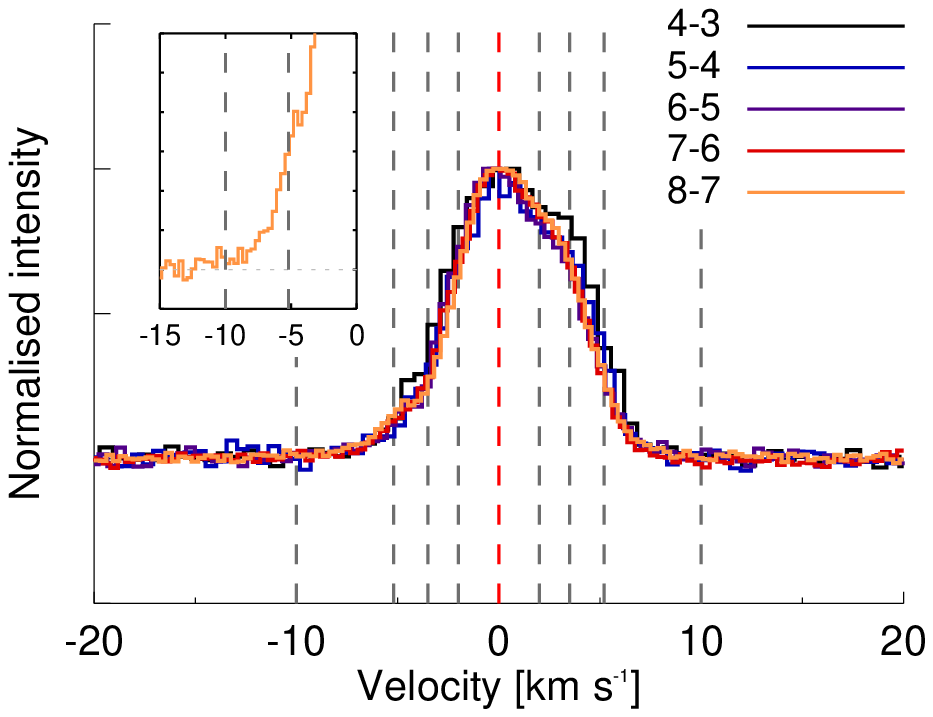}}%{30SiO_bump}}
\subfigure[SO (selection) \label{fig:bumps_SO}]{\includegraphics[width=.33\linewidth]{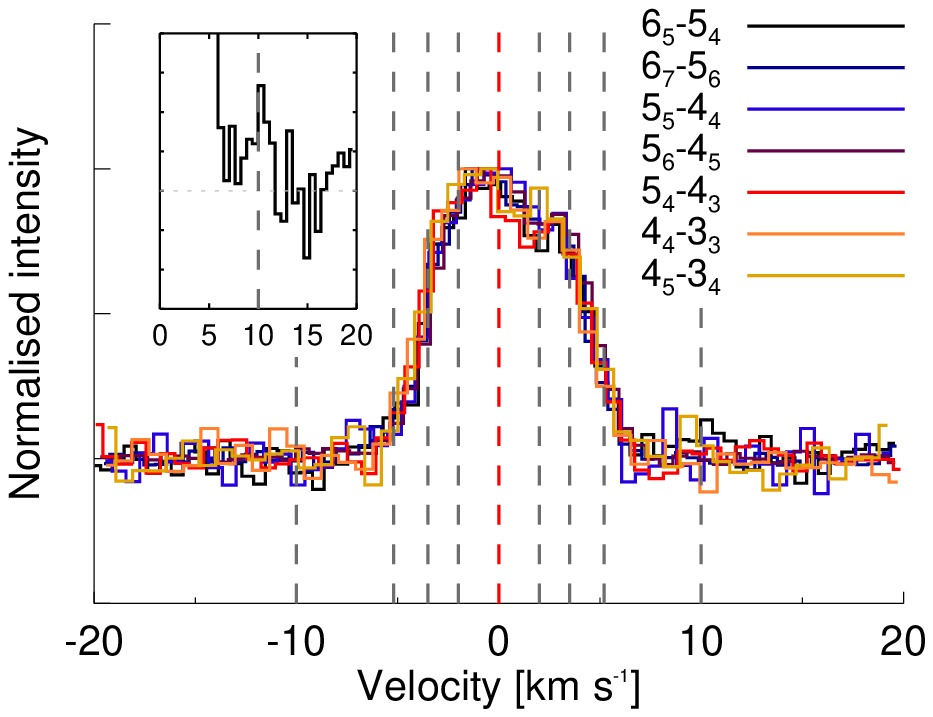}}%{SO_selected_bump}}
\subfigure[SO$_2$ (selection) \label{fig:bumps_SO2}]{\includegraphics[width=.33\linewidth]{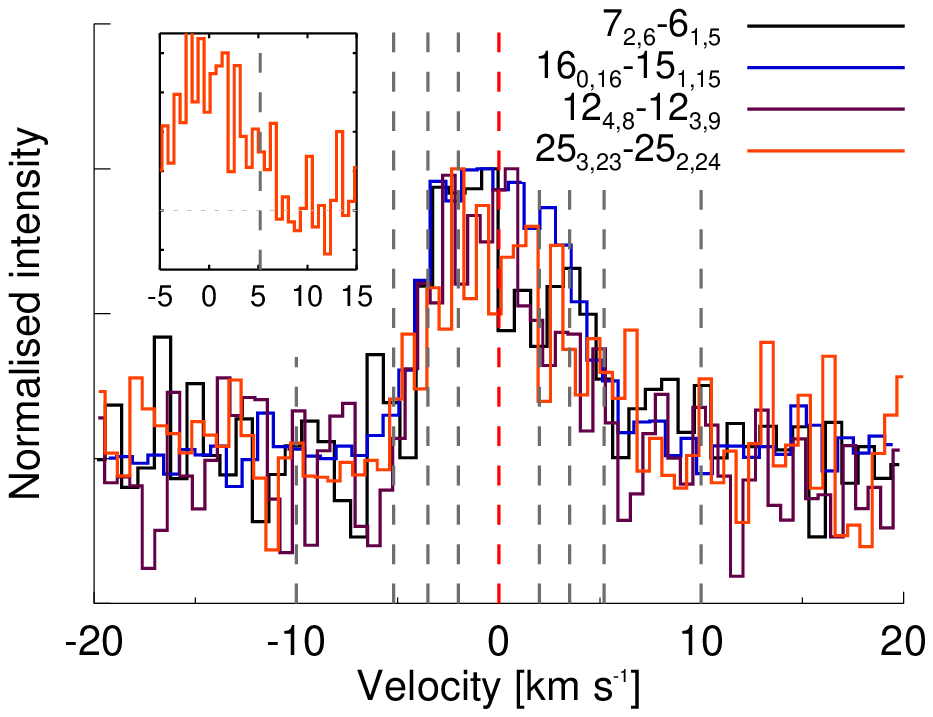}}%{SO2_selected_bump}}
\caption{Line shapes. Vertical dashed lines correspond to velocities $0$\,\kms, $\pm2.0$\,\kms, $\pm3.5$\,\kms, and $\pm5.2$\,\kms with respect to the stellar \vlsr, where distinct emission features appear in multiple of the presented lines. Insets (with the same colour coding as the main plots) are added to several of the panels for increased visibility of these substructures. \label{fig:bumps} }
\end{figure*}

\begin{figure}
\centering
\includegraphics[width=.9\linewidth]{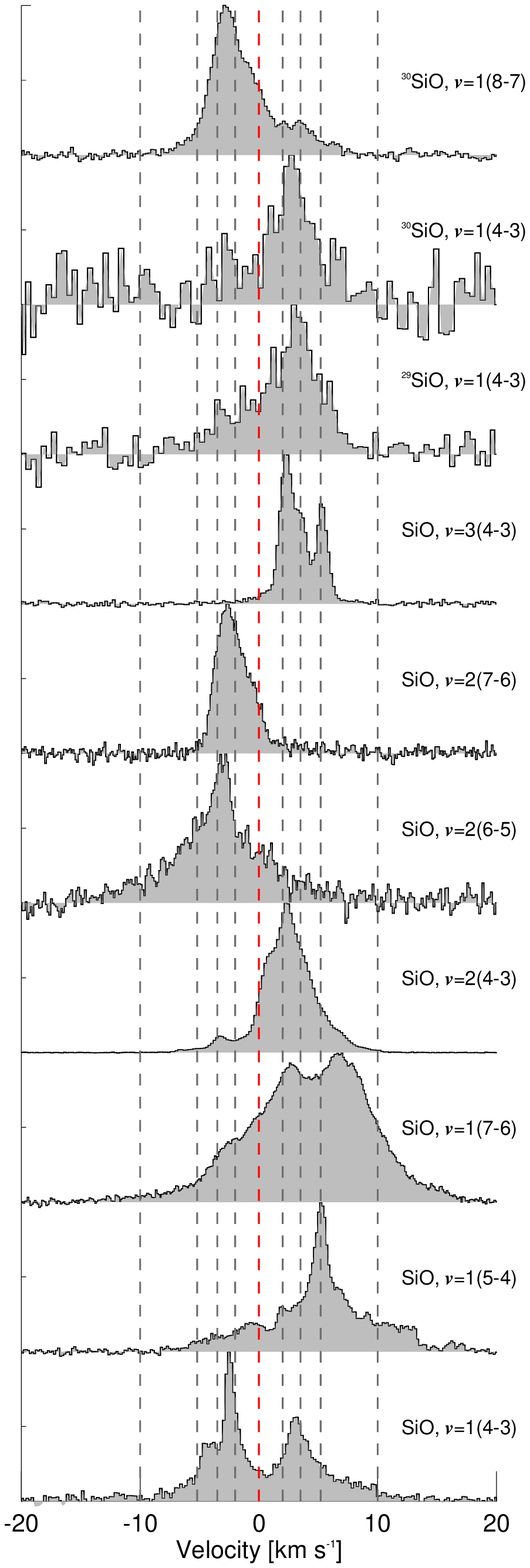}%{SiO_masers}
\caption{Selected SiO maser emission lines. Vertical dashed lines correspond to the velocities (relative to the stellar \vlsr) 0.0\,\kms, $\pm2.0$\,\kms, $\pm3.5$\,\kms, $\pm5.2$\,\kms, and $\pm10.0$\,\kms. Intensities are normalised to the peak and vertical offsets are added for visibility. We note that the observations are not coeval; see App.~\ref{sect:maserobs}.}\label{fig:SiOmasers}
\end{figure}

\section{Outflow kinematics}\label{sect:kinematics}

In case of a smooth, spherical wind described by a constant mass-loss rate, one expects a smooth line profile in the range flat-topped to parabolic for spatially unresolved emission, while spatially resolved emission leads to enhanced (in a relative sense) emission at the extreme velocities, where the lines may even become double-horned. The line profiles in the survey are largely represented by this range of profiles, indicating that the outflow of \rdor can, most likely, be approximated by a spherical wind. However, we do find multiple peaks in several line profiles, indicative of deviations from this simple structure. The CO emission lines observed with APEX and HIFI \citep{justtanont2012_orich} show a small bump at the stellar \vlsr and several features at other velocities.  In Fig.~\ref{fig:bumps} we show the presence of a distinct emission feature in several lines, pertaining to different species, at $-5.2$\,\kms with respect to the systemic velocity. This is clearest for the SiO lines, in particular for $^{29}$SiO and $^{30}$SiO ($J=4-3,\dots,8-7$; Figs.~\ref{fig:bumps_29sio} and \ref{fig:bumps_30sio}), and for $^{28}$SiO observed with HIFI ($J=12-11,\dots,28-27$; Fig.~\ref{fig:bumps_28sio_HIFI}). An analogous red-shifted component at $5.2$\,\kms is marginally visible in the HIFI observations of CO ($J=10-9$ and $16-15$; Fig.~\ref{fig:bumps_CO_HIFI}) and is more clearly seen in the emission of, for example, SO$_2$ ($7_{2,6-6_1,5},16_{0,16}-15_{1,15},12_{4,8}-12_{3,9},25_{3,23}-25_{2,24}$; Fig.~\ref{fig:bumps_SO2}). These lines have $E_{\mathrm{up}}/k$ of 36,  121, 111, and 326\,K, tracing a large range of excitation energies. Based on the above, the features at $\mbox{\vlsr}\pm5.2$\,\kms are probably real.  We additionally propose the existence of features at $\pm2.0$\,\kms and $\pm3.5$\,\kms with respect to the stellar \vlsr.  We show a selection of SiO maser emission lines in Fig.~\ref{fig:SiOmasers} and indicate the $\pm2.0$, $\pm3.5$\,\kms and $\pm5.2$\,\kms positions. These velocities in many cases correspond to individual maser peaks, or have peaks that fall exactly in between them, strengthening the idea that these velocities are linked to some physical structure in the outflow.

Furthermore, the o-H$_2$O($1_{1,0}-1_{0,1}$) line observed with HIFI and shown by \citet{maercker2016_water} shows clear bumps in the profile on both the red- and blue-shifted sides of the main emission component, at about $\pm 10$\,\kms. We see similar bumps for different lines in our survey, for example HCN($2-1$) and SO($6_5-5_4$), albeit at a low intensity level.  At $\pm10$\,\kms we also detect maser emission in, for example, SiO($v=1,J=5-4, 7-6$) and SiO($v=2, J=6-5$); see Fig.~\ref{fig:SiOmasers}. 
 
We conclude that the spectrally resolved emission lines of \rdor show signs of tracing up to five  components in the CSE of \rdor: \emph{(1)} a dominant component centred at the stellar \vlsr (6.5\,\kms), and components that arise at \emph{(2)}  $\pm$2.0\,\kms, \emph{(3)} $\pm3.5$\,\kms, \emph{(4)} $\pm5.2$\,\kms, and \emph{(5)} $\pm\sim10$\,\kms  with respect to the stellar \vlsr. From our spatially unresolved observations, we cannot readily draw conclusions on the geometry of these extra components. Their apparent symmetry around the \vlsr likely excludes that they are ``random'' inhomogeneities in the outflow and might imply that they are spatially confined in, for example, rings, shells, or spiral arms, but this is not an entirely straightforward explanation. The components are traced by line emissions of various molecular species that do not necessarily reside in the same parts of the CSE, and by transitions with very different excitation properties. Also, since the main wind has an expansion velocity of 5.7\,\kms \citep[e.g.][]{maercker2016_water} the extra emission at $\sim$10\,\kms poses a problem in a smooth, monotonously accelerating wind. We very carefully speculate that this could be a higher-velocity outflow component very close to the star, given that we see it in SiO maser line emission and \water line emission, both believed to originate close to the star. Observations that resolve the emission both spectrally and spatially are clearly necessary to constrain the kinematics and geometry of the outflow and its possible components.

\section{Conclusions}\label{sect:conclusion}
We present a spectral scan of the circumstellar environment of the oxygen-rich AGB star \rdor over the range $159.0-368.5$\,GHz, interrupted at $321.5-338.5$\,GHz. We carried out the observations with the APEX telescope, using the new SEPIA/Band-5 and the SHeFI facility instruments. This is the first survey of the circumstellar emission from this nearby, low-mass-loss-rate star over a large frequency range and is only the second such survey published in its entirety for an oxygen-rich AGB star after the one for \iktau by \citet{velillaprieto2017_iktau_iram}. Thus far, efforts have mainly been targeted at carbon-rich and high-mass-loss rate objects such as \irc or red supergiants such as \vycma.

The survey exhibits roughly 320 spectral features (omitting those linked to instrumental effects). The flux in the spectrum is heavily dominated by thermal and maser emission from the different SiO isotopologues. In numbers, the SO$_2$ emission lines dominate by far.  We detect several lines of CO ($^{12}$CO and $^{13}$CO), HCN (H$^{12}$CN and H$^{13}$CN), SiO ($^{28}$SiO, $^{29}$SiO, $^{30}$SiO, Si$^{17}$O, Si$^{18}$O), CN, H$_2$O, SO, and SO$_2$. We detect PO and PN, the latter through a stacked spectrum, for the first time towards this source, but cannot claim the conclusive detection of any other P-bearing molecules. We note that ALMA observations will be performed to study in more detail the phosphorous chemistry in the CSEs of \rdor and \iktau. We suggest the tentative detection of TiO$_2$, also from a stacked spectrum, AlO, and NaCl in the spectrum. These species are considered potentially important in the dust-condensation process. Sixteen features are currently still unidentified. Of these, one is a very strong maser at 354.2\,GHz, which is possibly identifiable as H$_2$SiO (silanone).  We could not confirm the presence of this emission in \rdor's spectrum at a later time, but are positive that we are dealing with a real spectral feature and not an instrumental effect of any type.

We present radiative transfer models for the thermal emission in the vibrational ground state ($v=0$) of five silicon monoxide isotopologues: $^{28}$SiO, $^{29}$SiO, $^{30}$SiO, Si$^{17}$O, and Si$^{18}$O.  Radiative transfer models of the SO and SO$_2$ emission in a large part of the survey was already presented by \citet{danilovich2016_sulphur}. 

We provide estimates for isotopic ratios for C, O, Si, and S, both from line-intensity ratios and from radiative transfer models. Using the derived circumstellar Si$^{17}$O/Si$^{18}$O as a proxy for the stellar $^{17}$O/$^{18}$O we constrain the initial mass of  \rdor to the range $1.3-1.6$\,\msun. 

We find detailed features in the emission line profiles that arise in many of the emission lines, both thermal and maser emission, spread throughout the full spectrum and also in several emission lines measured with Herschel/HIFI. We suggest that these could trace up to five components in the CSE of \rdor: \emph{(1)} a dominant smooth component centred at the stellar \vlsr, and components that arise at \emph{(2)}  $\pm2.0$\,\kms, \emph{(3)} $\pm3.5$\,\kms, \emph{(4)} $\pm5.2$\,\kms, and \emph{(5)} $\pm\sim10$\,\kms  with respect to the stellar \vlsr. The presence of these indicates possible deviations in the wind of \rdor from a smooth, spherical outflow. Spatially and spectrally resolved observations are needed to decipher what these components could be.

\begin{acknowledgements}
EDB acknowledges financial support from the Swedish National Space Board. HO acknowledges financial support from the Swedish Research Council. The APEX observations were obtained under project numbers O-087.F-9319A-2011,  O-094.F-9318A-2014, O-096.F-9336A-2015. The authors acknowledge John H. Black for his input to the molecular description of SiO used in the radiative transfer modelling.
\end{acknowledgements}

\addcontentsline{toc}{chapter}{Bibliography}
\bibliographystyle{aa}
\bibliography{32470.bib}

\begin{appendix}

\section{CO monitoring}\label{sect:COvariability}
\begin{figure*}
\centering
\begin{minipage}{0.3\linewidth}
\includegraphics[width=\linewidth]{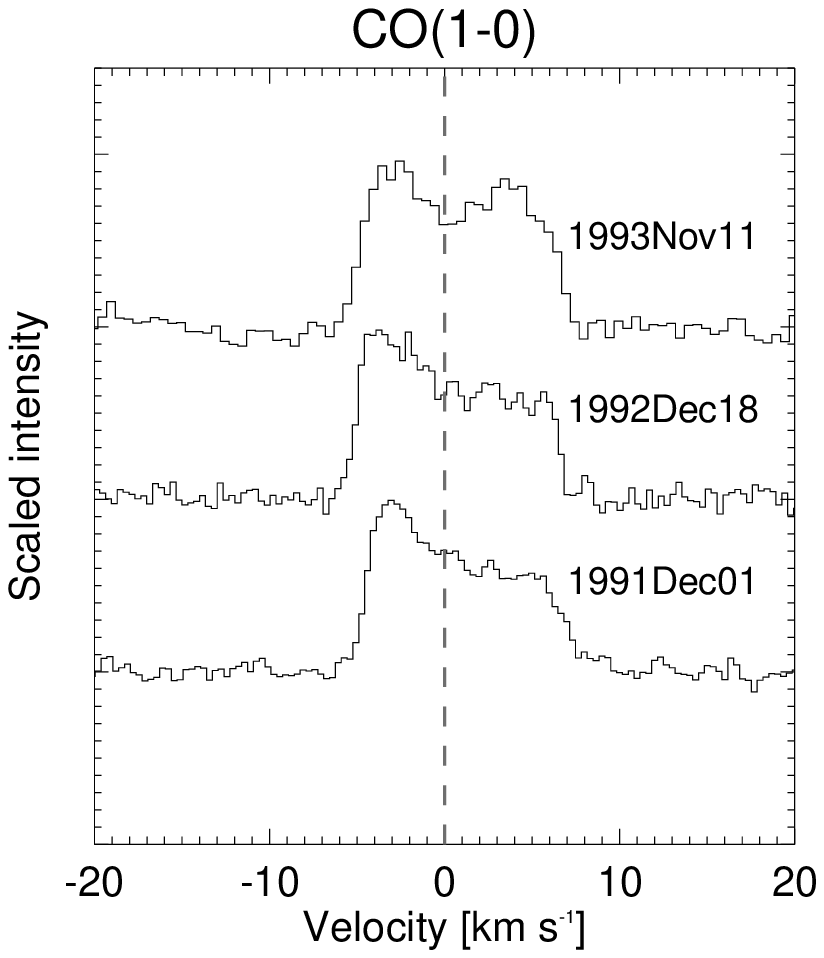}%{CO10_variations}
\end{minipage}
\begin{minipage}{0.3\linewidth}
\includegraphics[width=\linewidth]{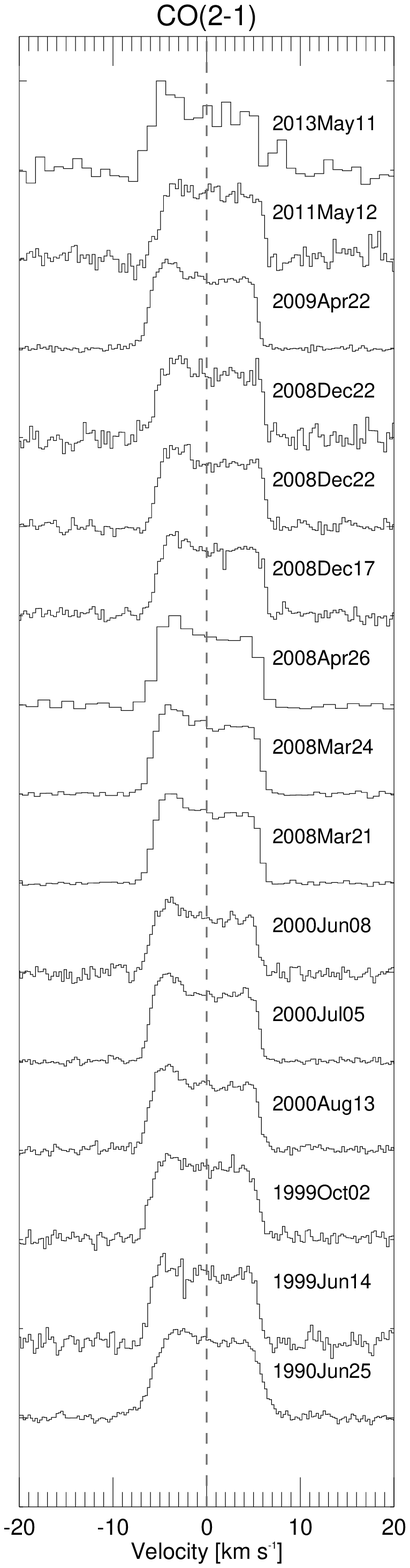}%{CO21_variations}
\end{minipage}
\begin{minipage}{0.3\linewidth}
\includegraphics[width=\linewidth]{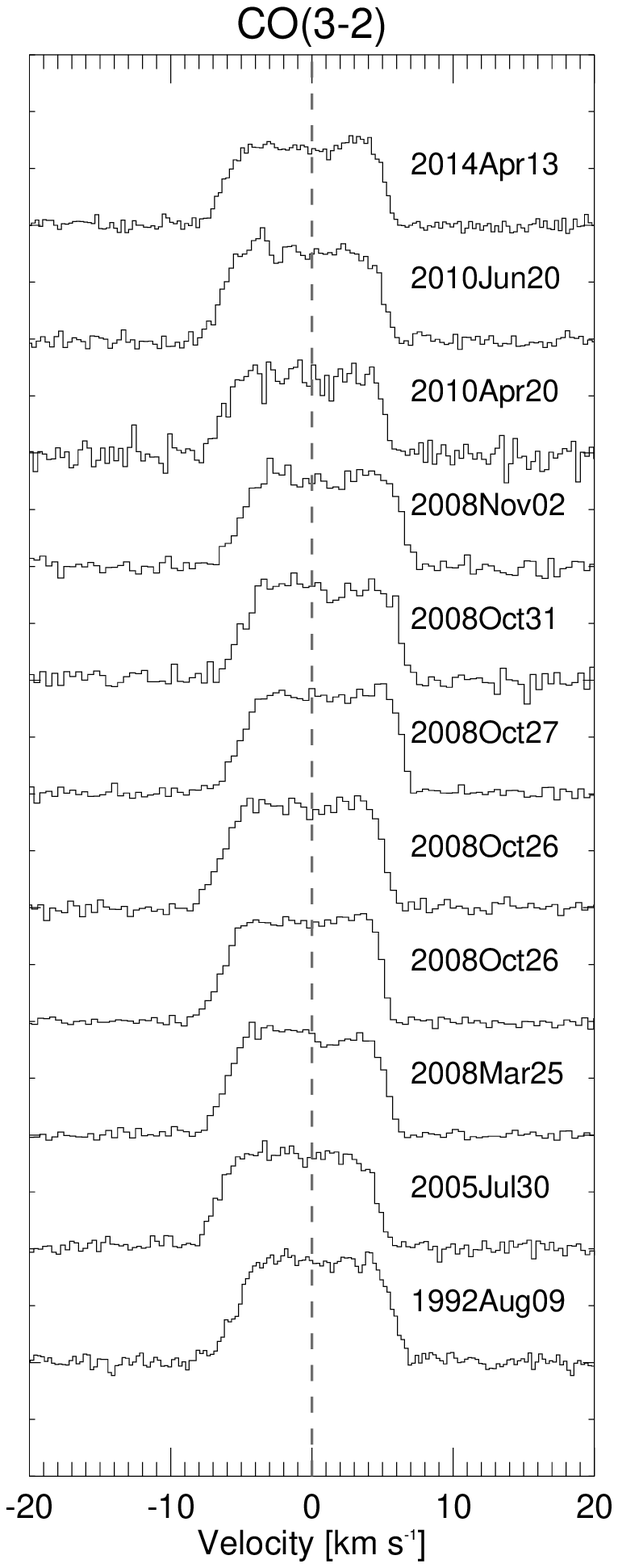}%{CO32_variations}
\\
\includegraphics[width=\linewidth]{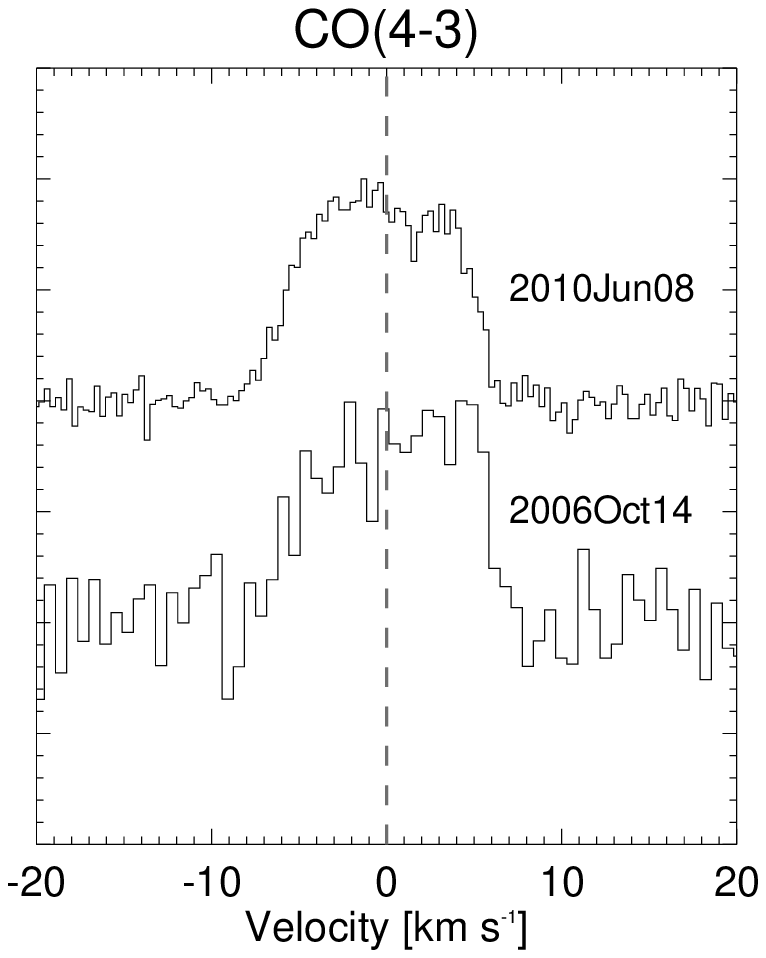}%{CO43_variations}
\end{minipage}
\caption{CO line emission from \rdor at different observing epochs: emission from $J=1-0, 2-1, 3-2$ measured by SEST in $1990-2000$, and from $J=2-1,3-2,4-3$ by APEX in $2005-2014$. The $y$-axis shows the line intensity at an arbitrary scale, chosen to optimise the visibility of the possible variations.}\label{fig:COvariability}
\end{figure*}

We show observations of CO emission with single-dish facilities in Fig.~\ref{fig:COvariability}. The CO($J=1-0, 2-1, 3-2$) emission was observed $1990-2000$ using the Swedish-ESO Submillimetre Telescope (SEST). The CO($J=2-1,3-2,4-3$) emission was repeatedly observed with APEX in the time frame $2005-2014$. The spectra at different epochs can be used to investigate possible variability in the CO emission of \rdor. Unfortunately, we do not have any recent observations of the $J=1-0$ line, whereas we mainly have recent observations for the higher-$J$ lines, and only a few or none from the 1990s, complicating a coherent time-variability study of the CO emission.

The $J=1-0$ spectra from the two earliest epochs (1991, 1992) agree very well. The third epoch (1993) shows the presence of an emission feature at $2-5$\,\kms red-shifted with respect to the systemic velocity, which is not seen in the earlier epochs. This "extra" peak matches quite well in velocity the features seen in, for example, the CO emission in our APEX survey data (Figs.~\ref{fig:CObumps} and \ref{fig:CO_zoom}) or in any of the $J=2-1,3-2,4-3$ spectra shown in Fig.~\ref{fig:COvariability}. We do not have a straightforward explanation for this change in the emission feature of the $J=1-0$ line. We remind the reader that the $e$-folding radius for the CO abundance distribution is about $1.6\times10^{16}$\,cm \citep[36\arcsec in diameter;][]{maercker2016_water} and that the emitting region of this transition covers a large part of the envelope. For this "extra" spectral feature to be a consequence of a morphological change in the emitting region, a large change within the emitting volume would be required when we assume collisional excitation of CO. The low expansion velocity of $\sim$6\,\kms leads to only $\sim$1\,AU radial motion over an entire year, the time between the second and third epoch of the observations, too little to cause a significant change at large radii. Unless a very strong, yet relatively small-scale, inhomogeneity started contributing significantly to the $J=1-0$ emission, a morphological argument seems hard to defend in the case of collisional excitation. However, as argued by  \citet{morris1980_co} and \citet{khouri2014_whya_co}, low-density winds like that of \rdor could give rise to a CO envelope that is not dominantly collisionally excited, but where the pumping of CO to its vibrationally excited state by the 4.6\,\um radiation from the star plays an important role. Another possible explanation would be that there is variable, weak maser emission in an inhomogeneous envelope \citep{morris1980_co}. However, as above, it seems unlikely that the properties of the emitting region would change very drastically within a year.

It is harder to make a case for any observed variability in the case of the $J=2-1,3-2,4-3$ lines. Although small changes do seem to occur between different epochs, it is less clear whether these are real or still within the general observational uncertainties. However, it does seem that the $J=3-2$ emission has been showing more prominent blue or red bumps depending on the epoch of observation. 

We cannot draw any firm conclusions on possible variability based on the single-dish spectra. Spatially resolved observations for one or multiple of these emission lines will be crucial to understand what could possibly be the cause of these changes, along with the origin of the components identified in Sect.~\ref{sect:kinematics}. 

\section{Maser observations}\label{sect:maserobs}
\begin{figure}
\centering
\subfigure[SiO, $v=2$ ($J=6-5$) at 256.9\,GHz]{\includegraphics[width=0.8\linewidth]{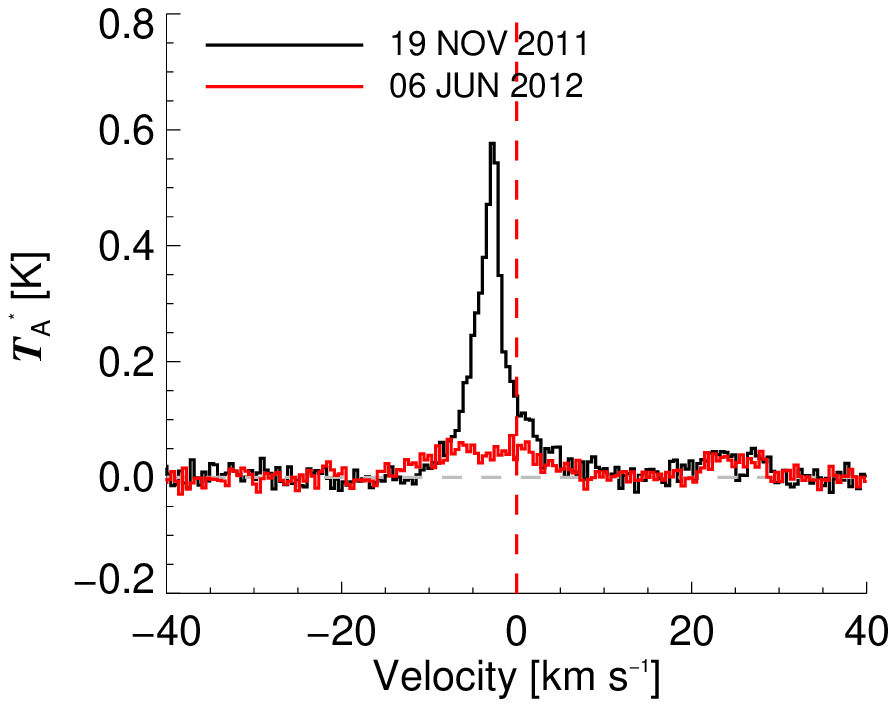}}%{maser_256898.391}}
\\
\subfigure[SiO, $v=1$ ($J=6-5$) at 258.7\,GHz]{\includegraphics[width=0.8\linewidth]{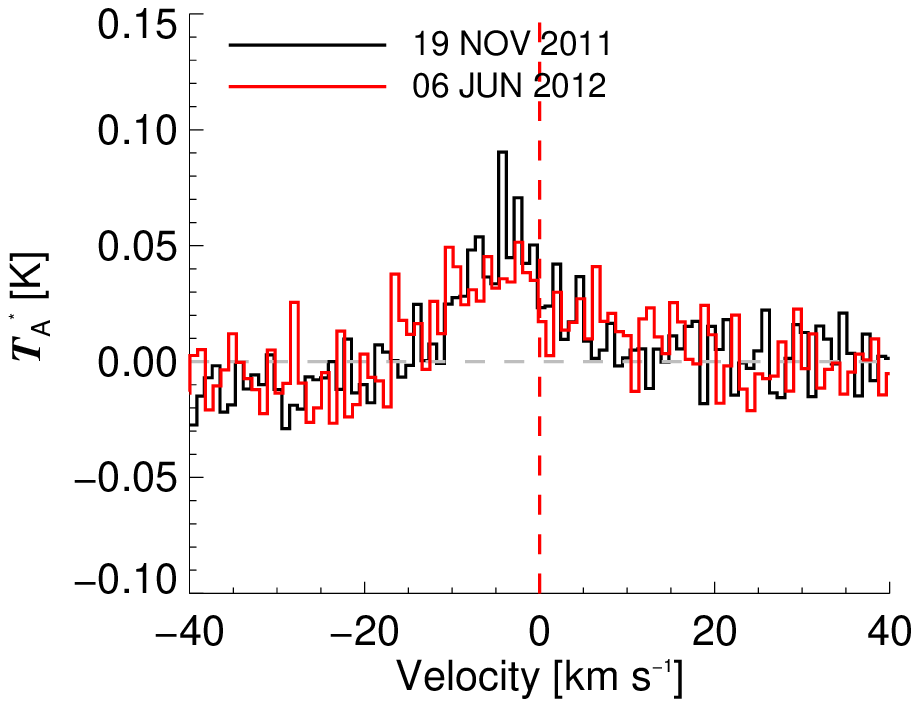}}%{maser_258707.328}}
\\
\subfigure[SiO, $v=1$ ($J=7-6$) at 301.8\,GHz]{\includegraphics[width=0.8\linewidth]{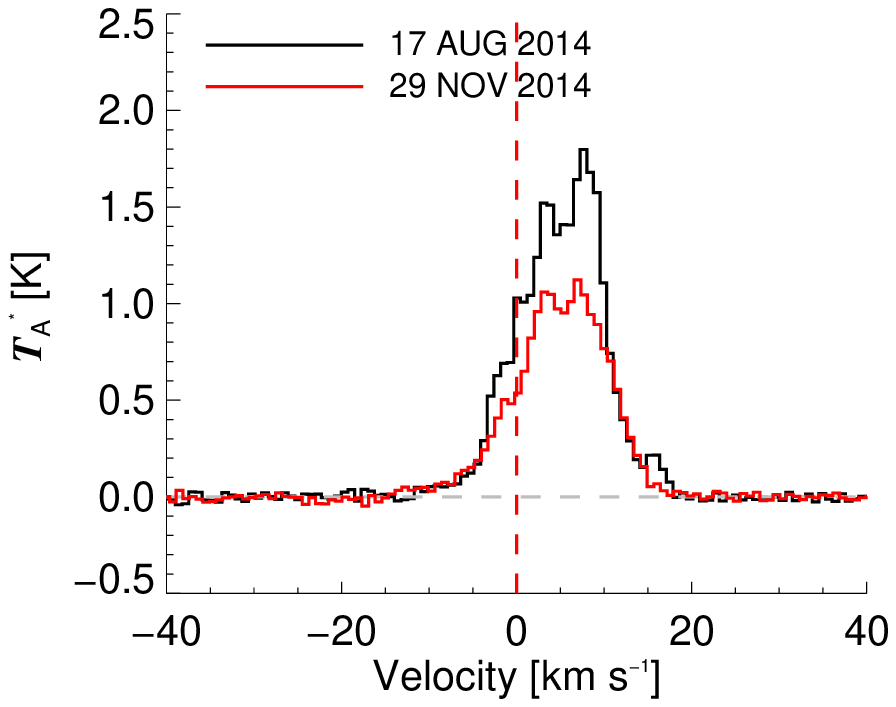}}%{maser_301814.344}}
\caption{Maser variability in our survey data. The rest frequency of each line agrees with a 0\,\kms velocity in these plots. \label{fig:maservariability}}
\end{figure}

With this paper we do not aim to study the maser excitation, but rather wish to report on the presence of the lines in the spectrum of \rdor. This is why all presented maser spectra in the main body of the paper are averages over all observations of that particular frequency.  However, maser emission around AGB stars is known to significantly vary with time and we did not obtain all observations simultaneously.   We therefore list the dates of observation for all maser lines in Table~\ref{tbl:maserdates}. Figure~\ref{fig:maservariability} presents the date-separated spectra of the maser lines observed on multiple days with significant S/N. We do not discuss any differences or implications of these in this paper. We note that we did not consider variations in those masers observed on consecutive days. If so wished, reduced spectra for separate dates of the other, much weaker, masers can be provided upon request.

\begin{table}
\caption{Observation dates for all maser lines in the survey, ordered according to increasing rest frequency.\label{tbl:maserdates}}
\centering
\begin{tabular}{p{1.75cm}p{1.45cm}rp{2.65cm}}
\hline\hline\\[-2ex]
Molecule        &       Transition      &       Frequency       &       Observation date    \\
&&(MHz) & \\
\hline\\[-2ex]
$^{30}$SiO, $v=2$       &       $4-3$   &       167160.943      &       22, 23 NOV 2015\tablefootmark{$\dagger$}    \\
$^{30}$SiO, $v=1$       &       $4-3$   &       168323.353      &       22, 23 NOV 2015\tablefootmark{$\dagger$}    \\
SiO, $v=3$      &       $4-3$   &       170070.348      &       22, 23 NOV 2015\tablefootmark{$\dagger$}   \\
$^{29}$SiO, $v=1$       &       $4-3$   &       170328.321      &       22, 23 NOV 2015\tablefootmark{$\dagger$}    \\
SiO, $v=2$      &       $4-3$   &       171275.165      &       22, 23 NOV 2015\tablefootmark{$\dagger$}   \\
SiO, $v=1$      &       $4-3$   &       172481.117      &       22, 23 NOV 2015\tablefootmark{$\dagger$}   \\
H$_2$O  &       $3_{1,2}-2_{2,0}$       &       183310.087      &       22, 23 NOV 2015\tablefootmark{$\dagger$}    \\
$^{30}$SiO, $v=2$       &       $5-4$   &       208946.055      &       22, 23 NOV 2015\tablefootmark{$\dagger$}    \\
SiO, $v=4$      &       $5-4$   &       211077.906      &       18 AUG 2014     \\
SiO, $v=3$      &       $5-4$   &       212582.550      &       18 AUG 2014     \\
SiO, $v=2$      &       $5-4$   &       214088.575      &       18 AUG 2014     \\
SiO, $v=1$      &       $5-4$   &       215596.018      &       15 MAY 2011     \\
H$_2$O, $v_2=1$ &       $5_{5,0}-6_{4,3}$       &       232686.700      &       03 SEP 2011        \\
SiO, $v=5$      &       $6-5$   &       251481.622      &       29 JUN 2015     \\
$^{30}$SiO, $v=1$       &       $6-5$   &       252471.372      &       29 JUN 2015        \\
$^{29}$SiO, $v=2$       &       $6-5$   &       253703.479      &       30 JUN 2015        \\
$^{29}$SiO, $v=1$       &       $6-5$   &       255478.495      &       30 JUN 2015        \\
SiO, $v=2$      &       $6-5$   &       256898.396      &       19 NOV 2011, \\
&&& 06 JUN 2012\tablefootmark{$\dagger$}        \\
SiO, $v=1$      &       $6-5$   &       258707.324      &       19 NOV 2011, \\
&&& 06 JUN 2012\tablefootmark{$\dagger$}        \\
H$_2$O, $v_2=2$ &       $6_{5,2}-7_{4,3}$       &       268149.117      &       13 JUN 2012, \\
&&& 22 NOV 2012\tablefootmark{$\dagger$}        \\
H$_2$O, $v_2=1$ &       $6_{6,1}-7_{5,2}$       &       293664.491      &       25 NOV 2011        \\
$^{29}$SiO, $v=3$       &       $7-6$   &       293907.859      &       25 NOV 2011        \\
H$_2$O, $v_2=1$ &       $6_{6,0}-7_{5,3}$       &       297439.276      &       25 NOV 2011        \\
SiO, $v=3$      &       $7-6$   &       297595.467      &       25 NOV 2011     \\
$^{29}$SiO, $v=1$       &       $7-6$   &       298047.637      &       25 NOV 2011        \\
SiO, $v=2$      &       $7-6$   &       299703.909      &       25 NOV 2011     \\
SiO, $v=1$      &       $7-6$   &       301814.332      &       17 AUG 2014,  \\
&&&29 NOV 2014\tablefootmark{$\dagger$} \\
H$_2$O  &       $10_{2,9}-9_{3,6}$      &       321225.677      &       24 NOV 2011        \\
$^{29}$SiO, $v=3$       &       $8-7$   &       335880.695      &       23 NOV 2011        \\
$^{30}$SiO, $v=1$       &       $8-7$   &       336603.002      &       23 NOV 2011        \\
SiO, $v=4$      &       $8-7$   &       337687.290      &       23 NOV 2011     \\
$^{29}$SiO, $v=2$       &       $8-7$   &       338245.183      &       23 NOV 2011        \\
SiO, $v=3$      &       $8-7$   &       340094.734      &       22 NOV 2011,  \\
&&&27 JUN 2015\tablefootmark{$\dagger$} \\
$^{29}$SiO, $v=1$       &       $8-7$   &       340611.884      &       22 NOV 2011,  \\
&&&27 JUN 2015\tablefootmark{$\dagger$} \\
SiO, $v=2$      &       $8-7$   &       342504.383      &       11 NOV 2011,  \\
&&&26, 27 JUN 2015\tablefootmark{$\dagger$}     \\
SiO, $v=1$      &       $8-7$   &       344916.332      &       11 NOV 2011, \\
&&& 26, 27 JUN 2015\tablefootmark{$\dagger$}    \\
u       &       ?       &       $\sim$354200    &       01 SEP 2011     \\
&&& 09 JUN 2016\tablefootmark{$\ddagger$}\\
\hline
\end{tabular}
\tablefoot{
\tablefoottext{$\dagger$}{Since we are currently not studying the detailed maser variability, observations carried out on multiple dates were averaged in the spectra presented in Fig.~\ref{fig:fullscan} and throughout the paper. If wanted, date-separated spectra can be provided by us or retrieved from the ESO archive. }
\tablefoottext{$\ddagger$}{Maser emission not detected on this date. Data not combined with original survey data.}
}
\end{table}

\section{APEX survey data}\label{sect:overview}
\begin{figure*}
  \includegraphics[width=.95\linewidth]{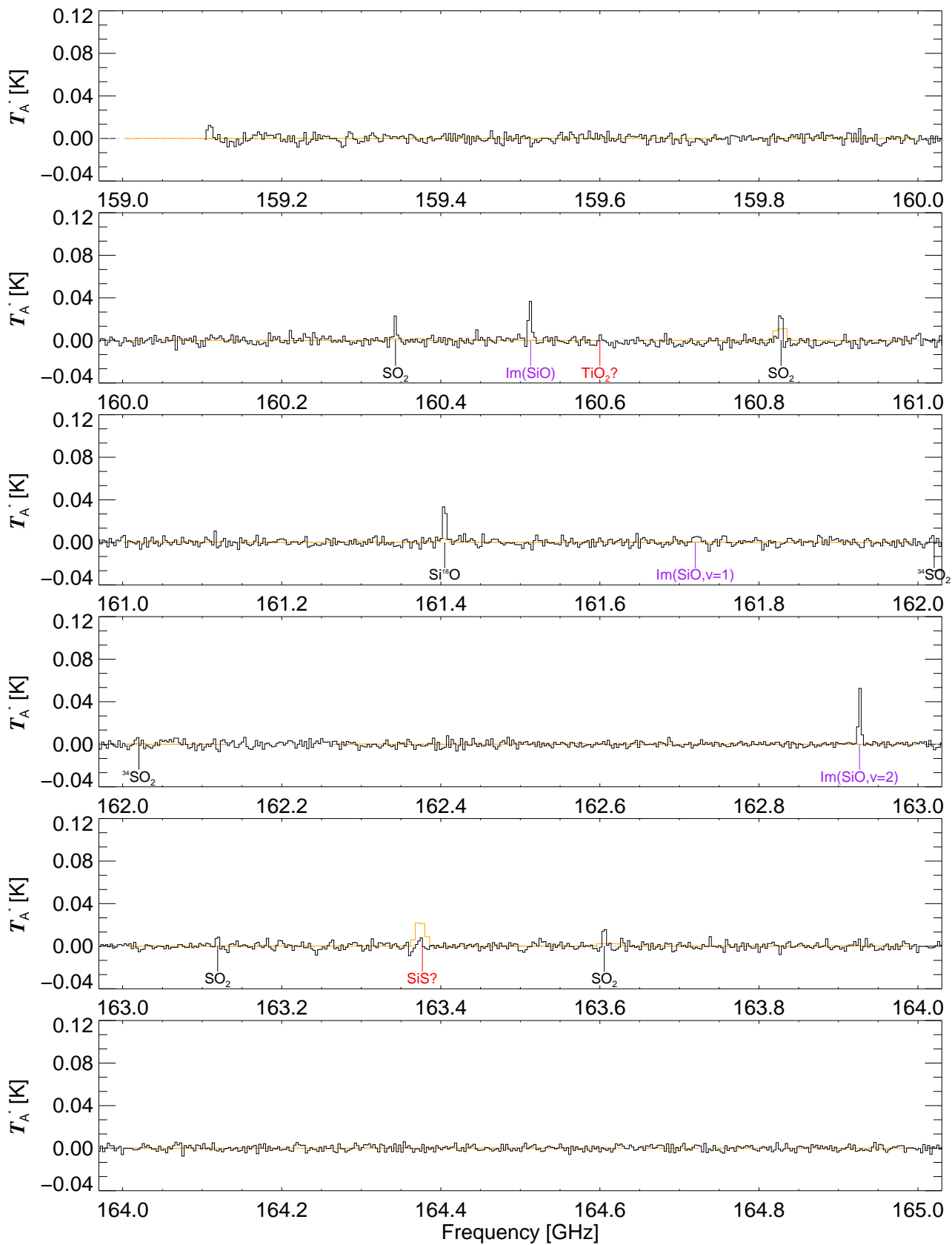}\par%{RDOR_IKTAU_comparison_1}
  \caption{APEX line survey of \rdor \emph{(black)}. Labels show the carrier molecule of the indicated emission. Red labels indicate tentative or unidentified detections. Purple labels (Im(...)) pertain to emission contaminating the signal from the image sideband (see Sect.~\ref{sect:observations}).   For visibility, some parts of the survey were rescaled to fit the vertical scale.  The colour coding of the spectrum corresponds to the following scale factors: \emph{(black)} 1; \emph{(green)} 1/5; \emph{(blue)} 1/25; \emph{(red)} 1/125.  Note the gap in the range $321.5-328.0$\,GHz.  We also show the  \irames survey of \iktau \citep[\emph{orange};][]{velillaprieto2017_iktau_iram} for direct comparison of the two data sets. The \iktau spectrum has been scaled according to the description in App.~\ref{sect:overview}. \label{fig:fullscan}}
\end{figure*}

We provide a complete overview of the survey in Table~\ref{tbl:lineID} and Fig.~\ref{fig:fullscan}.

To facilitate a direct comparison to the spectrum of \iktau, Fig.~\ref{fig:fullscan} also shows the \irames observations (in orange) of \citet{velillaprieto2017_iktau_iram}. We multiply the \iktau spectrum with a factor $A$ to account for the difference in distance ($d$) and mass-loss rate ($\dot{M}$) between the two targets, and for the frequency-dependent difference in point-source sensitivity ($\sigma_{\nu}$) between the two telescopes: 
\begin{equation} 
A = \frac{\dot{M}_{\rm R Dor}}{\dot{M}_{\rm IK Tau}}\times\left(\frac{d_{\rm R Dor}}{d_{\rm IK Tau}}\right)^{-2} \times \left(\frac{\sigma_{\nu,\mathrm{APEX}}}{\sigma_{\nu,\mathrm{IRAM}}}\right)^{-1} = 0.65 \times \left(\frac{\sigma_{\nu,\mathrm{APEX}}}{\sigma_{\nu,\mathrm{IRAM}}}\right)^{-1}.
\label{eq:iktau_scaling}
\end{equation}
The distances and mass-loss rates are those reported by \citet{maercker2016_water}. The point-source sensitivities $\sigma_{\nu,\mathrm{APEX}}$ are those listed in Sect.~\ref{sect:observations}; the values  for $\sigma_{\nu,\mathrm{IRAM}}$ are taken from \citet{velillaprieto2017_iktau_iram}.

\tiny
\longtab[1]{
\begin{landscape}
\begin{longtable}{rlcrrcrrrrrp{5cm}} 
\caption{Overview of lines in the APEX survey of \rdor. The columns list the line number $N_{\mathrm{line}}$, the molecule, the transition, the theoretical rest frequency, the energy of the upper level of the transition, a reference to the catalogue used for the identification, the velocity resolution at which the identification was made, the minimum and maximum velocities of the line, $\varv_{\mathrm{min}}$ and $\varv_{\mathrm{max}}$ (w.r.t. the line center), the peak flux value, the integrated flux value over thelisted $\varv$-range, and remarks.Catalogues are (1) CDMS and (2) JPL}\label{tbl:lineID}\\
\hline\hline\\[-2ex]
$N_{\mathrm{line}}$ & Molecule	&Transition 	& \multicolumn{1}{c}{$\nu_{\mathrm{lab}}$} & \multicolumn{1}{c}{$E_{\mathrm{up}}/k$} & \multicolumn{1}{c}{Cat.} 	& \multicolumn{1}{c}{$\Delta\varv$} &  $\varv_{\mathrm{min}}$ &  $\varv_{\mathrm{max}}$ &  \multicolumn{1}{c}{Peak} 	&\multicolumn{1}{c}{$I$}& Remarks \\
 & 			&            	& \multicolumn{1}{c}{(MHz)}                & \multicolumn{1}{c}{(K)}                  & 								&\multicolumn{3}{c}{(\kms)}    &  \multicolumn{1}{c}{(K)} 	&\multicolumn{1}{c}{(K\,km\,s$^{-1}$)}&  \\
\hline\\[-2ex]
\endfirsthead
\multicolumn{3}{l}{\textbf{{\tablename} \thetable{}.} Continued} \\[0.5ex]
\hline\hline\\[-2ex]
$N_{\mathrm{line}}$ & Molecule	&Transition 	& \multicolumn{1}{c}{$\nu_{\mathrm{lab}}$} & \multicolumn{1}{c}{$E_{\mathrm{up}}/k$} & \multicolumn{1}{c}{Cat.} 	& \multicolumn{1}{c}{$\Delta\varv$} &  $\varv_{\mathrm{min}}$ &  $\varv_{\mathrm{max}}$ &  \multicolumn{1}{c}{Peak} 	&\multicolumn{1}{c}{$I$}& Remarks \\
 & 			&            	& \multicolumn{1}{c}{(MHz)}                & \multicolumn{1}{c}{(K)}                  & 								&\multicolumn{3}{c}{(\kms)}     	&  \multicolumn{1}{c}{(K)} 	&\multicolumn{1}{c}{(K\,km\,s$^{-1}$)}&  \\
\hline\\[-2ex]
\endhead
\hline\\[-2ex]
\multicolumn{3}{l}{Continued on next page\ldots} \\[0.5ex]
\endfoot
\\[-1.8ex]
\endlastfoot
\footnotesize
  1  &                                  SO$_2$  &                   $18_{2,16}-18_{1,17}$  &          160342.990  &  170.8  &1  & 2.24  & -6.9  &  5.8  &   0.02  &   0.19  &                                                                                                                                                                                                                                                                                                                                                                                                                                                                                                                    \\
  2  &                                 Im(SiO)  &                                   $4-3$  &          160513.000  &  170.5  &1  & 2.24  &-24.8  & 15.5  &   0.04  &   0.31  &                                                                                                                                                                                                                                                                                                                                                                                                                                                                               Intensity at 2\% of signal intensity.\\
\end{longtable}
\normalsize%{APEX_all_lineIDs}
\end{landscape}
}
\normalsize

\newpage

\clearpage
\foreach \index in {2, ...,34}
{
\begin{figure*}
  \includegraphics[width=.95\linewidth]{32470_fgC1_\index.eps}\par %{RDOR_IKTAU_comparison_\index.eps}\par 
  {\textbf{Fig.~\ref{fig:fullscan}.} Continued.}
\end{figure*}
\clearpage
}

\end{appendix}

\end{document}